\documentclass[12pt]{book}
\usepackage{supertabular}
\usepackage{suthesis}
\usepackage{graphicx}
\usepackage{makeidx}
\usepackage{url}
\renewcommand{\thefootnote}{\fnsymbol{footnote}}
 
\newcommand{\cosb}{{\em COS-B}\/}
\newcommand{\oso}{{\em OSO-3}\/}
\newcommand{\sas}{{\em SAS-2}\/}
\newcommand{\cgro}{{\em CGRO}\/}
\newcommand{\egret}{{\em EGRET}\/}
\newcommand{\glast}{{\em GLAST}\/}
\newcommand{\batse}{{\em BATSE}\/}
\newcommand{\comptel}{{\em COMPTEL}\/}
\newcommand{\osse}{{\em OSSE}\/}

\newcommand{\einstein}{{\em Einstein}\/}
\newcommand{\sax}{{\em BeppoSAX}\/}

\newcommand{\likeprog}{{\tt LIKE}\/}
\newcommand{\unlikeprog}{{\tt unlike}\/}

\newcommand{\gismo}{{\tt gismo}\/}
\newcommand{\egs}{{\tt EGS}\/}
\newcommand{\gheisha}{{\tt GHEISHA}\/}
\newcommand{\glastsim}{{\tt glastsim}\/}

\newcommand{\gammaray}{$\gamma$-ray}
\newcommand{\gammarays}{$\gamma$-rays}

\newcommand{\perareasec}{~cm$^{-2}$~s$^{-1}$}

\newcommand{\perareasecsr}{~cm$^{-2}$~s$^{-1}$~sr$^{-1}$}

\newcommand{\etal}{et~al.}

\newcommand{\BO}{B$\emptyset$}

\newcommand{\eacc}{\'{e}}

\newcommand{\dbyd}[1]{\frac{\partial}{\partial #1}\hbox{}}
\newcommand{\deriv}[2]{\frac{\partial #1}{\partial #2}}

\def\cerenkov{\v{C}erenkov}

\def\bfxx{{\bf x}}
\def\bfx0{{\bf x}_0}
\def\bfxM{{\bf x}_M}
\def\bfxi{{\bf x}_i}

\def\like{\hbox{${\cal L}$}}

\newcommand{\chapt}[1]{Chapter~\ref{#1}}
\newcommand{\eq}[1]{equation~(\ref{#1})}

\newcommand{\fig}[1]{Figure~\ref{#1}}
\newcommand{\sect}[1]{\S\ref{#1}}
\newcommand{\tbl}[1]{Table~\ref{#1}}

\def\deg{\hbox{$^\circ$}}

\def\micron{\hbox{$\mu$m}}

\def\onehalf{1/2}

\def\onequarter{1/4}

\begin{document}
\widowpenalty=10000
\clubpenalty=10000

\title{Applications of Likelihood Analysis in Gamma-Ray Astrophysics}
\author{William Tompkins}
\dept{Physics}
\principaladviser{Peter Michelson}
\firstreader{Vah\eacc\ Petrosian}
\secondreader{Robert Wagoner}
\submitdate{March 1999}
\copyrightyear{1999}
 
\beforepreface
\prefacesection{Abstract}
The field of \gammaray\ astronomy relies heavily on the statistical
analysis of data.  Because of the paucity of data, and the often large
errors associated with detecting \gammarays, analysis and interpretation 
of the data require sophisticated statistical techniques.
Techniques for extending the currently used maximum likelihood
technique to more complicated data sets are presented.  Similarly
presented are methods of calculating the distribution and
behavior of the maximum
likelihood statistic used to measure source significance.

A new method for calculating source variability is also proposed,
and used to examine the sources found by the Energetic Gamma Ray
Experiment Telescope (\egret) on board the Compton Gamma Ray
Observatory (\cgro).  The results
show that the Active Galactic Nuclei (AGN), pulsars, and unidentified sources
have markedly different
variability, and that the unidentified sources fall into at least
two classes, differing in variability and spatial distribution.
A class composed of possible Supernova Remnant associations (SNR) is
distinctly more variable than the pulsars, but has a variability
consistent with that of the other low latitude unidentified sources.

Finally, the results of the Fall 1997 \glast\ prototype beam test
are presented.  The results of the beam test are compared with
simulated results, and found to be in remarkably good agreement.

\prefacesection{Acknowledgements}
As the eighth and final Stanford Ph.D. produced by the Stanford \egret\ 
project, I once again repeat our gratitude to those in the
collaboration who have made our work possible.  I am deeply indebted to
all of those people at
Goddard Space Flight Center, the Max Plank Institut f\"{u}r
Extraterrestriche Physik, Hampden-Sidney College, Grumman Aerospace,
and Stanford University who worked so hard for so many years to 
design, build, fly and support the \egret\ instrument.
Similarly, I would like to thank those whose analytical work I have tried
to build upon, specifically Tom Loredo, for his papers on Bayesian methods in Astrophysics,
and John Mattox, for his \likeprog\ program.

I have learned a great deal from many of those on the \egret\ team,
as well as the outside scientists who work in the high energy \gammaray\ 
region; too many people to list individually here.  But specifically,
I would like to thank Dave Thompson and Bob Hartman at Goddard for their
help and guidance, and their interest in my research.

The \glast\ collaboration is similar in scope, and again there are too
many to thank individually.  I would like to mention 
Steve Ritz, Toby Burnett, and Eric Grove as some of those who always
surprised me by listening as if I knew what I were talking about, and
who made me feel like a scientist rather than a grad student lackey.

Closer to home are the people I have
worked with at Stanford and SLAC.  Bill Atwood taught me a great deal
about particle physics, but more important to me was the implicit trust
he showed in my abilities as a physicist during the beam test.  No
mention of the beam test can be made without a reference to Chris
Chaput as well, without whom the whole process would have taken much
more time, and been much less fun.

I am frequently amazed at the freedom given me by Peter Michelson, my
advisor, to pursue those research areas which I have found most interesting.
In addition to finding the funding for all of my research, Peter has
always supported my efforts, even when they threatened to veer from
astrophysics into statistics.  Helping keep the Stanford group focused
on the astrophysics has been Y.C. Lin, who always seems to stay focused
on the science, even when the rest of us get bogged down in the systematics.

Pat Nolan wears many different hats in the Stanford group, and I owe him
thanks in each of his different roles.  As the institutional memory of the
Stanford \egret\ group, he has taught a succession of graduate students
about the instrument, about the analysis, and about the science.  He has
always been willing to help when I have difficulty, whether that is with
understanding a theory or proofreading a paper.  I knew that I was finally
ready for my Ph.D. when he actually asked {\em me} for advice.  In addition
to his role as a research scientist, Pat is invaluble as the person who
keeps the \egret\ computers running.  Without him, there would be chaos.

Each graduate student learns a tremendous ammount from the preceeding
grad students, and I am no exception.  Joe Fierro and Tom Willis were both
far more patient than I would have been, as they explained elementary
astronomy or the mysterious workings of \egret\ to me.  Mallory Roberts
was good to talk to for a non-\egret-team perspective.
The fellow graduate student to whom I owe the most, however, is Brian
Jones, a friend for nearly ten years, a roommate for three, and an
officemate for over five.  In our constant discussions and arguments
we managed to learn a great deal of physics and statistics.

Other friends in the Stanford Physics department have also been great
friends, most particularly Doug Natelson.  It was always refreshing to
hear the perspective of a non-astrophysicist.  Plus, he can make a good
curry.

My family has been tremendously supportive during my long trek towards a
doctorate.  Even if they weren't sure if I knew what I was doing, my
parents have always been great.

I end the list with both the most important scientist in my life, and
my wife-to-be: Barb Reisner.  I owe her more than words can express.

\clearpage
\null\vskip 3in
\centerline{For my parents}

\afterpreface
\chapter{The Sky in Gamma-Rays \label{sky}}

In the quest to understand the nature of the more enigmatic
astrophysical objects, astronomers have constantly sought new
sources of data.  One means of seeing the different aspects of
these objects is to look at them in different wavelengths, as was first done in
the 1930s when the radio range was first explored.  Later, with
the advent of spaceborne detectors, the X-ray, infrared, and ultraviolet regions
began to be explored in detail as well.
One of the last wavelength bands to be opened was the
\gammaray\ region, due to the low fluxes of cosmic \gammarays\ and the
high backgrounds with which the instruments must cope. A summary
of the instrumental history is given in \chapt{instruments}.

Each new band has presented a unique way of studying the universe,
and the high energy \gammaray\ region (20 MeV -- 100 GeV) is no
exception.  Many of the most energetic sources in the sky emit most of
their observable energy output in \gammarays.  To understand the 
source, one must understand the energy production.  And to measure the
energy production, one must look at the \gammarays.
In addition, \gammarays\ from distant objects generally proceed
unimpeded from their source to the earth.  A visible photon has a 99.995\%
chance of being absorbed as it passes through the central disk of our
galaxy, while a \gammaray\ photon has about a 1\% chance of interacting
\cite{fichtelbook}.
Thus, \gammarays\ come to us directly from their production site, allowing
astrophysicists access to the inner details of the objects they study.

\begin{figure}[tbhp]
\centering
\includegraphics[width = 4 in]{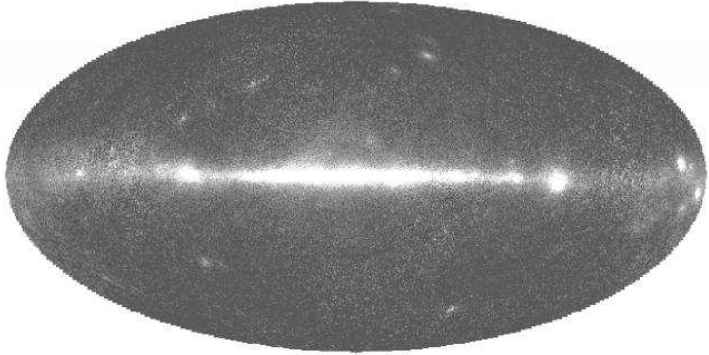}
\caption[The \gammaray\ sky above 100 MeV]{\label{allsky}
The \gammaray\ sky above 100 MeV, as measured by \egret.}
\end{figure}

\section{Diffuse Emission}
\subsection{Galactic Emission}

While dust and gas in our galaxy does not absorb the \gammarays\ from
distant objects, it does obscure them in a different way.  Cosmic rays
interacting with galactic gas make the galactic disk the brightest
\gammaray\ source in the sky, and can make it difficult to find
fainter sources.  But one person's background can be another person's treasured
signal:  this galactic emission (in conjunction with radio maps indicating
where the gas is) can be used to infer the cosmic ray density
function~\cite{bertsch_diffuse, hunter_diffuse, diffuse4compt}.

Because of the large point spread functions on \gammaray\ telescopes,
it is difficult to resolve
many of the weaker sources.  Some of the apparently diffuse galactic
emission will no doubt turn out to be from unresolved \gammaray\ 
pulsars (\sect{pulsars}).  In turn, some of the \gammaray\ sources 
currently in catalogs are probably due to enhanced diffuse emission,
rather than being from a compact object.

There is also a \gammaray\ source at (or near) the center of our galaxy
\cite{hanscenter}.
The strong diffuse emission makes it difficult to study, but it does
have a pulsar-like (and not diffuse-like) energy spectrum,
leading to the hypotheses that it might be multiple pulsars, or perhaps
diffuse emission from a region with a peculiar cosmic ray spectrum.
Other possibilities have been suggested, including that of emission from
a central black hole, or the possibility of dark matter annihilation
at the galactic core.  The merits of these various hypotheses are
discussed at length by Mayer-Hasselwander \cite{hanscenter}.

The Large Magellanic Cloud (LMC) is the only normal galaxy other than
our own which is visible in high energy \gammarays.
The emission from this dwarf companion galaxy to the Milky Way is
consistent with the expected diffuse emission from the same
cosmic ray interactions which produce the diffuse background in our
own galaxy \cite{kumar_magellanic, lin_southpole}

\subsection{The Extra-Galactic Background}

In addition to the diffuse Galactic \gammaray\ background, there is
an apparently isotropic background flux of \gammarays.
This presumably extragalactic
background has an energy spectrum consistent with that of the AGN, and
is probably composed mainly (or fully) of sub-threshold
AGN~\cite{kumar_extragalactic,tomthesis}.  It is interesting to speculate
on sources of a truly isotropic \gammaray\ background, but experimental
results will have to wait for the next generation of \gammaray\ telescopes.

\section{Point Sources}

\subsection{Pulsars \label{pulsars}}

The only identified compact sources of \gammarays\ in our galaxy are the
six \gammaray\ pulsars~\cite{joethesis,fierro_pulsarsII,nel_pulsarsIII,
nolan_survey,thompsonreview,thompson_pulsarsI}.  They exhibit fairly
hard spectra in the \egret\ energy range, and show little sign of
variability (\sect{variresults}).  Sophisticated models of the \gammaray\ 
production have been developed, notably
the outer gap~\cite{outergap,outergapII,romanimodel,romani}
and the polar cap~\cite{polarcap,polarcapII} models, but current
observations are not sufficiently precise to distinguish them.

\subsection{Active Galactic Nuclei}

Ninty-three of the 271 \egret\ sources are BL Lacertae (BL Lac),
Optically Violent Variable (OVV) Quasi-Stellar Objects (QSO), or high-polarization QSO,
which are collectively termed blazars, and one is the radio galaxy
Cen A~\cite{cat3}.  These objects are all termed Active Galactic Nuclei (AGN),
together with many objects which are not seen in \gammarays\ such as the Seyfert galaxies
(type I and II) and other types of QSO.
The unified AGN model indicates that all of the above, observationally different, objects
are in fact
same type of object: a supermassive black hole at the center of a galaxy, 
surrounded by a hot accretion disk, and producing
relativistic jets of charged particles.  This system is surrounded by a
torus of gas which obscures the central disk from many angles.  The observational
differences between the classes of AGN stem from the orientation of the jets compared
to the line of sight, as well as the size of the central black hole, and its
accretion rate.

The blazars are those AGN which have their jets pointed at us, and the
high energy emission is produced by the relativisic jet.  The most
plausible mechanism for the power-law emission seen at \egret\ energies
is the ``synchrotron self-Compton'' (SSC) process, whereby synchrotron photons
emitted by the electrons in the jet at radio to X-ray energies are scattered
by other electrons in the jet into \gammaray\ energies.

AGN are dramatically variable in \gammarays.  Flares typically last
tens of days, but variability is seen from day to day as well~\cite{bloom}.
The energy spectrum is typically a power law with index $\approx -2.1$.

\subsection{The Unidentified Sources}

Well over half of the \egret\ sources do not have identified counterparts
in other wavelengths.  This is primarily due to the large error regions
obtained: there are frequently multiple possible counterparts within the
95\% error ellipse.  Many of the high latitude unidentified sources will
no doubt be identified as AGN, but there is also a population of galactic
unidentified sources~\cite{kanbach_characteristics, grenier}.
Some of these are probably pulsars, but
many are not, as evidenced by their variability (\sect{variresults}),
energy spectrum~\cite{merck_unid, mukherjee_unid}, and the sheer number of
sources.
The puzzle of the nature of these unidentified sources is an
exciting challenge to the next generation of \gammaray\ telescopes.

\subsection{Gamma-Ray Bursts}

Since 1967, flashes of \gammarays\ have been observed from the sky.  These
bursts are often much brighter than the rest of the sky at their
peak, and show variation on a time scale of a few milliseconds.

\subsubsection{Soft Gamma Repeaters}
A few of these bursts, with softer energy spectra, were seen to burst
repeatedly from the same spot on the sky.
Four different ``Soft Gamma Repeaters'' have
been identified, three of which are identified with supernova remnants
\cite{sgr}.  The inferred neutron star origin of these burst has been
confirmed with X-ray observations.

\subsubsection{Classical Gamma-Ray Bursts}

More puzzling are the so-called ``classical'' gamma ray bursts, which
are much more numerous than the SGRs.  These are apparently
extragalactic in origin, implying that enormous energies are involved.
Most of the science on \gammaray\ bursts has been done at lower energies,
specifically by the \batse\ \cite{firstbatse, thirdbatse} and \sax\ \cite{frontera97} instruments,
but very large fluxes
above 30 MeV have been observed with \egret\ as
well~\cite{egret910601, catelli, dingus, egret940301,egret930131, egret910503}.  In addition,
\egret\ made the remarkable discovery of delayed high energy emission
from an apparently brief burst.  The observation
by \egret\ of an 18 GeV photon over an hour after a burst was seen
by \batse\ was quite unanticipated by theory~\cite{egret940217}.

\subsection{Local Gamma Ray Sources}
\subsubsection{The Sun}
The quiescent sun is not visible in \gammarays, although there
is presumably a low \gammaray\ flux from interactions between cosmic rays
and gas near the surface of the sun~\cite{moon}.  Solar flares, however, can be
extremely bright in high-energy \gammarays\ \cite{solarflare}.
Because the large flux of X-rays accompanying a flare
trigger the anti-coincidence
shield of a \gammaray\ telescope, it is difficult to measure the 
true flux from a bright flare.

\subsubsection{The Moon and Planets}

Cosmic ray interactions with the lunar surface produce a low flux of
high energy \gammarays,
which has been detected by \egret\ \cite{moon}.  In addition, lower energy
\gammaray\ 
spectrometers have been part of missions to the Moon, Mars, and a near
Earth asteroid.  The \gammaray\ lines detected have been used to measure the
isotopic and elemental abundances in these objects, and
thus yield clues to their formation.

\subsubsection{The Earth}
The material at the surface of the Earth produces these same
low energy \gammaray\ lines.  In addition, flashes of low energy \gammarays\ from
lightning have been observed by \batse.  The largest source of terrestrial source of
\gammarays\ for an orbiting telescope, however, is the atmosphere.  Cosmic ray
interactions with the Earth's atmosphere produce showers containing
a very large flux of high energy \gammarays.  These showers proceed
in the same direction as the cosmic ray, and must originate in the
atmosphere: thus these \gammarays\ are seen to come from the edge of the
Earth's disk, rather than uniformly across the disk \cite{sas2albedo}.

The rejection of this \gammaray\ background is based on direction.  The
combination of directional cuts, together with the significant width of
the point spread function for \gammaray\ telescopes can make analysis
difficult, and lead to spurious detections if one is not
careful~\cite{tomthesis}.

\chapter{Spaceborne Gamma-Ray Telescopes \label{instruments}}
\section{Principles of Gamma-Ray Detection}

There are a number of different methods for detecting \gammarays.
At lower energies (100 keV -- 30 MeV),
either crystal scintillators or semiconductors are typically used.
Both types of detectors rely on the ionization caused by the
passage of the gamma ray: in a semiconductor the electrons and holes
are collected with an electric field, while in a scintillator they produce 
visible photons which are then amplified with a photomultiplier tube.
The energy of the gamma ray can be measured by the ionization response.
A number of different means can be used to measure the direction
of the gamma ray, including the use of a collimator, a coded aperture
in conjunction with a position sensitive detector, multiple position
sensitive layers (measuring two successive Compton scatters at the higher energies),
or time of flight information between physically separated detectors.

At higher energies ($>10$ MeV), the dominant interaction in a high-Z
material is the conversion of the \gammaray\ into an $e^+/e^-$ pair.
A succession of detectors which measure the ionization produced by the
pair of charged particles can reconstruct incident \gammaray\ direction.
The energy of the \gammaray\ can be measured by the scattering
of the $e^+$ and $e^-$ tracks in the detector material,
but better measurements can be obtained with a
calorimeter (generally a scintillator) positioned behind the tracking
detector.
\section{Early Gamma-Ray Instruments}
\subsection{Balloon experiments}
In the 1940's, experiments began to be flown on balloons, airplanes,
and rockets to determine the nature of cosmic rays, using
Geiger-M\"{u}ller tubes \cite{schein}, ionization chambers \cite{hulsizer},
and cloud chambers \cite{crichfield}.  Although experimentalists searched
for a primary \gammaray\ component, the difficulties presented by
the low rate of celestial \gammarays\,, the high backgrounds,
and immature balloon technology defeated experimentalists for twenty
years.
After the first satellite detections, a balloon-borne experiment
did confirm the Galactic plane emission in 1970 \cite{kniffensatellite}.
Another balloon experiment provided the first evidence of pulsed
\gammarays\ from the Crab pulsar in 1971 \cite{browning}.

\subsection{Early satellites}

The first detection of astrophysical \gammarays\ was in 1962, when
an instrument on board the moon probe {\em Ranger 3}
detected an apparently isotropic flux of \gammarays\  \cite{arnold}.  In 1968,
an experiment on board the third Orbiting Solar Observatory (\oso)
detected a total of 621 events above 50 MeV, with an apparent excess
along the Galactic plane and at the Galactic center \cite{clark}.  

\begin{figure}[tbhp]
\centering
\includegraphics[width = 4in]{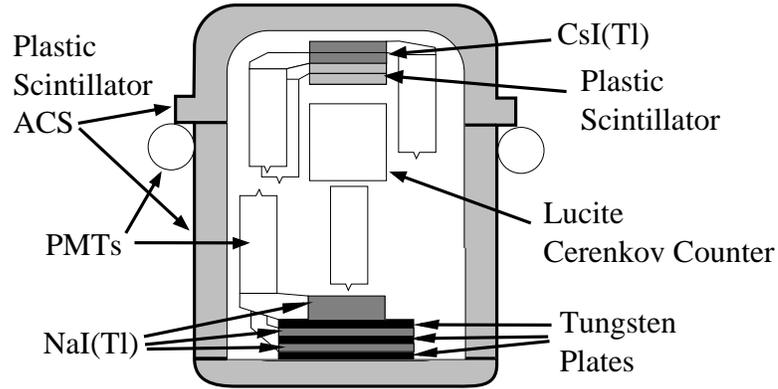}
\caption[\oso]{\label{oso}
Schematic diagrams of the \oso\ instrument.
}
\end{figure}

The \oso\ instrument
consisted mainly of scintillation detectors read out by photomultiplier
tubes.
The initial conversion was detected
with four successive plates of scintillator: two of CsI followed
by two plastic planes (\fig{oso}).
The pair then passed through a \cerenkov\ 
telescope composed of a lucite block and a photomultiplier tube, which
ensured that the particles were directed down, rather than up.
A calorimeter composed
of a block of NaI followed by alternating layers of tungsten and NaI
was used for an energy measurement.  The entire apparatus was enclosed
in plastic scintillator used as an anti-coincidence shield.  Because the
telescope did not track the positions of the $e^+/e^-$ pair, it did not
make any measurement of the direction of the \gammarays.
The limited directional response of about $15\deg$, however, acted as a collimator.

The instrument suffered from calibration problems which were resolved
with the aid of the following balloon-borne instruments \cite{kraushaar}.
The effective area at multi-GeV energies was approximately 9 cm$^2$.

\subsection{\sas\ and \cosb}
\begin{figure}[tbhp]
\centering
\includegraphics[width = 4in]{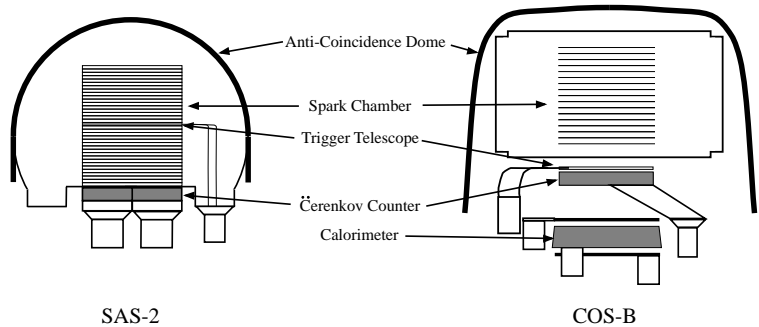}
\caption[\sas\ and \cosb]{\label{sascos}
Schematic diagrams of the \sas\ and \cosb\ instruments.
}
\end{figure}

In 1972, the second Small Astronomy Satellite (\sas), was launched, 
containing a \gammaray\ telescope with a peak effective area of
about 115 cm$^2$.  As important as the increased effective area was the
directional resolution of the on-board spark chamber (\fig{sascos}): about
$5\deg$ at 30 MeV and $1\deg$ at 1 GeV.

The \sas\ detector measured the $e^+/e^-$ tracks in the spark chamber
to provide both the direction of the incoming \gammaray\,, as well as
the energy (by measuring the scatter of the track).  A set of scintillator
tiles in the middle of the spark chamber, together with the \cerenkov\ 
detector below, provided the trigger for the spark chamber.

Due to the failure of a power supply, \sas\ only took data for slightly
over six months.  During its operation, it detected over 8000 \gammaray\ events,
and identified emission from the Crab \cite{sascrab} and Vela pulsars
\cite{sas_vela}.
In addition, it found the 
the first \gammaray\ source which was not immediately associated with 
an object detected in other wavelengths:
Geminga~\cite{gem_discoveryI, gem_discoveryII}.

The \cosb\ instrument, launched in 1975, was similar in design and
capability to the \sas\ mission.  It did incorporate a calorimeter
(\fig{sascos}), and thus
provided improved energy resolution.  The seven year lifetime of the
\cosb\ instrument provided much improved measurements of the previously
detected sources, as well as establishing many new detections, resulting
in 17 sources detected along the galactic plane~\cite{cosb,hanscosb}.  

\section{\cgro: a New Era in Gamma-ray Astronomy}

\begin{figure}[tbhp]
\centering
\includegraphics[width = 4in]{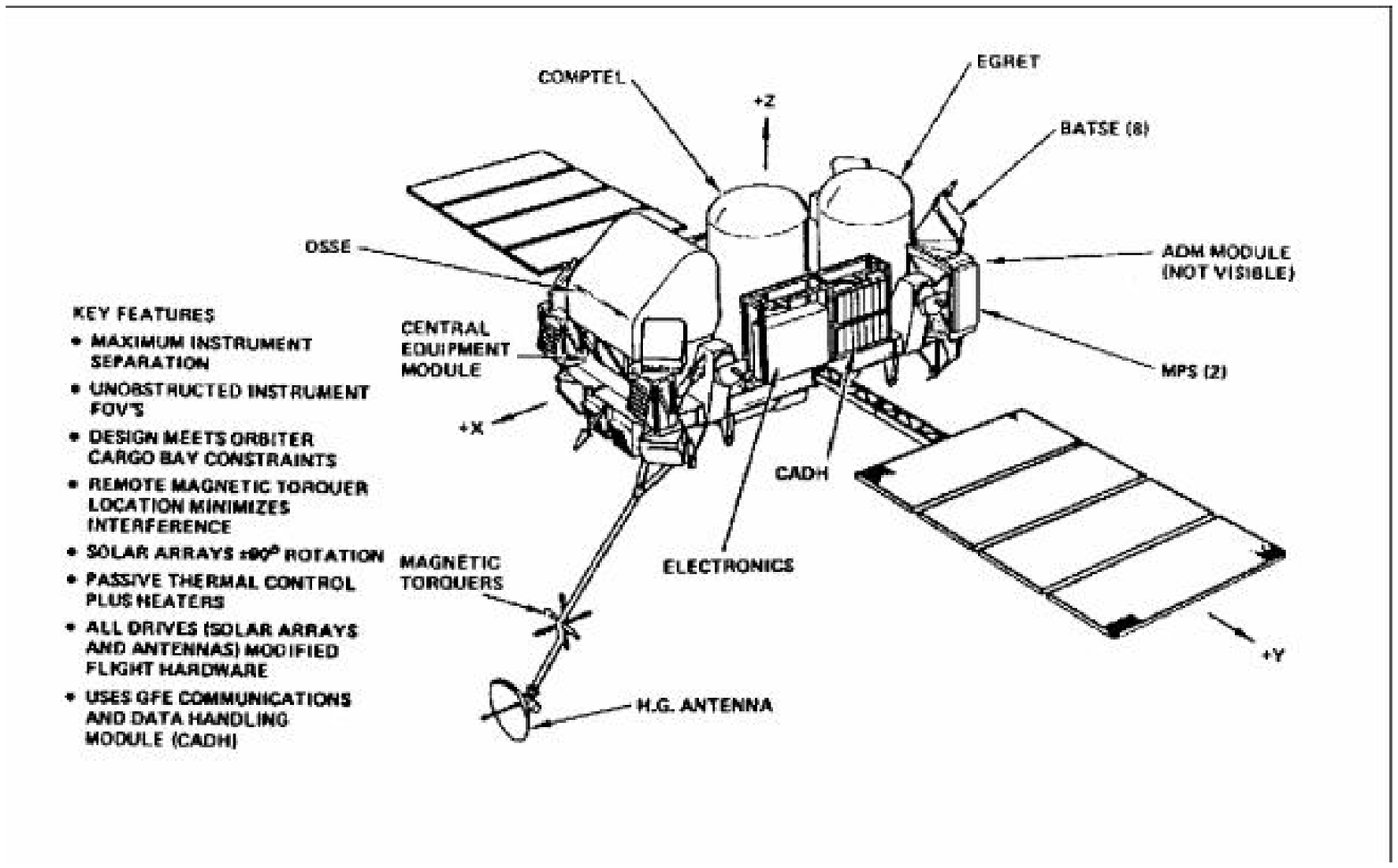}
\caption[\cgro\ satellite]{\label{cgro}
The \cgro\ satellite.  The \batse\ detector is comprised of eight distinct
pieces on the corners of the spacecraft bus, while along the top of
the satellite are \egret\ and \comptel, pointing in the Z direction, and
\osse, which has its own orientation capability.
}
\end{figure}

In April 1991, the Compton Gamma Ray Observatory (\cgro) was launched
on board the Space Shuttle Atlantis (STS-37).  A
product of the NASA {\em Great Observatories for Space Astrophysics} program,
the instruments on board were designed to study the \gammaray\ sky from
 25 keV to more than 10 GeV.  Because of the large energy coverage, four
different \gammaray\ telescopes were put on board, each designed for a different
energy range and purpose (\fig{cgro}).

\subsection{\batse}
The Burst And Transient Source Experiment (\batse) is a unique
\gammaray\ telescope.  Designed to measure quickly varying signals in the
25 keV -- 2 MeV region, the \batse\ detector consists of eight
scintillation counters.  By comparing the measurements of the different
detectors, the position of a flaring source can be ascertained to within
about $8\deg$.

\subsection{\osse}
The Oriented Scintillation Spectrometer Experiment (\osse) measures from 
50 keV to 10 MeV.  With an energy resolution of about 8\% and a field of
view of 3.8$\deg$ x 11.4$\deg$, it is designed to detect emission lines.

\subsection{\comptel}
The Compton Telescope (\comptel) measures double Compton scattering events
in two scintillators separated by 1.5 m.  In this way, it obtains both the
direction and energy of the incident \gammaray.  Its energy range is governed
by the region where Compton scattering is a significant interaction process
of \gammarays: from 1 -- 30 MeV.

\section{High Energies with CGRO: \egret}

\begin{figure}[tbhp]
\centering
\includegraphics[width = 4in]{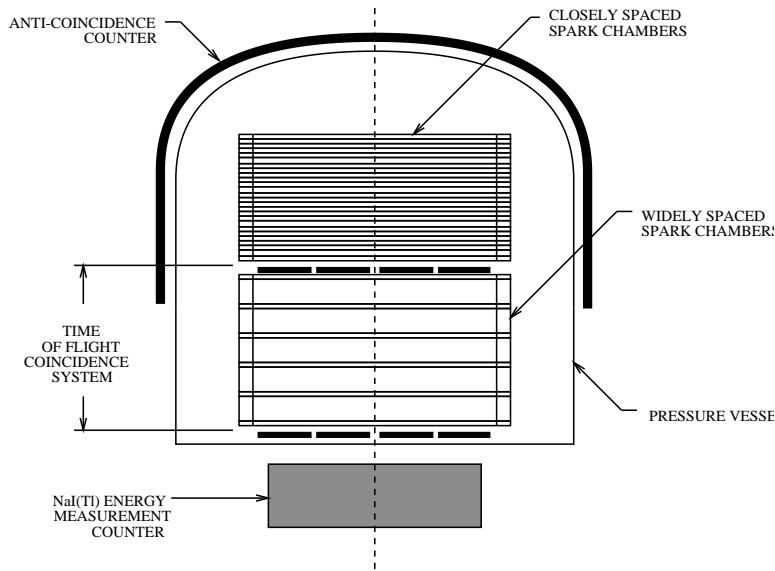}
\caption[\egret\ schematic]{\label{egret}
Schematic diagram of the \egret\ instrument.
}
\end{figure}

The Energetic Gamma Ray Experiment Telescope (\egret) is the successor to
the earlier \sas\ and \cosb\ instruments (\fig{egret}).
Similarly designed with a magnetic
core spark chamber, it differs mainly from these previous instruments in
its size (\fig{all}).  The triggering of \egret\ is somewhat different as
well:  the requirement that the secondary particles have downward trajectories
is enforced by a time of flight (TOF) system rather than a \cerenkov\ telescope.
The segmentation of the two TOF planes into two 4x4 arrays of scintillator
permits directional triggering as well \cite{egretcalibrate93}.

Of the 256 possible upper-tile/lower-tile combinations, 96 are able to trigger 
\egret.  These correspond to those directions from which the particle would
have left a track in the spark chamber.  These 96 combinations were grouped
into 74 instrument configuration modes \cite{egretcalibrate93}.  In addition to the typical
mode which triggers on any \gammaray\ which converted in the tracker,
there were other modes designed to limit exposure in a given
direction.  Thus, the telescope can be pointed near the horizon without
being overwhelmed by triggers from Earth albedo \gammarays.  Because of
failures in the photomultiplier tubes, extra configuration modes have been
added during the \egret\ mission, resulting in 87 different configurations \cite{inflightcalibrate}.

The \egret\ tracker consists of 28 closely spaced spark chamber modules,
interleaved with sheets of tantalum (Ta).  The Ta (Z = 73) provides the
most likely interaction point for an incoming \gammaray; above the
critical energy of 10.7 MeV, this is most likely to be a pair conversion event.
Each spark chamber module consists of two orthogonal sets of wires.  When
triggered by the TOF system (and not vetoed by the anti-coincidence dome),
a high voltage is applied across the sets, leading to a spark where the
the interior gas has been ionized.  The spacing of the wires (0.8 mm) sets
the nominal accuracy of the track measurement.

The Total Absorption Shower Counter (TASC) sits below the TOF system, and
is comprised of approximately 8 radiation lengths of NaI read out by
photomultiplier tubes.  It measures the
energy of the particle pair with a resolution of about 20\% FWHM when both
the e$^+$ and e$^-$ are intercepted.  At low energies, this measurement
needs to be corrected for the energy lost by the pair as they traverse the
spark chamber.  At the highest energies, leakage from the TASC becomes
an important issue.

Events in \egret\ are time-tagged with an accuracy of 100 $\mu$s, which
permits the study of the lightcurve shape of the \gammaray\ pulsars.
The instrumental dead time of 100 ms after each event is rarely a concern,
given the low flux of high energy \gammarays.  During extremely bright
solar flares and bursts, however, \egret\ has been dead
time-limited~\cite{solarflare, egret930131}.

The \egret\ instrument was calibrated before launch at both the
Stanford Linear Accelerator Center
and the MIT Bates Linear Accelerator \cite{egretcalibrate93},
resulting in extensive tables of the instrument response to
\gammarays\ with various energies and incident directions for
all of the various triggering modes.
Further calibration studies have been performed in orbit to
ensure that degradation over its nearly eight year lifetime (much longer than
the planned mission length) does not affect the accuracy of the
scientific results~\cite{inflightcalibrate}.

The scientific results from the \egret\ mission are numerous, including
the first all-sky survey at high energy \gammaray\ wavelengths, and
comprise nearly all of the high energy results discussed in \chapt{sky}.

\section{The Future: \glast \label{glastintro}}

\begin{figure}[tbhp]
\centering
\includegraphics[width = 4in]{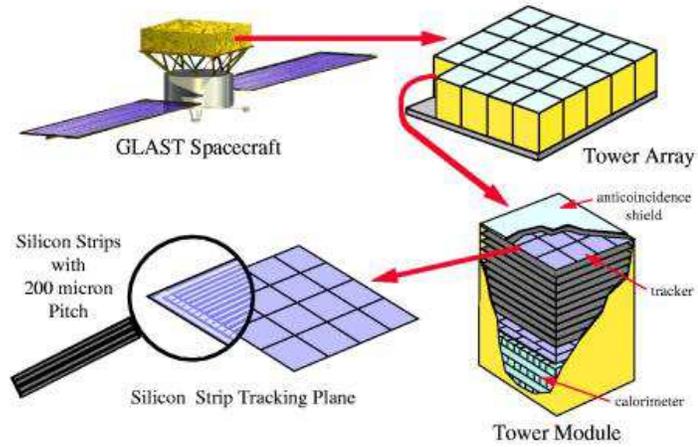}
\caption[\glast]{\label{glast}
Artist's illustration of the \glast\ instrument and spacecraft.
}
\end{figure}

\begin{figure}[tbhp]
\centering
\includegraphics[width = 4in]{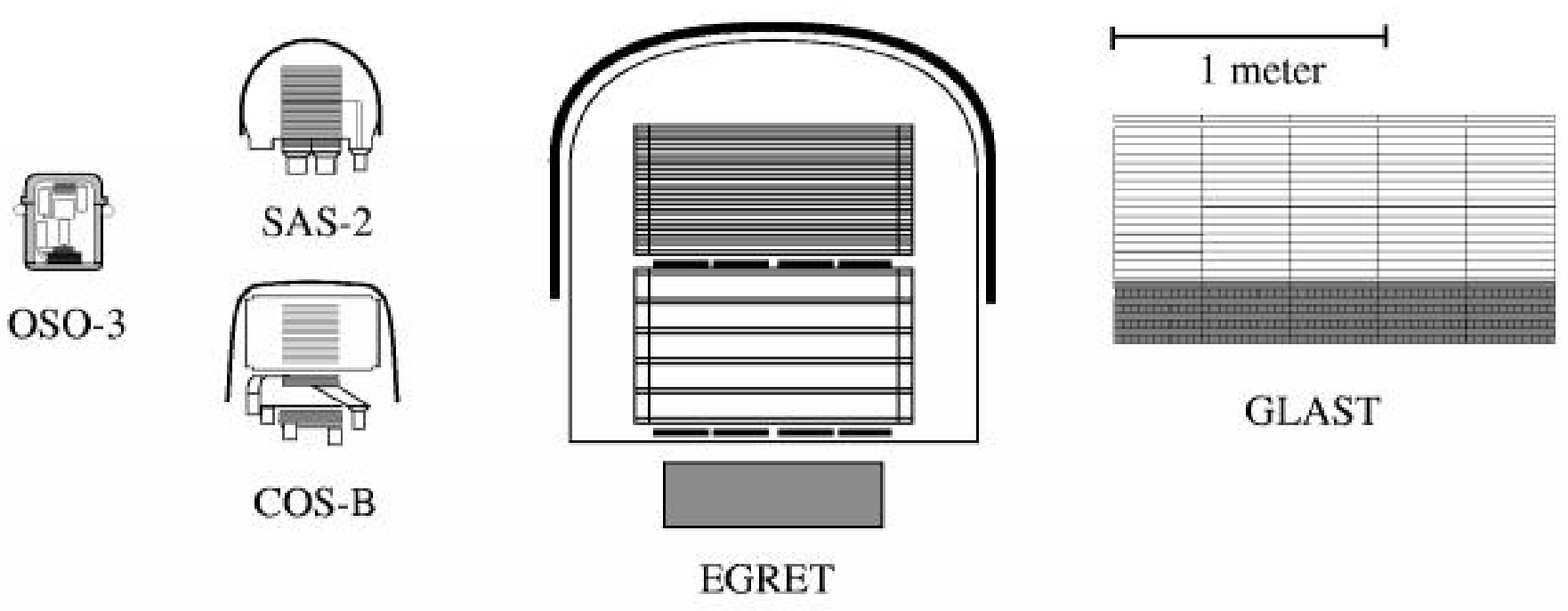}
\caption[\sas, \cosb, \egret, and \glast]{\label{all}
Schematic diagrams of the \sas, \cosb, \egret, and \glast\ instruments,
to scale.
}
\end{figure}

While the \egret\ instrument improved on the previous \gammaray\ 
telescopes mainly in its size, the proposed Gamma-ray Large Area Space
Telescope (\glast) gains primarily through improved technology.
The proposed \glast\ baseline instrument\footnote{Alternative technologies
have been proposed for the telescope, including scintillating fiber and
gaseous detectors.  These technologies are somewhat less developed than the
silicon strip and CsI detectors discussed here, and thus are not used in
this baseline design.} is composed of a 5x5 array of towers (\fig{glast}).
Each tower
has 16 planes of silicon strip detectors, used to track the e$^+$/e$^-$
pair, a hodoscopic calorimeter with eight layers of CsI logs, and an
anti-coincidence detector (ACD) composed of plastic scintillator 
\cite{glast:proposal95,glast:atwood94,glast:bloom,glast:doe_proposal}.
The striking
differences between the \glast\ instrument and the previous telescopes are
the flattened aspect ratio due to the lack of a time of flight or \cerenkov\ 
detector, and the lack of a monolithic anticoincidence shield (\fig{all}).

These differences are due to the on-board computing power available
to the \glast\ instrument.  The discrimination between upward- and downward-going
events is handled by on board event reconstruction.  Similarly, the rejection
of the hadronic background is accomplished by projecting the track back into
the segmented ACD, and checking for a hit there.  The segmentation of the ACD
prevents a veto of the event by backsplash X-rays from the calorimeter triggering
the ACD.
These structural changes permit a
much wider field of view for the instrument, and a higher sensitivity,
particularly at high energies where low energy photons produced by the shower
in the calorimeter can trigger an anti-coincidence shield.

The technological improvements are not limited to the computers, however.
The silicon strip detectors, with a strip pitch of 200 \micron, provide a
much more accurate measurement of the particle tracks than did the
spark chambers, resulting in a significantly improved point spread function.
The use of these detectors also eliminates the need for the gas used in a spark chamber,
as well as the need for high voltages on the instrument, both of which are
difficult to deal with on board a spacecraft.
The partition of the calorimeter into 2.75 cm wide logs, in alternating directions,
permits particles to be tracked in the calorimeter as well.  This will
permit the study of \gammarays\ which convert in the calorimeter, and
thus greatly increase the sensitive area at high energies.

\glast\ will have an on axis sensitive area of approximately 6000 cm$^2$ at 100
MeV, over seven times the sensitive area of \egret.  This factor of improvement
increases significantly at higher and lower energies: \glast\ will be have an
energy range from approximately 20 MeV to several hundred GeV.  The point spread
function (PSF) at low energies will be less than half as wide as
the \egret\ PSF , and at
high energies less than one sixth as wide.  Similarly, the energy resolution
will be improved by a factor of two.  As significant as these improvements is
the much wider field of view, with more than 5 times the solid angle
coverage of \egret.

The combination of these improvements will certainly revolutionize the field
of \gammaray\ astronomy.  \glast\ will detect several thousand AGN
\cite{tomthesis}, and with
its large sky coverage can monitor them for flares.  \glast\ will also 
detect many new \gammaray\ pulsars, while the much improved high energy performance will
permit exciting new studies of the Crab, Vela, and Geminga pulsars.  Perhaps
most exciting are the new classes of sources which \glast\ will see: with
its smaller point spread function, \glast\ will be able to see some of the
structure of extended sources, and will be able to pinpoint the locations of
the \egret\ unidentified sources, enabling their identification with sources
at other wavelengths.

\chapter{Likelihood Analysis of \egret\ and \glast\ data \label{likelihood}}
\section{Principles of Likelihood Analysis}
The use of the likelihood function in data analysis extends back through
the work of Bayes \cite{bayes} in 1763, while extensive development of the
Maximum Likelihood statistic began with Fisher~\cite{fisher21} in
1921.  Detailed explanations
of both the Bayesian and Frequentist uses of likelihood can be found
elsewhere~\cite{kendall,eadie}, and there will not be an attempt to detail
the myriad results here.  Rather there will a small introduction, suitable
for understanding the other results presented here.

In most places in this thesis, the Maximum Likelihood statistic
is used, rather than
a Bayesian analysis.  This is generally because it is easier computationally
to maximize a function of many variables than to integrate it.  In addition,
the use of a flat prior (which a Bayesian would claim is implicit in the
use of Maximum Likelihood) seems reasonable for the analysis in question.

\subsection{The Likelihood Function}
Given a data set $D$ and a parametrized model $M({\bf x})$,
 the likelihood of the model is simply the probability of obtaining the
data set if the model is true: 
\[ \like (M({\bf x})) = P(D | M({\bf x})). \]
In general, the value of the likelihood itself is not meaningful,
but the ratio of two likelihood values does have meaning, and
can be used to determine which of the two models is more likely to be correct.

Thus the value of ${\bf x}$ which maximizes the likelihood function is
taken to be the estimate of the true value,  and the more quickly the
likelihood function falls off from this maximum, the smaller the confidence
regions will be.

It is typical in Maximum Likelihood work to use logarithms of the
likelihoods, which renders some of the computations easier and makes them
less prone to numerical error.  Thus we have the log likelihood ratio
for comparison of two models:
\[  \ln \like \left(M_1 \right) - \ln \like \left( M_2 \right) .\]

When evaluating methods of estimation, the important properties of the
calculated estimate of a parameter (or ``estimator'') are
\begin{itemize}
\item{\em Consistency}-- That the estimator converges to the true value as more
data are accumulated.
\item{\em Efficiency}-- The rate at which the estimator converges to the true
value.  If this is equal to the Rao-Cram\`er-Frechet bound (the maximum
rate of convergence), the estimator is said to be {\em efficient} \cite{eadie}.
\item{\em Bias}-- The expectation of the difference of the estimator and the
true value.  
\item{\em Robustness}-- The insensitivity of the estimator to errors in
the assumed probability distributions (such as noise).
\end{itemize}

It can be proved~\cite{kendall} that if a consistent, efficient estimator exists,
then the Maximum Likelihood estimate is both consistent and efficient.
The Maximum Likelihood estimate may be biased, however, and its robustness depends
on the models used.  If the estimate is biased, an unbiased estimator can be
constructed from the Maximum Likelihood estimate; however, the unbiased estimate is not
efficient.

\subsection{Confidence Regions \label{confregions}}

An easy way of determining a confidence region for the model parameter
${\bf x}$ is given in Eadie~\cite{eadie}.  We begin by noting the
distribution function of $\chi^2(N)$ (where N is the number of ``degrees
of freedom''): \[
f_{\chi^2(N)} (x) = \frac{ \left( \frac{x}{2}\right)^{(\frac{N}{2})-1} e^{-\frac{x}{2}}}{2 \Gamma\left(\frac{N}{2}\right)} ,\]
and defining $\chi^2_\alpha \left(N\right)$ through
\[ \alpha = \int_0^{\chi^2_\alpha(N)} f_{\chi^2(N)}
(x) dx , \]
where $0 \leq \alpha \leq 1$.

This function is used to define an $\alpha$--confidence region, or that region which
has probability $\alpha$ of containing the true value:
\begin{quote}
{ \em
  The $\alpha$--confidence region is the locus of values of ${\bf x}$ where
   $2 \ln \like (M({\bf x}))$ is within
  $\chi^2_\alpha \left(k\right)$ of the maximum likelihood value,
  where $k$ is the number of parameters (the dimensionality of ${\bf x}$).
}
\end{quote}

Strictly speaking, the above definition of the confidence region is an
approximation, true as the amount of data becomes large.  For small data
sets, Eadie, \etal~\cite{eadie} provide corrections to this interval.
There are other confidence regions which can be defined (see \sect{feldman}),
but this procedure provides the smallest possible region.

\subsection{Wilks' Theorem \label{wilks}}

An equivalent statement, known as Wilks' theorem~\cite{wilks,cash,eadie}
is:
\begin{quote}
{ \em
  If the model $M({\bf x_0})$ (termed the Null model) is correct,
  and the maximum likelihood value obtained by allowing the k parameters
  of $\bfxx$ to vary is $\like(\bfxM)$,
  then the statistic
  \[ C = 2 \left[ \ln \like(M(\bfxM)) - \ln \like(M(\bfx0)) \right] \]
  is asymptotically distributed like $\chi^2(k)$ as the size of the data
  set increases.
}
\end{quote}
There is an important caveat, however.  The proof of this theorem requires
the inverse of the matrix
$ \left( \frac{\partial^2 \ln \like(\bfxx)}{\partial x_i \partial x_j} \right)_{\bfxx = \bfx0} $, which may be singular (e.g. if $\frac{\partial \like(\bfxx)}{\partial x_i} = 0$ for some parameter $x_i$ ).  Thus there are cases where
an alternate distribution for the maximum likelihood ratio must be found
(c.f. chapter \ref{tsdist}).


\section{Calculations of Likelihood \label{calculation}}

Likelihood analysis has been used for analysis of \gammaray\ data for
both the \egret\,, \comptel\,, and \cosb\ instruments~\cite{like,comptel,pollock},
as well as X-ray instruments
like the \einstein\ observatory~\cite{pollockeinstein}.
Binned data has typically been used, and models (often quite elaborate)
of the non-source background have been constructed.  This section examines
the computations of likelihood which are used in such analysis, and develops
some mathematical tools which can be used to investigate the capabilities
of such analyses.

\subsection{Unbinned Likelihood}

The model predicts a rate of \gammarays\ as a function of direction (galactic
longitude and latitude, or $\ell$ and $b$), energy ($E$),
and time ($t$): $R(\ell, b, E, t)$.  If we construct bins of size $d\ell \times \,db \times \,dE \times \,dt$,
where each bin is sufficiently small, the rate will be roughly constant
across each bin, and there the chance of getting more than one photon in a bin
can be made arbitrarily small.  Using the properties of the Poisson distribution,
we can calculate
the likelihood of getting a photon in a bin:
\[ P_1 = R(\ell, b, E, t) \,d\ell \,db \,dE \,dt \times e^{-R(\ell, b,E,t) \,d\ell \,db \,dE \,dt}, \]
and the likelihood of not getting a photon in a bin:
\[ P_0 = e^{-R(\ell, b,E,t) \,d\ell \,db \,dE \,dt}. \]

If we write the set of bins with a photon as $P$, and the set with no
photons as $Q$, we can write the likelihood of the entire data set as
\begin{eqnarray}
\like(R) & = & \prod_{i \in P} R(\ell_i, b_i, E_i, t_i) \,d\ell \,db \,dE \,dt\,e^{- R(\ell_i, b_i, E_i, t_i) \,d\ell \,db \,dE \,dt\ } \nonumber \\
& & \times \prod_{i \in Q} e^{- R(\ell_i, b_i, E_i, t_i) \,d\ell \,db \,dE \,dt\ } \nonumber\\
& = &  \prod_{i \in P} R(\ell_i, b_i, E_i, t_i) \,d\ell \,db \,dE \,dt \times \prod_{i \in P,Q} e^{- R(\ell_i, b_i, E_i, t_i) \,d\ell \,db \,dE \,dt\ }.
\end{eqnarray}
Taking the logarithm of both sides, letting sums become integrals
where appropriate, and writing the number of photons as $N$, we have
\begin{eqnarray}
\ln \like(R) & = & \sum_{i \in P} \ln R(\ell, b, E, t) +
 \sum_{i \in P}\ln (d\ell \,db \,dE \,dt) \nonumber\\
& & - \sum_{i \in P,Q} R(\ell_i, b_i, E_i, t_i) \,d\ell \,db \,dE \,dt\ \nonumber \\
& = &  \sum_{i \in P} \ln R(\ell_i, b_i, E_i, t_i) +
 \left [ \ln \left( d\ell \,db \,dE \,dt \right) \right]^N \nonumber \\
& &  -  \int R(\ell, b, E, t) \,d\ell \,db \,dE \,dt . \label{infp}
\end{eqnarray}

The second term in \eq{infp} is independent of the model parameters: it
depends only on the data.  It will drop out when we calculate likelihood
ratios or look for the maximum likelihood model.  Thus we may discard it
and write simply
\begin{equation}
\ln \like(R)  =  \sum_{i \in P} \ln R(\ell_i, b_i, E_i, t_i)
 -  \int R(\ell, b, E, t) \,d\ell \,db \,dE \,dt \label{almostdone}.
\end{equation}

The integral term in \eq{almostdone} is just the total number of
photons predicted by the model, which we can write as $N_{pred}$.  We
are left with
\begin{equation}
\ln \like(R)  =  \sum_{i = 1}^N \ln R(\ell_i, b_i, E_i, t_i)
 -  N_{\rm pred}, \label{unbinnedeq}
\end{equation}
where $\ell_i, b_i, E_i,$ and $t_i$ denote the position, energy, and time of the
$i$'th photon.

The computation time needed to evaluate $\ln \like$ for a model is dominated by the
calculation of $R$, and will thus scale
linearly with the number of photons detected.  In general, a fixed amount of
time will also be required to calculate $N_{\rm pred}$ for a given model as well.

\subsection{Binned Likelihood}

If we do not make the bins of the previous section infinitesimal,
we must integrate to find the rate in each bin:
\[ r_j = \int_{{\rm bin}\,j} R(\ell, b,E,t) \,d\ell \,db \,dE \,dt . \]
Using the Poisson distribution
\[ P(n, \mu) = \frac{ e^{-\mu} \mu^n}{n!} \]
as the probability of observing $n$ counts when the rate is $\mu$, we can write
the likelihood
\begin{equation} \like = \prod_j \frac{ e^{-r_j} r_j^{n_j}}{n_j!} ,
\end{equation}
where $n_j$ is the number of photons in bin $j$.

Taking the logarithm, we have
\[ \ln \like = \sum_j {-r_j} + n_j \ln r_j - \ln n_j!. \]
As in the previous section, we may omit the last term, as it is model independent,
and note that the first term is the total number of counts predicted:
\begin{equation} \ln \like = \sum_j n_j \ln r_j - N_{\rm pred}. \label{binnedeq}
\end{equation}

This result is quite similar to the unbinned result.  In fact, if we sum over
counts, rather than bins, the only difference is
the replacement of $ \ln R(\ell_i, b_i, E_i, t_i) $ in \eq{unbinnedeq} with
$ \ln r_i $ in \eq{binnedeq}, where $r_i$ is the rate in the bin containing photon $i$.
  Thus the difference is in replacing the modeled rate for the exact direction, energy,
and time of each photon with the integrated rate over the bin where the photon was detected.

The computational cost here scales with the total number of bins, unless $n_i$ is
generally zero (which reduces to the unbinned case of scaling with N).  Thus
this calculation can be significantly faster than the unbinned computation in
\eq{unbinnedeq}, albeit at the cost of lost information.  We will return the comparison
of unbinned and binned likelihood in \sect{comparison}, but first we must develop some
of the machinery necessary for such a comparison.  The reader interested in applying this
machinery, but who dislikes gory math, is invited to skip from the motivation (\eq{begingore})
 to the result (\eq{avgdelta}).

\subsection{Expectation Value of the Likelihood Ratio \label{expect}}

It is useful to be able to calculate the expected value of the likelihood ratio
of different models, in order to calculate the power of the likelihood ratio test.
We can do this by noting that the probability of a data set is given by the
 likelihood function itself.  Thus, if the model $M_{\rm T}$ is true, 
the average log likelihood ratio of two models
 $M_1$ and $M_2$ is given by the sum over all
 possible data sets D
\begin{equation} \label{begingore}
\langle \Delta \ln \like(M) \rangle = \sum_{D} \like(D | M_{\rm T}) \left[
\ln \like(D | M_1) - \ln \like(D | M_2) \right]
\end{equation}

Let us write the data set as a collection of $N$ vectors, ${\bf x}_i$, each of which
represents an observed photon.  The components of each vector will thus be the
observed direction, observed energy, and time of detection.  We can write the
true model prediction of the rate of photons with observed values $\bfxx$ as
$N_{\rm T} R_{\rm T}({\bfxx})\,d \bfxx$, where $R_{\rm T}({\bfxx})$ is normalized
such that $\int R_{\rm T}({\bfxx})\,d\bfxx = 1$.  Similarly, we can write the
predictions of $M_1$ and $M_2$ as $N_1 R_1({\bfxx})$ and $N_2 R_2({\bfxx})$.
The probability of observing a given data set is then
\begin{equation}
 \like(\left\{N, \bfxx\right\} | M) = \frac{e^{-N_{\rm T}} N_{\rm T}^N}{N!}
\prod_{i=1}^N R(\bfxx_i)\,d\bfxx_i,
\end{equation}
and we can verify that it is appropriately normalized
\begin{eqnarray}
 \int_{\left\{N, \bfxx\right\}} \like(\left\{N, \bfxx\right\} | M_{\rm T}) & = & 
\sum_{N=0}^{\infty} \int d\bfxx_1 \int d\bfxx_2 \dots \int d\bfxx_N
 \frac{e^{-N_{\rm T}} N_{\rm T}^N}{N!} \prod_{i=1}^N R_{\rm T}(\bfxx_i) \nonumber \\
& = & 1. \nonumber
\end{eqnarray}

The average log likelihood of data sets with $N$ photons is then
\begin{eqnarray}
\langle \Delta \ln \like( \left\{\bfxx\right\} |M) \rangle & = &
\sum_{N=0}^{\infty} \int d\bfxx_1 \int d\bfxx_2 \dots \int d\bfxx_N
 \frac{e^{-N_{\rm T}} N_{\rm T}^N}{N!} \prod_{i=1}^N R_{\rm T}(\bfxx_i) \times \nonumber \\
& & \left[
\left( \sum_{i=1}^N \ln N_1 R_1(\bfxx_i)\,d\bfxx_i - N_1 \right) -
\left( \sum_{i=1}^N \ln N_2 R_2(\bfxx_i)\,d\bfxx_i - N_2 \right) \right]
 \nonumber \\
& = & \left[ \sum_{N=0}^{\infty} \int d\bfxx_1 \dots \int d\bfxx_N
 \frac{e^{-N_{\rm T}} N_{\rm T}^N}{N!} \prod_{i=1}^N R_{\rm T}(\bfxx_i) 
 \sum_{i=1}^N \ln \frac{N_1 R_1(\bfxx_i)}{N_2 R_2(\bfxx_i)} \right]
 \nonumber \\
& & + N_2 - N_1
\label{uglyeq}
\end{eqnarray}
The normalization of $R_{\rm T}$ allows us to note that
\begin{equation}
\int \prod_i R_{\rm T}(\bfxi) \ln R_1(\bfxx_j) \,d \bfxx_k =  \left\{
\begin{array}{ll}
\prod\limits_{i\neq k}  R_{\rm T}(\bfxi) \ln N_1 R_1(\bfxx_j) & j \neq k \\
\prod\limits_{i\neq k}  R_{\rm T}(\bfxi) \int  R_{\rm T}({\bf y}) \ln N_1 R_1({\bf y})\,d{\bf y} &  j=k
\end{array}
\right. ,
\end{equation}
and thus \eq{uglyeq} simplifies to
\begin{eqnarray}
\langle \Delta \ln \like( \left\{\bfxx\right\} |M) \rangle & = &
\sum_{N=0}^{\infty} 
 \frac{e^{-N_{\rm T}} N_{\rm T}^N}{N!}  \sum_{i=1}^N \int d{\bf y}  R_{\rm T}({\bf y}) 
 \ln \frac{N_1 R_1({\bf y})}{N_2 R_2({\bf y})}  + N_2 - N_1 \nonumber \\
& = & \sum_{N=0}^{\infty} 
 \frac{e^{-N_{\rm T}} N_{\rm T}^N}{N!} N \int d{\bf y}  R_{\rm T}({\bf y}) 
 \ln \frac{N_1 R_1({\bf y})}{N_2 R_2({\bf y})}  + N_2 - N_1 \nonumber \\
& = & \sum_{N=1}^{\infty} 
 \frac{e^{-N_{\rm T}} N_{\rm T}^{N-1}}{(N-1)!} N_T \int d{\bf y}  R_{\rm T}({\bf y}) 
 \ln \frac{N_1 R_1({\bf y})}{N_2 R_2({\bf y})}  + N_2 - N_1 \nonumber \\
& = &  N_T \int d{\bf y}  R_{\rm T}({\bf y}) \left(\ln R_1({\bf y}) - 
\ln R_2({\bf y}) \right) \nonumber \\
& & + N_T \ln N_1 - N_T \ln N_2  + N_2 - N_1 \label{simplifyme}
\end{eqnarray}

When we are discussing models which are maximum likelihood models, a further simplification
can usually be made.  If we write the model parameters as
 $\left\{ {\bf \theta } , N_{\rm pred} \right\} $, so that our normalized
 $R = R({ \bf\theta})$, the conditions of maximum likelihood will generally be
\[ \deriv{\ln \like(M({\bf \theta }, N_{\rm pred}))}{\theta_k} = 0 \ ,\ 
 \deriv{\ln \like(M({\bf \theta }, N_{\rm pred}))}{N_{\rm pred}}=0.
\]
Writing the second condition in full, we see
\begin{eqnarray}
0 & = & \dbyd{N_{\rm pred}}
 \sum_{i=1}^N \ln N_{\rm pred} R(\bfxx_i) - N_{\rm pred} \nonumber \\ 
& = &
\dbyd{N_{\rm pred}} \left[
\sum_{i=1}^N \ln N_{\rm pred} + \sum_{i=1}^N \ln R(\bfxx_i) - N_{\rm pred}
\right] \nonumber \\ 
& = &
\dbyd{N_{\rm pred}} \left[
N \ln N_{\rm pred} + \sum_{i=1}^N \ln R(\bfxx_i) - N_{\rm pred}
\right] \nonumber \\ 
& = &
\frac{N}{N_{\rm pred}} -1 \nonumber, 
\end{eqnarray}
or $N = N_{\rm pred}$.

This will not be true if constraints on model parameters are enforced (i.e. if
a source flux is required to be non-negative, and the unconstrained
maximum likelihood model would have a negative source flux).  Usually, however,
such constraints are not applicable, and \eq{simplifyme} reduces to
\begin{equation}
\langle \Delta \ln \like( \left\{\bfxx\right\} |M) \rangle_{\bf x} =  
  N_T \int d{\bf y}  R_{\rm T}({\bf y}) \left(\ln R_1({\bf y}) - 
\ln R_2({\bf y}) \right) \label{avgdelta}
\end{equation}

This rather simple result can be used to calculate how much our test statistic
will decrease if a binned model is used rather than an unbinned model
(\sect{comparison}), the optimal sizes for bins, and (with some extensions)
the threshold intensity for the detection of a source (\sect{calcthresh}).


\section{Models Used for Likelihood Analysis \label{models}}

\subsection{An ``Exact'' Model}

The models for which we calculate likelihoods must give probabilities of
a given data set being seen.  Thus, the models must include both the
astrophysical reality (that there is a point source with a certain flux at
a certain location) and the instrument response to the incoming \gammarays.

The instrument response to an incoming gamma ray can be characterized
rather simply: for a given incoming photon direction and energy
($\ell_0, b_0, E_0$), we want to know the probability of a detected
direction and energy ($\ell, b, E$).  This will be a function of the
state and orientation of the instrument ($I$), and so we can write this
as $P(\ell, b, E | \ell_0, b_0, E_0, I(t))$. 

There are three principal sources of astrophysical gamma rays.  There are
point sources, which typically have a power law energy spectrum, perhaps
with a spectral break at high energy, and which may exhibit time
variability on millisecond time scales.  There is galactic diffuse emission
due to interaction of high energy cosmic rays with the diffuse gas
in our galaxy.  This has a
spatial form which can be estimated from radio maps of the galactic gas,
and the spectrum over the \egret\ energy range
can be approximated as a power law with index of
 $\alpha_M \approx -2.1$ \cite{hunter_diffuse}.  Finally, there
is a diffuse extra-galactic background, probably due to
unresolved AGN \cite{kumar_extragalactic,tomthesis},
which is isotropic and has a power law index
$ \alpha_B \approx -2.22$.
We will assume that both diffuse backgrounds are time independent.

We can write an $N$ source model, then, as being the sum of these three
components:
\[
P(\ell_0, b_0, E_0| t) = \sum_{i=1}^N S_i(E_0,t)\, \delta(\ell_0 - \ell_i)\, \delta(b_0 - b_i) +
G_{\rm M} M(\ell_0, b_0) E_0^{-\alpha_m} +
G_{\rm B} E_0^{-\alpha_B},
\]
and combine it with the instrument response function to find the probability
of detecting a photon with a measured position and energy:
\begin{eqnarray}
P(\ell, b, E | t) & = & \int d\ell_0 \int db_0 \int dE_0\, P(\ell_0, b_0, E_0 | t)
P(\ell, b, E | \ell_0, b_0, E_0, I(t)) \nonumber \\
& = &  \int dE_0 \sum_{i=1}^N S_i(E_0, t) P(\ell, b, E | \ell_i, b_i, E_0, I(t))
 \nonumber \\
& & + \int d\ell_0 \int db_0 \int dE_0 \left[ G_{\rm M} M(b_0, E_0) E_0^{-\alpha_M} +
G_{\rm B} E_0^{-\alpha_B} \right]  \nonumber \\
& & \cdot P(\ell, b, E | \ell_0, b_0, E_0, I(t)) \label{impractical}.
\end{eqnarray}

It is impractical to evaluate \eq{impractical} as written, for several reasons.
Evaluation of the triple integral term for all necessary positions, energies,
and times is computationally infeasible.  In addition, it is impractical to
measure the instrument response function adequately as a function of so many
parameters.

\subsection{A Calculable Model \label{calculable}}

The first simplifying assumption is that the instrument response is
separable into three different functions:
\begin{eqnarray}
P(\ell, b, E | \ell_0, b_0, E_0, I(t)) & = & SA(\ell_0, b_0, E_0, I(t)) \times PSF(\ell, b | \ell_0, b_0, E_0, I(t)) \nonumber \\
& & \times ED(E | \ell_0, b_0, E_0, I(t)), 
\end{eqnarray}
where $SA$ is the sensitive area of the instrument, $PSF$ is the normalized
point spread function, and $ED$ is the normalized energy dispersion function.

If the spectral index of the source is known in advance, the $PSF$ and $SA$
can be ``pre-dispersed''.  To do this, we construct $\widetilde{PSF}$ and
$\widetilde{SA}$ such that
\begin{eqnarray}
 \widetilde{SA}(\ell_0, b_0, E, I(t)) \widetilde{PSF}(\ell, b | \ell_0, b_0, E, I(t))
E^{-\alpha} = \nonumber \ \ \ \ \ \ \ \ \ \ \ \ \ \ \ \ \ \ \ \ \ \ \ \ \ \ \ \ \\
\ \ \ \ \ \int dE_0 \left[ SA(\ell_0, b_0, E_0, I(t)) \times
    PSF(\ell, b | \ell_0, b_0, E_0, I(t)) \times  \nonumber \right. \\
\left. ED(E | \ell_0, b_0, E_0, I(t)) E_0^{-\alpha} \right].
\end{eqnarray}
This can be done by the definitions
\begin{equation}
\widetilde{SA}(\ell_0, b_0, E, I(t)) =  \frac{\int dE_0 SA(\ell_0, b_0, E_0, I(t))
\times ED(E | \ell_0, b_0, E_0, I(t)) E_0^{-\alpha}}{E^{-\alpha}}
\end{equation}
and
\begin{equation}
\widetilde{PSF}(\ell_0, b_0, E, I(t)) =  \frac{P(\ell, b, E | \ell_0, b_0, E_0, I(t))
E_0^{-\alpha}}{\widetilde{SA}(\ell_0, b_0, E, I(t)) E^{-\alpha}}
\end{equation}

Calculations using $\widetilde{SA}$ and $\widetilde{PSF}$ can be performed
as if there were no energy dispersion (and using measured rather than true
energy).  It should be remembered, however, that the spectral index of the
\gammarays\  was assumed in the construction of these functions.

The next simplification is the pre-calculation of the convolution integral
in \eq{impractical}.  Because our background model $M(\ell_0, b_0)$ is fixed,
the integrals
\begin{eqnarray}
\tilde{M}(\ell, b, E, I(t)) & = & \int d\ell_0 \int db_0 \int dE_0\, M(b_0, E_0) E_0^{-\alpha_M}  SA(\ell_0, b_0, E_0, I(t)) \nonumber \\
& & \times PSF(\ell, b | \ell_0, b_0, E_0, I(t))
 ED(E | \ell_0, b_0, E_0, I(t))
\end{eqnarray}
and
\begin{eqnarray}
\tilde{B}(\ell, b, E, I(t)) & = & \int d\ell_0 \int db_0 \int dE_0\, E_0^{-\alpha_B}  SA(\ell_0, b_0, E_0, I(t)) \nonumber \\
& & \times PSF(\ell, b | \ell_0, b_0, E_0, I(t))
 ED(E | \ell_0, b_0, E_0, I(t))
\end{eqnarray}
can be precalculated.

This is a lot of data to precalculate, however.  It is usual for \egret\
analysis to use $0.5\deg \times 0.5\deg$ bins, about 25,000 of which are
needed for the for the full field of view.  The energy variation is obtained
by interpolation between 20 measured points.  Thus, to calculate this
for 78 instrument modes would require the integrals to be evaluated
over 30 million times.  These calculations would need to be redone whenever
the pointing of the instrument changed.

By averaging over the instrument modes, and by assuming that the convolved
energy spectrum is still a power law with index $\approx 2$, the calculations
may be sped up considerably.  If the instrument response is averaged over
the incident photon direction as well, the calculations do not have to be
redone for the different instrument pointings: the computations can be
performed once and stored for future use.

\subsection{\likeprog\ Model}

The \likeprog\ program \cite{like} performs a likelihood analysis in one
energy band, using the average instrument parameters for that band, and
the counts and exposure maps which are the standard \egret\ data products.
Thus, it is a binned analysis using one energy bin, and spatial bins of
$0.5\deg \times 0.5\deg$ in $\ell$ and $b$.

The \likeprog\ program can be used to analyze several different energy bands
sequentially.  This is then equivalent to having multiple energy bins, but
with the caveat that source and background fluxes are allowed to vary
independently in the different bins.  

The source and null models used by \likeprog\ generally have other known
sources included at fixed positions (the same position in both models),
with the source model having one additional source at a trial location.
The comparison of two models which have multiple sources differing
between them must be done by hand.  Thus a ``two vs. one'' test, 
comparing a model with one source to a model with the same source split
into two sources, is difficult to perform.

\subsection{\unlikeprog\ Model}

Although it would be desirable to perform an unbinned analysis on the \egret\ 
data, the computational difficulties shown in \sect{calculable} make it
impractical.  In addition, the two quantities in \eq{unbinnedeq} can both
become very large.  The fact that we are differencing these two large numbers
to get the log likelihood and then  differencing two log likelihoods to get the
log likelihood ratio means that very small errors can cause large changes
in the test statistic.  Although similar concerns exist for the binned case,
in practice there are fewer problems, since the differencing can be done
for each bin, rather than after the sum in \eq{binnedeq}.

Thus, the \unlikeprog\ program uses a binned model, but incorporates more
flexibility than does \likeprog.  It will simultaneously analyze several
different maps (using the same model), and so allows multiple energy bins
in a more natural way than
does \likeprog, which analyzes only one map at a time.  Because of this,
the flux of a source is not correlated between different \likeprog\ maps, while
in \unlikeprog\ either correlated or uncorrelated fluxes can be used.
The maps in \unlikeprog\ can have different spatial bin sizes, so that high
energy maps are binned on a finer scale than low energy.  The models
assume a power law spectrum for the backgrounds, and allow either a power
law or an arbitrary spectrum for the sources.

The \unlikeprog\ program is, however, not fully developed, and is thus not
suitable for everyday analysis purposes.  Because the benefit gained by using multiple energy
bins is smaller for \egret\ data than \glast\ data (see \sect{implications}),
it did not seem worthwhile to produce a robust, user-friendly program.  The principal
use of \unlikeprog\ to date was for the generation of the likelihood
curves used in the variability analysis in \chapt{variegret}.
The code is available from the author
at {\tt http://razzle.stanford.edu/\~{ }billt}.  

Because of the
\glast\ instrument's scanning mode, new analysis programs will have to be developed.
It is hoped that such programs can benefit from concepts used in the \unlikeprog\ analysis
system, particularly the simultaneous analysis of multiple energy bins, and the inclusion
of multiple sources in a model.

\subsection{Future Development}

Nothing so far has been said of different classes of events, either due to different
instrument modes, or to varying quality of the events themselves.  Such modes, and
any other experimentally determinable quality parameters, can and should be added
into the above analysis.  This is accomplished by letting our rate $R$ be a function
of these quantities as well as the photon direction, energy, and time.

In \egret, there are 78 different triggering configurations of the instrument, and
the sensitive area, energy dispersion, and point spread function were measured for
each \cite{egretcalibrate93}.  Events were also classified as Class A,
Class B, or Class C events, depending
on the quality of the electron tracks.  For all practical analysis purposes, however,
the rate was averaged over these parameters when finding the likelihood function.  This
averaging lost little information, however, since quantities like the
PSF did not very dramatically from class to class or mode to mode.

In \glast, it is conceivable that the events should be separated into separate categories
for the purposes of analysis.  For example, at high energies, the PSF might depend
dramatically on the point of conversion in the tracker.  Especially if events which
did not convert in the tracker at all (calorimeter-only events) are included in the
analysis, the events should be binned by separating them into classes which have similar
PSF's.  Calculations like those shown in \sect{comparison} can help
determine when such a separation of events into multiple classes will be worthwhile.

\section{Quantitative Comparison of Methods \label{comparison}}

We will now calculate how much information is being lost in the binning process, using 
the results of \sect{expect}.  We will take the situation of
a source on a flat background, where each has some spectral index ($\alpha_{\rm S}$ and
$\alpha_{\rm B}$).  The source will be at ($\theta = 0, \phi=0$), and we will only use
data with $\theta < R_{\rm anal}$ and $ E_{\rm min}<E_{\rm meas}<E_{\rm max} $.  This
corresponds to the usual case in \egret\ analysis:  cuts are made based on the measured
energy, and only photons within a fixed radius of the source position are used (to
speed the analysis, and allow for backgrounds which differ as a function of sky position).
First let us write the rate from the source and
from the background independently.  The unnormalized source rate will be
\begin{equation} \label{protosrcrate}
K_{\rm S}(\theta, \phi, E)\,d\theta\,d\phi\,dE = \int dE_0\, ED(E,E_0) PSF(\theta, E_0)\, SA(E_0)\, E_0^{-\alpha_{\rm S}} \,d\theta\,d\phi\,dE,
\end{equation} where $E_0$ is the true energy and $E$ is the measured energy.  The normalized source rate will be
\begin{equation} \label{srcrate}
R_{\rm S}(\theta, \phi, E) = \frac{K_{\rm S}(\theta, \phi, E)}{\int_0^{R_{\rm anal}}d\theta \int_0^{2\pi}d\phi \int_{E_{\rm min}}^{E_{\rm max}}dE\, K_{\rm S}(\theta, \phi, E) \sin \theta}.
\end{equation}

Similarly, the unnormalized background rate will be
\begin{equation}
K_{\rm B}(\theta, \phi, E)\,d\theta\,d\phi\,dE = \int dE_0\, ED(E,E_0) SA(E_0)\, E_0^{-\alpha_{\rm B}} \,d\theta\,d\phi\,dE
\end{equation}
leading to a normalized rate of 
\begin{equation}
\label{bgrrate}
R_{\rm B}(\theta, \phi, E) = \frac{K_{\rm B}(\theta, \phi, E)}{
\int_0^{R_{\rm anal}}d\theta \int_0^{2\pi}d\phi \int_{E_{\rm min}}^{E_{\rm max}}dE\,K_{\rm B}(\theta, \phi, E) \sin \theta} .
\end{equation}

Then if a fraction $f$ of the photons are from the source, we have
\begin{equation} \label{fraceq}
R(\theta, \phi, E) = f R_{\rm S}(\theta, \phi, E) + (1-f) R_{\rm B}(\theta, \phi, E)
\end{equation}

For a binned model, the rate will be piecewise constant, where the constant value
in each bin is the average of the unbinned rate over that bin:
\begin{equation} \label{binavgeq}
R_{{\rm bin}\, i}(\theta, \phi, E) = \int_{\theta_{\min_i}}^{\theta_{\max_i}}
\int_{\phi_{\min_i}}^{\phi_{\max_i}}
\int_{E_{\min_i}}^{E_{\max_i}} R(\theta, \phi, E)
\end{equation}

Application of \eq{avgdelta}, with $R_{\rm T}$ and $R_{1}$ equal to the unbinned
model and $R_2$ equal to the binned model, will now yield the expected decrease
in the log likelihood ratio when the binned model is used.  Similarly, use of
two different binned models for $R_1$ and $R_2$ can indicate which of two binned
models is better, and by how much.

\subsection{Implications for \glast \label{implications}}

The above calculations have been carried out in Mathematica to determine
how binning will affect the analysis of \glast\ data.  For the
purposes of this analysis, the instrument parameters were taken from
the Baseline II instrument (Feb 1998), and both the point spread function
and the energy dispersion function were assumed to be Gaussian.  Only
energy binning was considered: events were considered unbinned in position
(equivalent to the use of bins which are small compared to the point spread function), and
the dependence of the point spread function on factors like the incident
direction or conversion point was not taken into account.

The integrals in \eq{binavgeq}, \eq{protosrcrate}, and \eq{avgdelta} can
be very time consuming, so it was necessary to pre-disperse the point
spread function as described in \sect{calculable}, and approximate
$\widetilde{PSF}$ (and the appropriate energy averaged analogue) by
interpolation between previously calculated points.

The calculations were performed for models with up to six energy bins,
as well as one model with 18 energy bins, to approximate the unbinned case.
All results are normalized to the one-bin case, and thus represent the
factor of improvement gained by using multiple bins.  The model used a
source flux of $1 \times 10^{-8}$ \perareasec ($E_{\rm meas}>100$ MeV),
with an photon energy spectrum of $E^{-2}$,
and a flat background with flux
$5 \times 10^{-5}$ \perareasecsr ($E_{\rm meas}>100$ MeV) with a similar spectrum.
This is
equivalent to a source near the \glast\ threshold in a moderately strong
background.  For comparison, the same calculations were done for a very
bright source, with a source/background flux 1000 times higher.

\begin{figure}
\centering
\includegraphics[width = 4in]{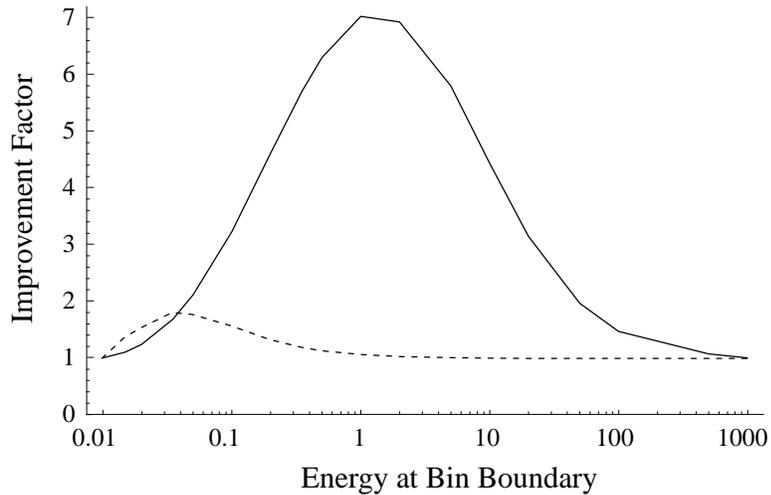}
\caption[Two energy bins: \glast]{\label{twobinpic}
Two-bin analysis of \glast\ data.  The improvement factor reflects the
change in $\Delta \ln {\cal L}$ vs. the one-bin model.  The solid curve
is for a source near threshold, the dashed curve is for a very bright source.}
\end{figure}

The two-bin results are shown in \fig{twobinpic}.  For a source near
threshold, the best division
between the energy bins is around 1 GeV, and the test statistic used to
measure the significance of the detection ($T_S = 2 \Delta \ln {\cal L}$)
is increased
by a factor of seven.  In the background dominated situation, $T_S$ is
proportional to the source strength.  Thus, an sevenfold increase in
$T_S$ is equivalent to a sevenfold decrease in the source threshold. 
For the very bright source, the best division
is around 50 MeV, since the source is detectable in even
the lowest energy photons.

\begin{figure}
\centering
\includegraphics[width = 4in]{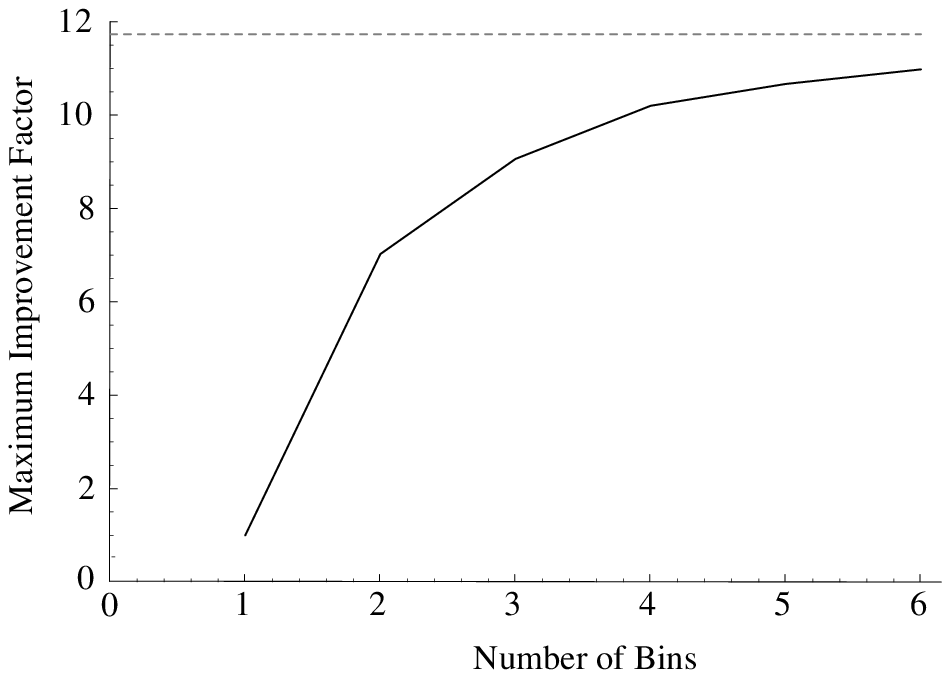}
\caption[Improvement vs Number of Bins: \glast]{\label{allbinpic}
The improvement over the one-bin model as a function of number of
bins.  The horizontal line is the improvement using an unbinned
model.}
\end{figure}

The results of the other binning systems are given in \tbl{engbintable}
 and \fig{allbinpic}.
As shown there, four or five energy bins are probably adequate for detection
analysis.  For spectral analysis of sources, of course, more bins would
be desirable, particularly at low energies.

\begin{table}
\centering
\begin{minipage}{4.5 in}
\footnotesize
\begin{tabular}{|c|l|r@{.}l|}\hline
Number & Breaks & \multicolumn{2}{|c|}{Improvement} \\ \hline
2 & 1 GeV & \ \ \ \ \ \ \ 7 & 0 \\
3 & 500 MeV, 5 GeV  & 9 & 1 \\
4 & 200 MeV, 1 GeV, 5 GeV & 10 & 2 \\
5 & 200 MeV, 500 MeV, 2 GeV, 10 GeV & 10 & 7 \\
6 & 100 MeV, 350 MeV, 1 GeV, 2 GeV, 10 GeV & 11 &0\\ \hline
\end{tabular}
\end{minipage}
\caption[Multiple energy bins for \glast]
{\label{engbintable}
Best energy binning schemes for a fixed number of bins, together with the
improvement vs. the one-bin model. 
}
\end{table}

Similar calculations for \egret\ indicate that the benefits from multiple
energy bins are much more modest.  Because the energy range of \egret\ is
smaller than \glast\,, and because the point spread function reaches
a limiting width at a lower energy for \egret\,, the gain from using
multiple bins ranges from 5\%--80\%, depending on the source strength.
The analysis technique used in the Third \egret\ Catalog \cite{cat3} of summing
likelihood maps for different energy bands yields results which are nearly
as good as a simultaneous fit.


\chapter{Thresholds \label{threshold}}

\section{Thresholds vs Upper Limits}

Before we begin calculating thresholds, it is useful to review the difference
between a threshold and an upper limit (see \chapt{upperlimit}).
We can define the $\alpha$-threshold (for
a given observation time, instrument sensitivity, etc...) as that source flux
which would result in a detection a fraction $\alpha$ of the time.  Note that the
threshold does
not depend on any observed data.  Typically, one uses the 50\% threshold, as the
average limiting source flux which will be detectable by the instrument.

Upper limits, on the other hand, do depend on the
data.  Although this definition might not always be meaningful, we would like to
define the $\alpha$-upper limit
by saying that the true source flux will be less than the $\alpha$-upper limit
a fraction $\alpha$ of the time.  Here, $\alpha$ is typically 90\% or 95\%:
as with confidence regions, one wants to avoid the situation where the true flux
is larger than the upper limit.
Clearly, an upper limit contains more information
than a threshold, since it incorporates the information in the data as well as
the models.  When a threshold is used in conjunction with data, the only information
used from the data is whether or not a source was actually detected. The
fact that a source was not detected does not neccesarily mean that the source flux
was less than the threshold, since 50\% of sources at threshold will not be
detected.   Because thresholds do not depend on the data, however, they can be easier
to calculate than upper limits,
and are useful when discussing the merits and projected results of a proposed
mission, like \glast.

\section{Calculation of Thresholds \label{calcthresh}}

One can calculate a threshold by using the results from \sect{expect} to
calculate the expected likelihood, given a source strength.  By varying that
source strength so that the expected likelihood is that required for a detection,
one obtains approximately the $50\%$ detection threshold.  By calculating the
moments of the likelihood distribution, the distribution itself can be found, and
other thresholds can be obtained.

The two rates in \eq{avgdelta} in this application will come from the two models
under consideration: the model with a source as well as background, and the null
model with background only.  The background in \egret\ is modeled as the sum of
a gas-map component and a constant, where the intensity of each is fit for any
given observation, and the energy spectrum of each\footnote{As noted in
\sect{models}, the spectral indices of the two background components are slightly
different.  Here, for ease of calculation, they are assumed to be equal.}
 is $\alpha = -2.1$.  For simplicity,
we will consider first the case where the gasmap is constant as well, and thus
there is really only one background component.  

We then can use \eq{avgdelta} with the model in \eq{fraceq}.  Our null model will
have $f$ (the fraction of photons coming from the source) set to 0, while
for our source model, it will be set to the fraction of photons within
$R_{\rm anal}$ coming from the source.  Because $\Delta \ln \like$ (and thus
the detection test statistic $T_s$) is linear with the number of photons,
the distribution of $T_s$ for the model
with the true source intensity when $i$ photons are observed will be the convolution
of $i$ copies of the one photon distribution, which we will call $T_s^1$.
The full distribution will then
be the sum of the distributions for $0, 1, 2\dots$ photons, each weighted by the
Poisson probability of getting that number of photons.  The moments of this
distribution are easily calculated.

\begin{eqnarray}
\left \langle T_s^{``true"} \right \rangle
& = & \sum_{N_T^{\rm obs} =0}^{\infty} P(N_T^{\rm obs}) \sum_{i=1}^{N_T^{\rm obs}}
 N_T^{\rm obs} \langle T_s^1 \rangle \nonumber \\
& = & N_T \langle T_s^1 \rangle
\end{eqnarray}
where $\langle T_s^1 \rangle$ is given by
\begin{eqnarray}
\langle T_s^1 \rangle & = & 4 \pi \int_0^{R_{anal}} d\theta \int_{E_{min}}^{E_{max}} dE\, 
\left( f R_{\rm S}(\theta, E) + (1-f) R_{\rm B}(\theta, E) \right)
\times \nonumber \\
& & \left[ \ln \left( f R_{\rm S}(\theta, E) + (1-f) R_{\rm B}(\theta, E) \right)
- \ln R_{\rm B}(\theta, E) \right] \nonumber \\
& = &
4 \pi N_T \left[ \int_0^{R_{anal}} d\theta \int_{E_{min}}^{E_{max}} dE\, 
\left( f R_{\rm S}(\theta, E) + (1-f) R_{\rm B}(\theta, E) \right)
\times \nonumber \right.\\\
& & \left. \ln \left( \frac{f R_{\rm S}(\theta, E)}{R_{\rm B}(\theta, E)} + (1-f)
\right) \right] \label{firstts}
\end{eqnarray}

The calculations in \eq{firstts} are calculable as written, but can be simplified
if the spectral index of the source and background are the same:

\begin{eqnarray}
\langle T_s^1 \rangle & = & 
\frac{4 \pi}{\int_{E_{min}}^{E_{max}} dE\, 
\widetilde{SA}(E) E^{-\alpha}}
 \left[ \int_0^{R_{anal}} d\theta \int_{E_{min}}^{E_{max}} dE\, 
\widetilde{SA}(E) E^{-\alpha} \times \nonumber \right. \\
& & \left. \left( f \widetilde{PSF}(\theta, E) + \frac{1-f}{A} \right)
\ln \left( f A \widetilde{PSF}(\theta, E) + (1 - f) \right) \right] \label{simplts}
\end{eqnarray}

Similar computations show that
\begin{equation}
\left \langle (T_s^{``true"})^2 \right \rangle
 =  N_T \left \langle (T_s^{1})^2 \right \rangle + \left( \left \langle
 (N_T^{\rm obs})^2 \right \rangle - N_T \right)
 \left \langle T_s^{1}\right \rangle^2, \label{tssqr}
\end{equation}
where 
\begin{eqnarray}
\langle (T_s^1)^2 \rangle & = & 
\frac{8 \pi}{\int_{E_{min}}^{E_{max}} dE\, 
\widetilde{SA}(E) E^{-\alpha}}
 \left[ \int_0^{R_{anal}} d\theta \int_{E_{min}}^{E_{max}} dE\, 
\widetilde{SA}(E) E^{-\alpha} \times \nonumber \right. \\
& & \left. \left( f \widetilde{PSF}(\theta, E) + \frac{1-f}{A} \right)
\left[\ln \left( f A \widetilde{PSF}(\theta, E) + (1 - f) \right) \right]^2 \right]
\end{eqnarray}

Because $N_T^{\rm obs}$ has a Poisson distribution, 
$ \left \langle (N_T^{\rm obs})^2 \right \rangle = N_T (N_T+1)$,
and \eq{tssqr} becomes
\begin{eqnarray}
\left \langle \left( T_s^{``true"} \right)^2 \right \rangle
& = & N_T \left \langle (T_s^{1})^2 \right \rangle +  N_T^2
\left \langle T_s^{1} \right \rangle^2.
\end{eqnarray}

The variance of $T_s^{``true"}$ is then
\begin{eqnarray}
\left \langle (\Delta T_s^{``true"})^2 \right \rangle
& = & \left \langle (T_s^{``true"})^2 \right \rangle - 
\left \langle T_s^{``true"} \right \rangle^2 \nonumber \\
& = & N_T \left \langle (T_s^{1})^2 \right \rangle
\end{eqnarray}

\section{Allowance for Extra Degrees of Freedom}

As discussed earlier in \sect{wilks}, if the parameters of the model are allowed
to vary from the true value, the distribution of the change in $T_s$ will be
distributed like $\chi^2$, with the number of degrees of freedom equal to the
number of parameters which may vary.  Because there are 5 parameter in the typical
\egret\ source model (one for source intensity, and two each for source position
and background level), and 2 parameters in the null model, we know that
$T_S - T_S^{``true"}$ is distributed like $\chi^2_3$.  This results in
\begin{eqnarray} \mu_{T_s} & = &
\left \langle T_s \right \rangle \nonumber \\
& = & \left \langle T_s^{``true"} \right \rangle + 3
\end{eqnarray}
and
\begin{eqnarray} \sigma_{T_s}^2 & = &
\left \langle \left(\Delta T_s\right)^2 \right \rangle \nonumber \\
& = & \left \langle \left(\Delta T_s^{``true"}\right)^2 \right \rangle + 6.
\end{eqnarray}

\subsection{The Full Distribution of $T_s$}

Unfortunately, the distribution of $T_s$ is quite skew, and thus the
Gaussian with mean $\mu_{T_s}$ and standard deviation $\sigma_{T_s}$
is not a good approximation to the distribution of $T_s$.  A much better
approximation is that $\sqrt{T_s}$ has a Gaussian distribution.  The
mean and standard deviation of  $\sqrt{T_s}$ can be written in terms of
$\mu_{T_s}$ and $\sigma_{T_s}$ (calculated above):
\begin{equation}
\mu_{\sqrt{T_s}} = \root 4 \of{\mu_{T_s}^2 - \frac{\sigma_{T_s}^2}{2}},
\label{musqrtts}
\end{equation}
and
\begin{equation}
\sigma_{\sqrt{T_s}} = \sqrt{\mu_{T_s} - \mu_{\sqrt{T_s}}^2}.
\label{sigsqrtts}
\end{equation}

The calculation of $\langle T_s^1 \rangle$ and $\langle (T_s^1)^2 \rangle$
must be done numerically because of the form of the point spread function.
For \egret\, sufficiently accurate empirical approximations are:
\begin{equation}
\langle T_s^1 \rangle =  18.556
\left( x^{0.7602} + 1.77 \right)^{-2.63}
, \label{empiricalF}
\end{equation}
and
\begin{equation}
\langle (T_s^1)^2 \rangle = 73.424
\left( x^{0.94} + 1.568 \right)^{-2.1277}
, \label{empiricalG}
\end{equation}
where $x \equiv B_c / S_c$, $S_c \equiv S \times E$, the expected number
of photons received from the source, and
$B_c \equiv 214.09 \times B \times E$, the expected number of photons
which would be received within $15^\circ$ of the source
position if the background were uniform.

Monte Carlo simulations indicate that \eq{musqrtts} and \eq{sigsqrtts}
are accurate to within $5\%$ for flat backgrounds.

The threshold usually desired is the $50\%$ threshold, which will be the
source flux necessary to make $\mu_{\sqrt{T_S}} = 4$
(or $5$ for $\left|b\right| < 10^{\circ}$).  The full distribution of $T_S$
(and thus the probability that a source of a certain strength would be
detected) should be useful when analyzing source distributions, as sources
which are nominally below threshold do have some chance of being detected
(as well as the reverse).

\subsection{Background Structure}

Unfortunately, the presence of background structure can have dramatic
effects on the instrumental threshold.  It is not feasible to perform
the above calculations with every possible shape of background.  Rather,
the general effects of the varying background were modelled by multiplying
the expected $T_S$ by a compensation factor:
\begin{equation}
C = .37 + .0061 \left| b \right| \label{correctionfactor},
\end {equation}
where $b$ is the galactic latitude in degrees.  Although there is significant
longitude dependence near the plane, this dependence could not be easily modelled
in this way.

The expected $T_s$ from above should be multiplied by this correction factor
for a more realistic estimate of the source significance; unfortunately the
result is correct only on average, and additional structure of the diffuse
emission make it accurate only far from the plane.

For an estimation of the probability that the $T_s$ is greater than a
certain level, one must change variables from $T_s$ to $\sqrt{T_s}$, and use
\eq{musqrtts} and \eq{sigsqrtts}, multiplying by the compensation
factor from \eq{correctionfactor}

Thresholds calculated by the above technique may still have significant
errors.  Pointlike structure in the background can change the expected
$T_S$ by up to a factor
of 2.  Nonetheless, this technique does capture the average behavior of
the instrumental threshold, and is a significant improvement to the
assumption that the threshold is constant.  If desired, accurate determination
of the threshold at a given location can be found by repeating the calculation
with the appropriate background included.

\chapter{Upper Limits \label{upperlimit}}

The concept of an upper limit is often ambiguous.
The average graph reader might think that the region below a
``$1 \sigma$'' upper limit is equivalent to a ``$1 \sigma$'' confidence region:
implying that there is a 68\% chance that the true value is less than the upper limit.
Unfortunately, if one sets out to create an upper limit, there is no such value from the
frequentist point of view.
The various means of calculating something to call an upper limit are discussed below.

The Bayesian analyst has no such problems: the interval estimation done in Bayesian
analysis (which does not strictly correspond to a confidence region) is directly
applicable to upper limits, that is, situations where the interval includes 0.

To demonstrate the problem with defining a upper limit which is a confidence region,
we will begin by reviewing the definition of a confidence region.

\section{Confidence Regions}

According to the frequentist philosophy, one can make no probabilistic statements
about the truth of a model.  There is a model which is correct, and others which
are incorrect: there is no probability about it.
This is somewhat different than what one might say
in everyday conversation, and thus can lead to some confusion for the statistics
student, who might say something like: ``There is a 95\% chance that this is a
a two-headed coin,'' when faced with a series of all-heads coin flips.
Strictly speaking (according to the frequentist) the coin is
either two-headed or not, there is no probability about it.  The only probability
is that related to the data set: the probability of getting a data set with that
many heads in a row might be less than 5\%.  This is in keeping
with Bernoulli's definition of probability, relating the probability of an event
to the frequency of that outcome in repeated trials: thus the appellation ``frequentist''.

How then is a confidence region defined?  Well, the first step is to note that the
estimate $\hat{x}$ of a model parameter $x$ depends on the data.  So we are permitted
to make probabilistic statements about $\hat{x}$, if not about $x$ itself.  In fact,
we can make probabilistic statements about $x - \hat{x}$: again a data-set dependent
quantity.  If there is a way to calculate $C$ such that $P(|x - \hat{x}| < C) = 95\%$,
for any value of x, then the region $[\hat{x} - C, \hat{x} + C]$ will contain the
true value 95\% of the time (that is, for 95\% of the data sets).
To deal with non-symmetric confidence regions, one may simply change variables in
order to make the confidence region symmetric.

\subsection{The problem with upper limits}
The difficulty with the notion of an upper limit is the requirement that the
confidence region contain the true value of the model parameter a fixed fraction
of the time,  {\em for any true value of the model parameter}.  Because the region
which an upper limit defines always includes zero, this region will always
contain the true value if the value is zero.  And thus, no confidence region
can be defined by the upper limit.  The possible solutions involve relaxing the
requirement that the upper limit be greater than the true value
 exactly some fraction of the time, or by giving up on the idea of constructing
an upper limit, and simply constructing a confidence region.

\section{Definitions of Upper Limits}

Given that it is not possible to define a confidence region which always contains
the value zero (as we would like the region bounded by the upper limit to do), it
remains to determine what sort of upper limits one can define and use in a 
statistically meaningful way.

\subsection{Statistically Rigorous ``Upper Limit'' \label{rigor}}

The way in which one can define a statistically exact upper limit is by
constructing the appropriate confidence region, and quoting the top end of
the confidence region as the upper limit only if the lower end of the confidence
region is less than zero.  Because we are dealing with a confidence region, it
will contain the true value the stated fraction of the time.  But, by the nature
of the construction, we are not guaranteed an upper limit.  This might not seem
to be a problem, but in fact can lead to difficulties.

\begin{figure}[tbhp]
\centering
\includegraphics[width = 4in]{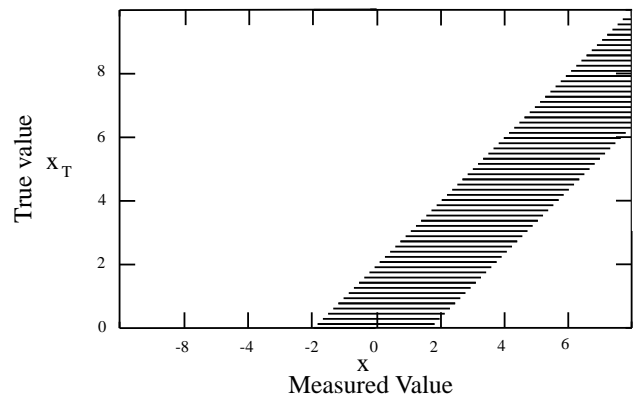}
\caption[Construction of confidence regions]{\label{confbelt1}
Construction of confidence regions.  The shaded region is formed by picking a
true value $x_{\rm T}$, then finding an interval in $x$ which will contain
the measured value 95\% of the time.
}
\end{figure}

Constructing the confidence region is done by means of
a confidence belt.  For each value of the true parameter $x_{\rm T}$, a range of
the measured parameter $x$ is constructed, such that in a fixed fraction of the
data sets, the measured value $x$ will be in that range.  This is illustrated in
\fig{confbelt1}.  For concreteness, let us examine the use of a 95\% frequency band.

The confidence belt is constructed horizontally then (for each value of $x_{\rm T}$,
a range in $x$ is constructed), but then read out vertically.  The justification
is simple:  for any value of $x_{\rm T}$, the probability of a measurement in the
shaded region is 95\%.  Thus, the probability of a pair $(x_{\rm T}, x)$ being
in the band is also 95\%.  And thus, for a measured value of $x$, the probability
that the band contains the true value $x_{\rm T}$ is 95\%.

\fig{confbelt1} shows a standard, symmetric, confidence belt.  Reading vertically
from a measured value of $x = -1$, an upper limit of slightly less than $1$ is
obtained.  What, though, if a measurement of $x = -3$ is obtained?  Certainly this
is unlikely, but it will occasionally happen, and one is left with no region at all.
One might be tempted to extend the confidence belt to negative values of $x_{\rm T}$,
resulting in a negative upper limit, but this is meaningless from both a statistical
point of view (if a model with $x_{\rm T} < 0$ is impossible, it makes no sense to
discuss the probability of data sets arising from it), and an analyst's point of
view (what does a negative upper limit mean?).  The one possibly useful thing about such
negative upper limits is if they are combined with the negative point estimate of
$x_{\rm T}$ (the value of $x_{\rm T}$  with the maximum likelihood) 
in an effort to rate physically meaningful values of $x_{\rm T}$.
Better than this approach, however, is to use the likelihood function itself
in the physically meaningful range.

\subsection{The Feldman-Cousins Method \label{feldman}}

In the 1998 Review of Particle Properties \cite{pdg98}, the prefered method listed
is that of Feldman and Cousins \cite{feldman}.  This unified approach to the
calculation of upper limits and confidence regions is similar to that of \sect{rigor},
but has the benefit that there are no measured values for which one does not obtain
a confidence region.

\begin{figure}[tbhp]
\centering
\includegraphics[width = 4in]{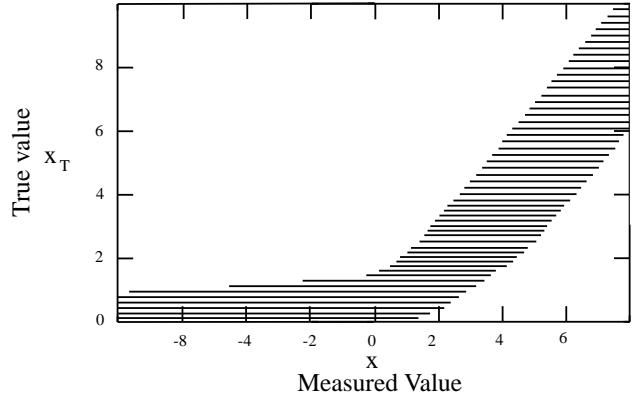}
\caption[Construction of confidence regions]{\label{confbelt2}
Construction of confidence regions.  The shaded region is formed by picking a
true value $x_{\rm T}$, then finding an interval in $x$ which will contain
the measured value 95\% of the time.  If this interval extends to $-\infty$
for some values of $x_{\rm T}$, this confidence region will be an upper
limit when appropriate.
}
\end{figure}

This is done with confidence regions that look like those in \fig{confbelt2}.
By using non-central confidence regions for the lower values of $x_{\rm T}$,
non-negative confidence regions will be obtained for negative measurements.

There is some difficulty in deciding how to construct the bands for \fig{confbelt2}.
Feldman and Cousins present a method for this construction, and apply it to
Gaussian and Poisson data.  For \egret-style data, the method would be
rather complex, but with some simplifications could probably be implemented.

In this way, one can avoid the problem of not being able to construct upper limits
which are confidence regions.  This problem (that the upper limits always include
zero) is avoided because 5\% of the time (for our 95\% confidence region) when the
true flux is zero, we will not have an upper limit.  Instead, there will be a
positive flux estimate and a confidence region which does not include zero.

The chief problem with this method is that some fraction of the time, one will
obtain a confidence region which does not include zero (and is thus not an upper
limit).  This would not be a problem, except that this will happen even when
a source detection is not claimed.  For example, suppose that the one trial
significance of an intensity measurement is 99.5\%.  This will not be accepted
as a real detection at the $5 \sigma$ level (standard for \egret\ analysis near
the galactic plane).  But the $95\%$ confidence region will not include zero,
and thus one cannot claim a $95\%$ upper limit which is statistically rigorous.
The solution to this problem is typically to quote the upper end of the $95\%$
confidence region as the upper limit, with the knowledge that it is a conservative
upper limit.

\subsection{The ``PDG Method''}

Prior to 1998, the Particle Data Group suggested several alternate
methods for calculating upper limits to avoid unphysical values\cite{pdg94}.
One, with
a decidedly Bayesian flavor, was the sole suggestion in early PDG publications
\cite{pdg88}, and thus has been referred to as the ``PDG method'', to the
consternation of some in the PDG\cite{pdg94}.
This method involves renormalizing the likelihood curve so that the
integral over the physical region is $1.0$.  For an upper limit, this corresponds
to the requirement
\[\int_0^\infty {\cal L}(x) \,dx = 1. \]
The upper limit $x_{\rm UL}$ is then determined by
\[\int_0^{x_{\rm UL}} {\cal L}(x) \,dx = \alpha. \]
This is equivalent to a Bayesian analysis where the prior probability is
\[ P(x) = \left\{\begin{array}{ll}
1 & \mbox{$x \geq 0$} \\
0 & \mbox{$x < 0$}
\end{array} \right. \]

This prior is generally conservative, as there tend to be more dim sources than
bright ones (and for most locations, no source at all).

\subsection{Ad Hoc Methods \label{adhoc}}

Other methods are presented in the 1994 Review of Particle Properties \cite{pdg94}
as well.  One, which is clearly conservative, is to find the central interval from
\sect{rigor} (where the confidence belt was extended to negative $x_{\rm T}$)
but to shift the interval so that the point estimate (the value of the parameter
with the peak likelihood) is at zero, rather
than the non-physical negative value.  Because this upper limit is always greater
than one which would contain the true value the stated fraction of the time, it
is necessarily conservative.  This is the method used by \likeprog, and thus
investigators should be aware of this built-in conservatism in the upper limit.

Another method is to pretend that the likelihood function has a maximum for the
zero value of the intensity, and to generate the upper limit by finding where
the likelihood has decreased a set amount from this ``maximum'', just as if
it were a true maximum.  This is, again, necessarily conservative, because the
true maximum likelihood is higher, and thus the upper limit from \sect{rigor} must
be lower than this one.

As pointed out by Feldman and Cousins \cite{feldman}, the post-hoc choice to
calculate an upper limit rather than a confidence region can bias the results
in strange ways. In particular, both overcoverage of the confidence belt (Conservatism)
and undercoverage can result.


\chapter{Distribution of $T_s$ \label{tsdist}}

As discussed in \sect{wilks}, the distribution of the maximum log likelihood ratio is generally
the $\chi^2$ distribution with some number of degrees of freedom.  As noted there,
however, there are situations where the proof of this does not hold, and we
must look for alternate sources for the distribution.  In the case of the maximum \egret\ 
test statistic in a patch of sky, Wilks' theorem does not hold: here an alternate
distribution for the largest $T_s$ in a survey (given that there are no real sources) is derived.

\section{An Empirical Distribution}
The \egret\ null model is equivalent to the one--source model where the source
intensity has been set to zero; the position of the ``source'' then does not matter.
Then, since $\left[\frac{\partial \like(S, \ell, b)}{\partial \ell} \right]_{S=0}= 0$,
the proof of Wilks' theorem breaks down.  It is clear by inspection that the theorem
cannot hold:  the distribution of the largest $T_s$ we see (if there is no source)
should depend on how large an area of sky we survey.  If the maximum $T_s$ were
distributed as $\chi^2$, then this could not be true.

One attempt to provide this area dependence has been by acting as though 
the maximum $T_s$ in a region is the maximum of some (discrete) number of
independent random
variables, where each of these random variables has a $\chi^2$ distribution.
Empirically, it is seen that the $\chi^2_3$ with some number of
trials gives a good fit to the Monte Carlo distribution~\cite{like}.  An alternate
distribution, with a somewhat more convincing derivation is presented
here~\cite{aldous,russian}.

\section{Distribution at a fixed point}
Wilks' theorem does provide the distribution of $T_s$ at a fixed source position: the
degeneracy discussed above is then removed.  As has been pointed out several
times, and verified by Monte Carlo~\cite{like,tomthesis}, this
distribution is $\chi^2_1$,
 that is, $\chi^2$ with one degree of freedom, with one
complication.  Since negative source fluxes are not permitted, half of our
$T_s$ values will be zero.  Thus the probability distribution of the fixed $T_s$
is:

\[
P_{fixed}(T_{s}) = \left\{
\begin{array}{ll}
\frac{1}{4 \pi} e^{- T_s^2} & \mbox{$T_s > 0$} \\
\frac{1}{2} \delta (T_s) & \mbox{$T_s = 0$} \\
0 & \mbox{$T_s < 0$}
\end{array} \right. \]

In practice, since we only wish to look for the maximum $T_s$ in a region, we
can take the distribution to be a full Gaussian instead, as it does not matter
whether half the values are negative or identically zero: only the positive
values have any chance of being the largest $T_s$ found.

Thus, we see that the problem of finding the maximum $T_s$ is equivalent to finding
the maximum value of a two dimensional Gaussian random field, where adjacent
values have some correlation.

\section{The One Dimensional Problem}

Before presenting the solution to our two dimensional problem, it is instructive
to look at the one dimensional case.  That is, we examine the distribution of
the maximum of a stochastic function $x = f(t)$ where the distribution of
$x$ at any point is Gaussian.  To account for the correlation in values, we
have a distribution of $v = \frac{df(t)}{dt}$.

The problem of finding the probability of a maximum is most easily approached by
finding the distribution of upcrossings of the function:  that is, where the
function crosses the line $x=b$ while increasing.  We start with the
probability of an upcrossing in the interval $[t, t +dt]$ as
\begin{equation}
p(b | t) dt = \int_0^\infty f(b, v | t) v\, dv\,dt
\end{equation}

This reduces, for a normal, stationary function with mean zero, to

\begin{equation}
p(b) dt= \frac{1}{2 \pi} e^{ - \frac{b^2}{2 \sigma_x^2}}
 \frac{\sigma_v}{\sigma_x}\, dt .
\end{equation}

This may be rewritten in terms of the correlation function $C(t)$, as
\begin{equation}
p(b) dt= \frac{1}{2 \pi C(0)} e^{ - \frac{b^2}{2 \sigma_x^2}}
 \frac{\partial^2 C(t)}{\partial t^2}\, dt .
\end{equation}

In a regime where upcrossings are uncommon, the probability that $\max f(t) < b$
in the interval [0, T] is equal to the probability that there are no upcrossings
of $b$: \begin{eqnarray}
P(\max f\, < b) & = & e^{- T p(b)} \nonumber \\
&=&  \exp \left[ - \frac{T}{2 \pi C(0)}
\, e^{ - \frac{b^2}{2 \sigma_x^2}}\, \frac{\partial^2 C(t)}{\partial t^2}\right].
\end{eqnarray}
Noting that $\frac{\partial P(\max f\ < b)}{\partial b} db
 = - P(\max f\, = b) db$, we can solve
for the distribution of $\max f$ in an interval of length $T$:
\begin{eqnarray}
P(\max f\,= b) db & = &- \frac{\partial}{\partial b} \exp \left[ - \frac{T}{2 \pi C(0)}
\, e^{ - \frac{b^2}{2 \sigma_x^2}}\, \frac{\partial^2 C(t)}{\partial t^2}\right] db
\nonumber \\
&=& - \exp \left[ - A e^{ - \frac{b^2}{2 \sigma_x^2}}\right] \left(A e^{ - \frac{b^2}{2 \sigma_x^2}}\right) \frac{b}{\sigma_x^2}
\, db,
\end{eqnarray}
where $A = \frac{T}{2 \pi C(0)} \frac{\partial^2 C(t)}{\partial t^2}$.

\section{The Multidimensional Problem}

By extension, in the multi-dimensional case we have a similar result~\cite{aldous}.
If our Gaussian stochastic function is $F({\bf x})$, we first must find the correlation
matrix of $F$, $\bf \Lambda$, defined through
\[
\left\langle  F({\bf x})  F({\bf x}+{\bf \delta}) \right\rangle
= 1 - \frac{1}{2} { \bf \delta }^T {\bf \Lambda} {\bf \delta} +
 O(\left| \delta \right|^4)
\]

The density of large values of $F$ is 
\[
\lambda_b = \frac{1}{2 \pi} \left| {\bf\Lambda } \right| ^{\frac{1}{2}} b \phi (b)
\]
where $\left| {\bf \Lambda } \right|$ is the determinant of $\bf \Lambda $,
and $\phi (b)$ is the standard Normal density:
\[
\phi (x) = \frac{1}{\sqrt{2 \pi}} e^{- \frac{x^2}{2}}
\]

By using $F({\bf x}) = \left[ T_s(\ell, b) \right]^2$, we find that the
distribution of the maximum $T_s$ in an area $A$ is given by
\begin{eqnarray}
P (\max_A T_s < k) & = & e^{- \lambda_{k^2} A} \nonumber \\
& = & e^{- \frac{1}{2 \pi} \left| {\bf \Lambda} \right| ^{\frac{1}{2}} k^2
 \frac{1}{\sqrt{2 \pi}} e^{- \frac{k^4}{2}} A} \nonumber \\
& = & \exp \left(- A (2 \pi)^{-\frac{3}{2}} \left|{\bf \Lambda }\right| ^{\frac{1}{2}} k^2
 e^{- \frac{k^4}{2}} \right).
\end{eqnarray}

The distribution can be calculated as before by taking the derivative, but this cumulative
distribution is most appropriate for finding the significance of an observed $T_s$.
The traditional~\cite{like} distribution to use for large values of $T_s$ has
been $\chi^2_3$ with some number of trials, where the number of trials was
found by Monte Carlo.  The method presented here also has one free parameter,
$\left| {\bf \Lambda } \right| $, which must be found by Monte Carlo if one cannot calculate
it a priori (which appears to be difficult).

One could calculate $\left| {\bf \Lambda } \right| $ from a simulation, but
without the large number of trials required for the Monte Carlo, however.
The easiest way to do this would probably be using the identity:
\[
{\Lambda_{ij} } = \left \langle \frac{\partial F({\bf x})}{\partial {x_i}}
\frac{\partial F({\bf x})}{\partial {x_j}}
\right \rangle
\] 

\chapter{Quantifying Source Variability \label{varimethod}}

\section{Introduction}

One of the biggest questions raised by the EGRET mission is that of the
nature of the unidentified sources.  Certainly some of these will turn out
to be AGN~\cite{tomthesis}, and it is likely that some of the ones
near the galactic
plane are associated with pulsars~\cite{romani,mukherjee_unid,merck_unid} 
and supernova remnants~\cite{sturner_snr,esposito_snr}.
A few, as well, may be discovered to be artifacts of the galactic diffuse
emission.
But as has been discussed elsewhere~\cite{merck_unid}, there are 
more unidentified sources than expected from these
populations.  The large error regions for the \egret\ sources often 
make it difficult to discern which of several candidates is the true
counterpart in another wavelength.

When studying the unidentified sources, there are several observables
which one might use to determine their nature.  One can look at the
spatial distribution~\cite{kanbach_characteristics,grenier} to
estimate what fraction
of the population is galactic in nature, and what scale height the galactic
fraction has.  One can look at positional correlations with other source
types~\cite{cat2,cat3}.  The energy spectrum is
another source characteristic which can be used to classify it as
``pulsar-like'', ``AGN-like'', or ``other''~\cite{merck_unid}.

In this same vein, one can examine the EGRET catalog sources for evidence
of time variability, in hopes of distinguishing the various source classes.
The known pulsars are seen to have a fairly constant flux (when averaged over
many pulse periods), as is expected from the nature of their energy production,
while AGN are seen to flare dramatically.  Certainly,
if an unidentified source has an obvious flare, we can rule out the
possibility that it is a pulsar, and raise the likelihood that it is an AGN or
another type of flaring source.
But, as with many aspects of the EGRET data, it is difficult to
characterize the time variability of most sources, because of the limited statistics
and the non-continuous observations.  Thus it is natural
to look for a statistic which will measure the variability in a rigorous way.

\subsection{Previous Methods}

The primary method used for estimating the variability of EGRET sources
has been that introduced by McLaughlin et al.~\cite{mclaughlin}.  This method 
finds the $\chi^2$ of the measured source intensities, and uses
$V = - \log Q$, where $Q$ is the probability of obtaining such a large
$\chi^2$ if the source flux were constant.
There are some
complications, such as what to do if there is only an upper limit in
a given observation, but this is the general approach.

There are two problems with this method: one statistical, and one
conceptual.  The statistical problem is that introduced when using upper
limits:  McLaughlin et al. use zero as the flux estimate when the measured
flux estimate would be negative, and use the \likeprog\ upper limits
as the error in that estimate.  As discussed in \sect{upperlimit}, these upper
limits (together with the flux estimate)
are not statistically meaningful as defining a confidence interval.
Thus the resulting ``$\chi^2$'' will not have a $\chi^2$ distribution.
Since the \likeprog\ results are statistically conservative, this means that
sources which have upper limits included in the analysis will have a lower
$V$ than implied by the data.

The conceptual problem with the method is the use of a $\chi^2$ statistic
to determine variability.  The $V$ statistic is a measure of how inconsistent
the data is with the model that the source flux is constant.  Thus, sources
with a large $V$ are inconsistent with being constant.  But we do not
know whether such a source has a large $V$ because of large intensity
fluctuations, or because of small error bars on the intensity measurements.
Similarly, a source with a small $V$ might truly be constant, or might
have just have very poor measurements of its flux.

As an example, take a source with an intrinsic variability of 10\%.  If
our measurements of the source flux have an error of 30\%, it will have
a small $V$.  If our measurements have an error of 1\%, it will have quite
a large $V$.  Thus, while the $V$ statistic is useful for excluding
possible pulsar candidates (by excluding those which are inconsistent
with being constant),
it is not useful for finding sources which have low intrinsic variability,
or for comparing variabilities of sources which have different errors in their
flux measurements.

\section{A Likelihood Approach}
\subsection {What Do We Want to Measure?}
The question remains: if we don't really want to measure the $\chi^2$ of
a source, what {\em do} we want to measure?  One possibility is the
standard deviation of the true flux of a source.  This has the following problem:
if one
views two identical sources, with one twice as far away as the other
(and with \onequarter\ the flux), then the further one will have
\onequarter\ the standard deviation of the closer one.

Another possibility is to estimate the standard deviation of the true flux
divided by the true flux.  This fractional variability is much closer to
what we really want to know about a source, particularly one which always has
the same level of variability.  There could be other statistics which
might be more meaningful for flaring sources, such as the peak flare flux
divided by the quiescent level.

In the spirit of measuring fractional variability, we can
define\footnote[1]{This ratio was defined as ``$V$'' by Raleigh~\cite{kendall}.  It is
defined here as ``$\tau$'' to minimize confusion with McLaughlin {\it et al.}'s
``$V$'' statistic.\cite{mclaughlin}}

\begin{equation}
\tau \equiv \frac{\sigma}{\mu},
\end{equation}
where $\mu$ and $\sigma$ are the average and standard deviation,
respectively, of the true flux of the source.  Since we have errors in
our flux measurements, we must now find a way to estimate $\tau$ given
the data.
\subsection{Modeling the Source Flux Distribution}
In the spirit of likelihood analysis, we begin by constructing models,
for which we can calculate the likelihood of obtaining our data.
Our model in this case is the distribution from which the true source 
fluxes are drawn.  Note that there are an infinite number of source flux
distributions which will yield the same $\tau$; we want to pick one that
is compatible with our notions of what the true source flux distribution
should be, but which is fairly general at the same time.

First, we will assume that the source flux is uncorrelated at different
times.  This can be, in fact, a very poor assumption: AGN flares can last
for weeks, or just days.  The presence of these correlations means that
our estimate of $\tau$ will depend on the time scales at which
we are looking.  If we want to look for long term variability, we
should average together observations which take place in the same short
time interval.  If we wish to examine short time scale variability, we
should take out the long time scale variability by normalizing observations
to the average within some longer interval.

The first source flux distribution one might try is a Gaussian.
Thus, we would find the likelihood of obtaining our data
given that the true flux $S$ was drawn from a Gaussian with mean $\mu$ and
standard deviation $\sigma = \tau \mu$.  We could then find
the maximum likelihood value of $\tau$, together with a confidence interval
for $\tau$ defined in the standard way (see \sect{confregions}).  Alternatively,
we could follow the Bayesian procedure: form priors for the distributions
of $\mu$ and $\tau$, and find an estimate of $\tau$, together with an
error region, by marginalizing over $\mu$.

The use of the Gaussian distribution for source fluxes has a flaw, however.
It allows the possibility of $\mu = 0$ ($\Rightarrow \tau = \infty$), as
well as the possibility of negative $\mu$.  Thus, we will instead use the
log-normal distribution; that is, that $\ln S$ has a Gaussian
distribution with 
\begin {equation}
\mu_{\ln S}= \ln \mu - \onehalf \sigma^2,
\end{equation}
 and
\begin {equation}
\sigma_{\ln S}= \sqrt{\ln{\frac{\sigma^2}{\mu^2} + 1}}.
\end{equation}
This distribution is generally appropriate for quantities which are bounded
below by zero and unbounded above, and for quantities where the scatter
is proportional to the value, as is the case here.

\subsection{Characterizing the Single-Measurement Likelihood \label{intcurve}}

It is standard in EGRET analysis to calculate the likelihood of the observed
data for a certain source flux (\sect{likelihood}).  Typically, the
values output by an analysis program are the most likely flux, and a
confidence interval.  In the case of upper limits, the most likely flux is
taken as zero, and the confidence interval might be defined in one of several
ways (\sect{upperlimit}).

For this work, we would like to be able to calculate the likelihood for
any source flux, not necessarily one near the maximum likelihood
value.  Thus it is useful to look for a parameterized family of curves
which will closely fit the full likelihood function.

If the photons had no spatial measurement, there would be an obvious
form of such a function.  We could simply write the expected number
of counts as $\mu = N_S + N_B$, and then find the Poisson likelihood
of observing $N$ photons:
\begin{eqnarray}
{\cal L}(N_S) & = &\frac{e^{\mu} (\mu)^N}{N!} \nonumber \\
& = &\frac{e^{-(N_S + N_B)} (N_S + N_B)^N}{N!}, \nonumber
\end{eqnarray}
giving us a two parameter fit ($N$ and $N_B$) to ${\cal L}(N_S)$.

When we seek to include the effects of the instrument point spread function,
we must replace the numbers of counts above with some ``effective'' number
of counts.  A photon closer to the source position will be weighted more than
one far away, in some sense.  To translate between the true source flux and
an effective number of counts, we can include an extra parameter, $S_M$, and
try
\begin{equation}
\mu = S \times S_M + S_0
\end{equation}
which yields a parameterization of the form
\begin{equation}
{\cal L}(S) = \frac{e^{-(S \times S_M + S_0)} (S \times S_M + S_0)^{N_{\rm eff}}}{N_{\rm eff}!}. \label{intcurvefit}
\end{equation}

\begin{figure}
\centering
\includegraphics[width = 5in]{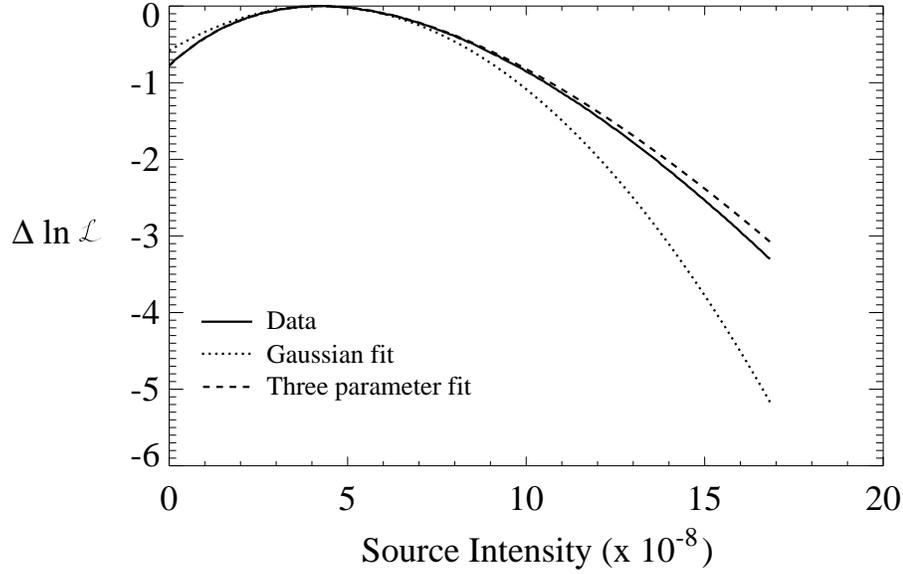}
\caption[Log Likelihood vs Intensity]{\label{intcurvefig}
The likelihood curve for a sample source, together with a Gaussian of the
same width, and a curve of the form in \eq{intcurvefit}
}
\end{figure}

This form fits the observed likelihood functions rather well
(\fig{intcurvefig})
, both in cases of large source flux and in upper limit situations.

Rather than providing $S_M$, $S_0$, and $N_{\rm eff}$ for source measurements,
it is convenient to transform to a set of variables more related to
the measured source flux and the error in that measurement.
It can be shown that the maximum likelihood value of $S$ occurs at
\begin{equation}
\mu_S = \frac{N_{\rm eff} - S_0}{S_M} ,
\end{equation}
and that the second derivative
of the likelihood function (which equals the standard deviation if the
distribution is Gaussian) is
\begin{equation}
\sigma_S = \frac{\sqrt{N_{\rm eff}}}{S_M}.
\end{equation}
Thus the likelihood as a function of source flux in a given viewing
period can also be characterized by $\mu_S$, $\sigma_S$, and $S_M$.

\subsection{Calculating $\tau$ \label{taucalc}}

With the parameterized distribution of source fluxes from the preceding
section, we can write the likelihood of observing a given viewing period's
worth of data ($D^i$) from a source with a certain $\mu$ and $\tau$:
\begin{eqnarray}
{\cal L}(\mu, \tau, D^i) & = & \int_0^\infty dS^i\, P(D^i | S^i) P(S^i|\mu, \tau) \\
& = &  \int_0^\infty dS^i\,  \frac{e^{-(S^i \times S_M^i + S_0^i)}
 (S^i \times S_M^i + S_0^i)^{N_{\rm eff}^i}}{N_{\rm eff}^i!} \nonumber \\
 & & \times \frac{1}{S^i \sqrt{2 \pi \ln(\tau^2 +1)}}
     \left[\frac{\mu}{S^i \sqrt{\tau^2+1}} \right]^{\frac{1}{2 \ln(\tau^2+1)}},
\label{tauonevp}
\end{eqnarray}
where the second term in \eq{tauonevp} is the Log-normal distribution
written in terms of $\mu$ and $\tau$.

The likelihood of a sequence of observations ($D^1 \dots D^M$) will then
be 
\[
{\cal L}(\mu, \tau) = \prod_{i=1}^M {\cal L}(\mu, \tau, D^i),
\]
or
\begin{equation}
\ln {\cal L}(\mu, \tau) = \sum_{i=1}^M \ln {\cal L}(\mu, \tau, D^i).
\end{equation}

\subsection{Dealing with Time Scales \label{timescales}}

Up to this point we have assumed that the source flux during each measurement
was drawn independently from the same distribution.  This can be problematic
if the measurements last for varying lengths of time.  For example, in the
\egret\ data, there are frequently several successive viewing periods of
a few days duration, pointing in nearly the same direction.  This is in
contrast to the normal mode of pointing in one direction for two weeks at
a time.  We do not want our results to depend on whether or not a two week
viewing period was chopped into several bits.

The simplest solution is to lump together successive measurements, when they
are on a short enough time scale, and treat them as one
measurement.  This simply involves replacing the first term in \eq{tauonevp}
with the product of several identical terms, one for each sub-measurement.

\section{Extensions of the Method \label{variextensions}}

There are several ways in which this method could be extended.  One of the
simplest would be to use a different distribution than the log-normal as
the distribution of source fluxes.  One could experimentally measure such
distributions in hopes of using a more accurate one.

A much more complex (but probably more useful) extension would be to
add time dependent correlations to the model.  Thus the true source flux
as a function of time would be modeled as a time series with a certain
spectrum in its correlation function.  
This would be much more difficult computationally, and one would have to
pick an appropriate spectrum for the correlation.  The benefit, however,
would be a much more realistic model of the variability.


\chapter{Time Variability of the \egret\ Sources \label{variegret}}

\section{Introduction}

In this chapter we apply the variability analysis described in
\chapt{varimethod} to the sources listed in the third \egret\ catalog~\cite{cat3}.
 Variability on time scales longer than one month
is examined for each source in the catalog, and the variability of
various source classes is compared. 

\section{Method \label{method}}

The first step in this analysis was the re-analysis of the third
\egret\ catalog with the \unlikeprog\ program.  For each Viewing Period,
any of the 271 catalog sources which were within the instrument's field
of view were added to the source list.  In addition, any of the 145 marginal
sources (which were used in the original analysis of the catalog but
were not significant enough to merit inclusion in the official list) which
were in the field of view were added as well.  The maximum likelihood
set of source fluxes was then found.  For each source, the parameters
characterizing ${\cal L}(S)$ described in \sect{intcurve} were found.
All source measurements were of flux ($> 100\ {\rm MeV}$), assuming a spectral
index of 2.0, using all data within $15 \deg$ of each source, and included
all nearby sources (both catalog and marginal) in the model.  All data
collected through VP6111 (corresponding to 18 Feb 1997) were used.

The likelihood of the sequence of observations as a function of $\tau$
was then calculated, as in \sect{taucalc}.  All observations within one
month of each other were lumped together, as per \sect{timescales}.

\section{Systematic Errors}

There are two main sources of systematic error in this analysis.  The
first is from the instrument itself.  As discussed
elsewhere~\cite{cat2,cat3,inflightcalibrate},
the sensitivity of the \egret\ instrument is time variable.  Despite
efforts to measure and compensate for long term drifts~\cite{inflightcalibrate},
 there is a residual variability of about 10\%.
In an effort to minimize this systematic error, only source observations
within $25 \deg$ of the instrument axis were used in the analysis.

Because of this variability, the background levels $G_M$ and $G_B$
were allowed to vary from viewing period to viewing period for a given
source.  Otherwise, a $5\%$ change in a strong background could lead to
a large change in a source flux.

The background levels were also allowed to float when ${\cal L}(S)$
was calculated.  Unfortunately, computational limitations precluded allowing
nearby sources to vary during this calculation as well.  Thus, nearby
sources were fixed at their maximum likelihood intensities (calculated
for that viewing period).  This will lead to an underestimation of the
error in flux, particularly for sources which have another source very close
by.

As both systematic effects will tend to exaggerate source variability,
it is expected that even constant sources will show some small variability.
To the extent that the systematics are dominated by the instrumental
effects, the measured variability will be equal to the true variability 
added in quadrature to a constant systematic variability of about 10\%.

\section{Results \label{variresults}}

A list of the Third \egret\ Catalog sources, together with
their measured variabilities, is given in \tbl{vartbl}.  For each source,
the 68\% upper and lower limits on $\tau$ are given as well, together
with the mean of the best fit log-normal flux distribution, and the number
of observations (within $25\deg$) used in the analysis.
 
{
  \footnotesize
  \renewcommand{\thefootnote}{\fnsymbol{footnote}}
  \tablefirsthead{\hline
                  3EG Name &$\ \ \ \ \ \ell$\ \ \ \ &\ \ \ \ b\ \ \ \ &
                  \ \ \ \ $\tau$\ \ \ \ \ & L limit& U limit &
                  Avg Flux\footnote[1]{$\times 10^{-7}$ \perareasec ($E>100 MeV$)} &
                  Num Obs \\ \hline\hline}

  \tablehead{\multicolumn{8}{l}{\small\sl \tbl{vartbl} -- {\it continued}}\\
             \hline
             3EG Name &$\ \ \ \ \ \ell$\ \ \ \ \ &\ \ \ \ b\ \ \ \ &
             $\ \ \ \ \tau$ \ \ \ &  L limit& U limit &
             Avg Flux & Num Obs \\ \hline\hline}
  \tabletail{%
             \hline
             \multicolumn{8}{r}{\tbl{vartbl} \ \ \ \ \ \ \ \ \ \ \ \ \ \ \ \small\sl {\it continued on next page}}\\}
  \tablelasttail{\hline}
  \bottomcaption{Variability $\tau$ of sources in the Third \egret\ catalog,
    together with the bounds of the 68\% confidence region.  The fitted average source
    flux and number of observations used in calculating $\tau$ are also given.
  }
\vspace{.25 in}
  \begin{mpsupertabular}[htb]{|l|r|r|r|r|r|r|r|}
  \label{vartbl}
  J0010+7309 & 119.92 &  10.54 &     0.31 &  0.16 &     0.55 &   3.99 & 4 \\
J0038-0949 & 112.69 & -72.44 &     0.00 &  0.00 &     0.89 &   1.24 & 3 \\
J0118+0248 & 136.23 & -59.36 &     5.17 &  0.90 & $\infty$\footnote[2]{All values larger than 10,000 are printed as $\infty$} &   0.92 & 4 \\ 
J0130-1758 & 169.71 & -77.11 &     0.00 &  0.00 &     0.38 &   1.19 & 4 \\
J0159-3603 & 248.89 & -73.04 &     0.00 &  0.00 &     1.16 &   0.79 & 4 \\
J0204+1458 & 147.95 & -44.32 &     6.14 &  1.08 & $\infty$ &   1.62 & 3 \\ 
J0210-5055 & 276.10 & -61.78 &     0.31 &  0.21 &     0.48 &   8.80 & 7 \\
J0215+1123 & 153.75 & -46.37 &    10.06 &  1.19 & $\infty$ &   1.39 & 3 \\ 
J0222+4253 & 140.14 & -16.77 &     0.00 &  0.00 &     0.23 &   2.02 & 4 \\
J0229+6151 & 134.20 &   1.15 &     0.39 &  0.16 &     0.74 &   3.57 & 7 \\
J0237+1635 & 156.77 & -39.11 &     1.16 &  0.62 &     3.88 &   2.98 & 3 \\
J0239+2815 & 150.21 & -28.80 &     0.00 &  0.00 &     0.24 &   1.60 & 5 \\
J0241+6103 & 135.87 &   0.99 &     0.49 &  0.33 &     0.79 &   6.44 & 7 \\
J0245+1758 & 157.62 & -37.11 &     2.63 &  0.73 &  2287.08 &   0.99 & 4 \\
J0253-0345 & 179.70 & -52.56 &    16.44 &  1.38 & $\infty$ &   2.74 & 2 \\ 
J0323+5122 & 145.64 &  -4.67 &     0.92 &  0.38 &     3.23 &   1.03 & 8 \\
J0329+2149 & 164.90 & -27.88 &     0.00 &  0.00 &     1.74 &   0.74 & 5 \\
J0340-0201 & 188.00 & -42.45 &     0.00 &  0.00 &     0.76 &   0.94 & 2 \\
J0348+3510 & 159.06 & -15.01 &     0.51 &  0.00 &     1.41 &   1.27 & 6 \\
J0348-5708 & 269.35 & -46.79 &     6.60 &  1.29 & $\infty$ &   0.51 & 10 \\ 
J0404+0700 & 184.00 & -32.15 &     0.34 &  0.00 &     1.65 &   0.94 & 7 \\
J0407+1710 & 175.63 & -25.06 &     1.33 &  0.27 &    19.75 &   0.59 & 11 \\
J0412-1853 & 213.90 & -43.29 &     0.00 &  0.00 & $\infty$ &   0.06 & 2 \\
J0416+3650 & 162.22 &  -9.97 &     0.63 &  0.32 &     1.17 &   1.74 & 9 \\
J0422-0102 & 194.88 & -33.12 &     0.00 &  0.00 &     0.57 &   1.03 & 5 \\
J0423+1707 & 178.48 & -22.14 &     0.51 &  0.10 &     1.08 &   1.35 & 12 \\
J0426+1333 & 181.98 & -23.82 &     0.29 &  0.00 &     0.71 &   1.71 & 14 \\
J0429+0337 & 191.44 & -29.08 &     0.00 &  0.00 &     0.45 &   1.54 & 9 \\
J0433+2908 & 170.48 & -12.58 &     0.39 &  0.21 &     0.63 &   2.70 & 13 \\
J0435+6137 & 146.50 &   9.50 &     0.00 &  0.00 &     0.30 &   1.57 & 9 \\
J0439+1105 & 186.14 & -22.87 &     0.00 &  0.00 &     0.42 &   1.06 & 14 \\
J0439+1555 & 181.98 & -19.98 &     1.70 &  0.74 &    12.73 &   0.69 & 14 \\
J0442-0033 & 197.20 & -28.46 & $\infty$ & 11.21 & $\infty$ &  92.90 & 9 \\ 
J0450+1105 & 187.86 & -20.62 &     1.60 &  0.99 &     3.50 &   1.78 & 14 \\
J0456-2338 & 223.96 & -34.98 &     0.16 &  0.00 &     1.07 &   0.95 & 5 \\
J0458-4635 & 252.40 & -38.40 &     0.50 &  0.00 &     1.26 &   0.83 & 7 \\
J0459+0544 & 193.99 & -21.66 &     0.76 &  0.08 &     2.67 &   0.55 & 12 \\
J0459+3352 & 170.25 &  -5.77 &     0.56 &  0.27 &     1.09 &   1.46 & 14 \\
J0500+2529 & 177.18 & -10.28 &     0.00 &  0.00 &     1.08 &   0.51 & 14 \\
J0500-0159 & 201.35 & -25.47 &     2.20 &  0.92 &    19.25 &   1.03 & 10 \\
J0510+5545 & 153.99 &   9.42 &     0.00 &  0.00 &     0.26 &   2.16 & 9 \\
J0512-6150 & 271.25 & -35.28 &     0.00 &  0.00 &     0.55 &   0.55 & 10 \\
J0516+2320 & 181.15 &  -8.77 &     9.43 &  1.49 & $\infty$ &   0.48 & 14 \\ 
J0520+2556 & 179.65 &  -6.40 &     0.00 &  0.00 &     0.31 &   1.33 & 14 \\
J0521+2147 & 183.08 &  -8.43 &     0.35 &  0.00 &     0.69 &   1.76 & 14 \\
J0530+1323 & 191.37 & -11.01 &     0.74 &  0.58 &     1.00 &   8.04 & 15 \\
J0530-3626 & 240.94 & -31.29 &     0.61 &  0.15 &     2.28 &   1.34 & 3 \\
J0531-2940 & 233.44 & -29.31 &     0.93 &  0.22 &     8.07 &   0.81 & 4 \\
J0533+4751 & 162.61 &   7.95 &     0.00 &  0.00 &     0.32 &   1.40 & 9 \\
J0533-6916 & 279.73 & -32.09 &     0.00 &  0.00 &     0.60 &   1.65 & 9 \\
J0534+2200 & 184.56 &  -5.78 &     0.00 &  0.06 &     0.10 &  22.47 & 14 \\
J0540-4402 & 250.08 & -31.09 &     0.85 &  0.56 &     1.48 &   3.32 & 7 \\
J0542+2610 & 182.02 &  -1.99 &     0.70 &  0.34 &     1.40 &   1.49 & 14 \\
J0542-0655 & 211.28 & -18.52 &    26.39 &  2.24 & $\infty$ &   2.39 & 8 \\ 
J0546+3948 & 170.75 &   5.74 &     0.11 &  0.00 &     0.47 &   1.71 & 12 \\
J0556+0409 & 202.81 & -10.29 &    50.00 &  0.00 & $\infty$ &   0.00 & 13 \\ 
J0613+4201 & 171.32 &  11.40 &     0.71 &  0.28 &     1.64 &   0.99 & 11 \\
J0616-0720 & 215.58 & -11.06 &     0.00 &  0.00 &     2.04 &   0.47 & 10 \\
J0616-3310 & 240.35 & -21.24 &     0.16 &  0.00 &     0.75 &   1.92 & 3 \\
J0617+2238 & 189.00 &   3.05 &     0.26 &  0.15 &     0.38 &   5.36 & 14 \\
J0622-1139 & 220.16 & -11.69 &     1.35 &  0.45 &    11.23 &   1.49 & 6 \\
J0628+1847 & 193.66 &   3.64 &    50.00 &  0.00 & $\infty$ &   0.00\footnote[3]{A fitted mean flux of zero indicates either an unphysically strict upper limit, due to instrument systematics, or numerical convergence problems with the method.  Such sources are most likely variable, but caution should be used in interpreting the results.} & 12 \\ 
J0631+0642 & 204.71 &  -1.30 &    75.89 &  7.89 & $\infty$ &   0.00 & 16 \\ 
J0633+1751 & 195.13 &   4.27 &     0.12 &  0.10 &     0.16 &  39.99 & 12 \\
J0634+0521 & 206.18 &  -1.41 &    72.05 &  5.15 & $\infty$ &   0.00 & 16 \\ 
J0702-6212 & 272.65 & -22.56 &     1.55 &  0.67 &     8.37 &   0.68 & 11 \\
J0706-3837 & 249.57 & -13.76 &     0.00 &  0.00 &  1731.83 &   0.36 & 4 \\
J0721+7120 & 143.98 &  28.02 &     0.08 &  0.00 &     0.28 &   1.89 & 7 \\
J0724-4713 & 259.00 & -14.38 &     2.53 &  0.87 &    91.40 &   1.14 & 6 \\
J0725-5140 & 263.29 & -16.02 &     1.09 &  0.51 &     4.05 &   0.87 & 8 \\
J0737+1721 & 201.85 &  18.07 &     0.00 &  0.00 &     0.49 &   1.53 & 5 \\
J0743+5447 & 162.99 &  29.19 &     2.46 &  0.99 &    32.90 &   1.10 & 7 \\
J0747-3412 & 249.35 &  -4.48 &     0.58 &  0.00 &     1.64 &   1.98 & 7 \\
J0808+4844 & 170.46 &  32.48 &     0.00 &  0.00 &     0.39 &   1.02 & 7 \\
J0808+5114 & 167.51 &  32.66 &     0.00 &  0.00 &     0.73 &   0.62 & 6 \\
J0808-5344 & 268.24 & -11.20 &     1.42 &  0.69 &     4.93 &   1.20 & 10 \\
J0812-0646 & 228.64 &  14.62 &     0.00 &  0.00 &     0.66 &   2.33 & 3 \\
J0821-5814 & 273.10 & -12.04 &     5.75 &  1.14 & $\infty$ &   0.59 & 10 \\ 
J0824-4610 & 263.28 &  -4.89 &     1.25 &  0.00 &    36.72 &   1.28 & 9 \\
J0827-4247 & 260.84 &  -2.46 &     0.00 &  0.00 &     0.54 &   1.88 & 8 \\
J0828+0508 & 219.60 &  23.82 &     0.00 &  0.00 & $\infty$ &   0.50 & 4 \\ 
J0828-4954 & 266.76 &  -6.45 &     0.62 &  0.28 &     1.43 &   1.92 & 9 \\
J0829+2413 & 200.02 &  31.87 &     0.54 &  0.28 &     1.27 &   4.48 & 2 \\
J0834-4511 & 263.55 &  -2.79 &     0.17 &  0.12 &     0.24 & 101.07 & 9 \\
J0841-4356 & 263.29 &  -1.10 &     4.03 &  0.83 & $\infty$ &   1.44 & 8 \\ 
J0845+7049 & 143.54 &  34.43 &     0.65 &  0.33 &     1.40 &   0.93 & 6 \\
J0848-4429 & 264.50 &  -0.46 &     1.31 &  0.19 &    33.43 &   1.26 & 8 \\
J0852-1216 & 239.06 &  19.99 &     2.06 &  0.74 &   147.20 &   1.39 & 5 \\
J0853+1941 & 206.81 &  35.82 &     0.00 &  0.00 &     0.37 &   1.11 & 4 \\
J0859-4257 & 264.57 &   2.01 &     0.23 &  0.00 &     0.73 &   2.36 & 8 \\
J0903-3531 & 259.40 &   7.40 &     0.52 &  0.19 &     1.20 &   1.69 & 7 \\
J0910+6556 & 148.30 &  38.56 &     0.49 &  0.00 &     1.14 &   0.79 & 6 \\
J0917+4427 & 176.11 &  44.19 &     0.00 &  0.00 &     0.34 &   1.63 & 6 \\
J0952+5501 & 159.55 &  47.33 &     0.46 &  0.00 &     0.84 &   1.16 & 10 \\
J0958+6533 & 145.75 &  43.13 &     1.17 &  0.55 &     5.06 &   0.64 & 8 \\
J1009+4855 & 166.87 &  51.99 &     0.00 &  0.00 &     0.60 &   0.51 & 9 \\
J1013-5915 & 283.93 &  -2.34 &     0.22 &  0.00 &     0.46 &   4.65 & 11 \\
J1014-5705 & 282.80 &  -0.51 &     0.65 &  0.33 &     1.13 &   4.17 & 12 \\
J1027-5817 & 284.94 &  -0.52 &     0.26 &  0.09 &     0.45 &   6.93 & 11 \\
J1045-7630 & 295.66 & -15.45 &     0.00 &  0.00 &     0.28 &   1.52 & 9 \\
J1048-5840 & 287.53 &   0.47 &     0.00 &  0.00 &     0.18 &   5.93 & 10 \\
J1052+5718 & 149.47 &  53.27 &     0.21 &  0.00 &     0.74 &   0.66 & 6 \\
J1058-5234 & 285.98 &   6.65 &     0.00 &  0.00 &     0.25 &   3.28 & 10 \\
J1102-6103 & 290.12 &  -0.92 &     0.00 &  0.00 &     0.90 &   2.78 & 9 \\
J1104+3809 & 179.83 &  65.03 &     0.33 &  0.05 &     0.63 &   1.41 & 8 \\
J1133+0033 & 264.52 &  57.48 &     0.71 &  0.16 &     2.00 &   0.58 & 11 \\
J1134-1530 & 277.04 &  43.48 &     2.85 &  1.11 &    51.53 &   1.12 & 10 \\
J1200+2847 & 199.42 &  78.38 &     2.36 &  1.16 &    11.36 &   1.32 & 10 \\
J1212+2304 & 235.57 &  80.32 &    78.82 &  0.00 & $\infty$ &   0.00 & 13 \\ 
J1219-1520 & 291.56 &  46.82 &     1.78 &  0.74 &    13.70 &   0.56 & 12 \\
J1222+2315 & 241.87 &  82.39 &    61.09 &  2.44 & $\infty$ &   0.48 & 12 \\ 
J1222+2841 & 197.27 &  83.52 &     0.58 &  0.31 &     1.02 &   1.51 & 9 \\
J1224+2118 & 255.07 &  81.66 &     0.47 &  0.20 &     0.90 &   1.27 & 11 \\
J1227+4302 & 138.63 &  73.33 &   136.23 &  2.90 & $\infty$ &   3.20 & 5 \\ 
J1229+0210 & 289.95 &  64.36 &     0.51 &  0.35 &     0.77 &   2.30 & 11 \\
J1230-0247 & 292.58 &  59.66 &     0.60 &  0.27 &     1.25 &   0.66 & 11 \\
J1234-1318 & 296.43 &  49.34 &     0.42 &  0.12 &     0.81 &   0.89 & 11 \\
J1235+0233 & 293.28 &  65.13 &     0.23 &  0.00 &     0.65 &   0.66 & 11 \\
J1236+0457 & 292.59 &  67.52 &     0.00 &  0.00 &     1.45 &   0.38 & 11 \\
J1246-0651 & 300.96 &  55.99 &     0.59 &  0.33 &     1.03 &   1.08 & 11 \\
J1249-8330 & 302.86 & -20.63 &     0.71 &  0.18 &     2.74 &   1.12 & 6 \\
J1255-0549 & 305.09 &  57.06 &     1.60 &  1.11 &     2.71 &   5.99 & 11 \\
J1300-4406 & 304.60 &  18.74 &     0.48 &  0.00 &     1.57 &   0.70 & 8 \\
J1308+8744 & 122.74 &  29.38 &     0.00 &  0.00 &     0.57 &   0.59 & 9 \\
J1308-6112 & 305.01 &   1.59 &     0.72 &  0.38 &     1.53 &   3.07 & 10 \\
J1310-0517 & 311.69 &  57.25 &     2.94 &  1.69 &     7.92 &   4.32 & 11 \\
J1314-3431 & 308.21 &  28.12 &     0.00 &  0.00 &     0.25 &   1.51 & 6 \\
J1316-5244 & 306.85 &   9.93 &     0.39 &  0.00 &     0.81 &   1.83 & 9 \\
J1323+2200 & 359.33 &  81.15 &     2.69 &  0.93 &    46.77 &   0.79 & 9 \\
J1324-4314 & 309.32 &  19.21 &     0.00 &  0.00 &     0.30 &   1.42 & 8 \\
J1329+1708 & 346.29 &  76.68 &     0.60 &  0.00 &     5.51 &   0.21 & 13 \\
J1329-4602 & 309.83 &  16.32 &     0.00 &  0.00 &     0.48 &   1.02 & 8 \\
J1337+5029 & 105.40 &  65.04 &     0.54 &  0.00 &     1.35 &   1.08 & 6 \\
J1339-1419 & 320.07 &  46.95 &     0.68 &  0.17 &     1.70 &   0.69 & 10 \\
J1347+2932 &  47.32 &  77.50 &     0.48 &  0.00 &     1.45 &   1.14 & 7 \\
J1409-0745 & 333.88 &  50.28 &    14.43 &  3.39 &   917.88 &   6.16 & 11 \\
J1410-6147 & 312.18 &  -0.35 &     0.33 &  0.16 &     0.55 &   8.95 & 8 \\
J1420-6038 & 313.63 &   0.37 &     1.22 &  0.51 &     7.43 &   3.52 & 8 \\
J1424+3734 &  66.82 &  67.76 &     0.01 &  0.00 & $\infty$ &   1.50 & 1 \\ 
J1429-4217 & 321.45 &  17.27 &     1.11 &  0.58 &     3.20 &   1.31 & 10 \\
J1447-3936 & 326.12 &  17.96 &     0.27 &  0.00 &     0.85 &   1.13 & 11 \\
J1457-1903 & 339.88 &  34.60 &     0.42 &  0.00 &     3.64 &   0.68 & 6 \\
J1500-3509 & 330.91 &  20.45 &     0.00 &  0.00 &     0.61 &   1.03 & 11 \\
J1504-1537 & 344.04 &  36.38 &    10.33 &  1.21 & $\infty$ &   1.47 & 5 \\ 
J1512-0849 & 351.49 &  40.37 &     0.00 &  0.00 &     0.32 &   2.35 & 5 \\
J1517-2538 & 339.76 &  26.60 &     0.00 &  0.00 &     8.90 &   0.43 & 8 \\
J1527-2358 & 342.97 &  26.50 &     4.04 &  0.00 & $\infty$ &   0.49 & 11 \\ 
J1600-0351 &   6.30 &  34.81 &    68.17 &  0.00 & $\infty$ &   0.00 & 4 \\ 
J1605+1553 &  29.18 &  43.84 &     0.00 &  0.00 &     0.56 &   1.45 & 4 \\
J1607-1101 &   0.91 &  29.05 &  2370.92 &  6.19 & $\infty$ &  16.37 & 11 \\ 
J1608+1055 &  23.03 &  40.79 &     1.92 &  0.45 &  4993.42 &   1.91 & 2 \\
J1612-2618 & 349.76 &  17.64 &     1.78 &  0.76 &    11.74 &   1.02 & 15 \\
J1614+3424 &  55.15 &  46.38 &     0.73 &  0.45 &     1.36 &   4.03 & 5 \\
J1616-2221 & 353.00 &  20.03 &     0.00 &  0.00 &     0.33 &   1.56 & 15 \\
J1621+8203 & 115.53 &  31.77 &     0.00 &  0.00 &     0.29 &   1.16 & 5 \\
J1625-2955 & 348.82 &  13.32 &     4.56 &  2.23 &    18.18 &   8.80 & 15 \\
J1626-2519 & 352.14 &  16.32 &     1.64 &  0.79 &     6.02 &   1.65 & 16 \\
J1627-2419 & 353.36 &  16.71 &     0.00 &  0.00 &     0.35 &   2.10 & 15 \\
J1631-1018 &   5.55 &  24.94 &     0.38 &  0.00 &     0.89 &   1.34 & 13 \\
J1631-4033 & 341.61 &   5.24 &     0.62 &  0.23 &     1.28 &   1.76 & 16 \\
J1633-3216 & 348.10 &  10.48 &     0.51 &  0.00 &     1.09 &   1.30 & 16 \\
J1634-1434 &   2.33 &  21.78 &     0.00 &  0.00 &     0.55 &   1.08 & 13 \\
J1635+3813 &  61.09 &  42.34 &     0.35 &  0.13 &     0.71 &   2.28 & 5 \\
J1635-1751 & 359.72 &  19.56 &     2.38 &  0.00 & $\infty$ &   0.30 & 16 \\ 
J1638-2749 & 352.25 &  12.59 &     0.45 &  0.23 &     0.76 &   2.03 & 17 \\
J1638-5155 & 334.05 &  -3.34 &     0.00 &  0.00 &     0.69 &   4.55 & 15 \\
J1639-4702 & 337.75 &  -0.15 &     0.00 &  0.00 &     0.38 &   6.63 & 14 \\
J1646-0704 &  10.85 &  23.69 &     0.68 &  0.28 &     1.42 &   1.55 & 13 \\
J1649-1611 &   3.35 &  17.80 &     0.54 &  0.00 &     1.40 &   1.28 & 17 \\
J1652-0223 &  15.99 &  25.05 &     0.00 &  0.00 &     0.62 &   1.27 & 8 \\
J1653-2133 & 359.49 &  13.81 &     1.34 &  0.65 &     4.65 &   0.91 & 17 \\
J1655-4554 & 340.48 &  -1.61 &     0.91 &  0.46 &     2.27 &   2.94 & 15 \\
J1659-6251 & 327.32 & -12.47 &     0.75 &  0.00 &     3.31 &   1.20 & 8 \\
J1704-4732 & 340.10 &  -3.79 &    29.90 &  2.67 & $\infty$ &   2.93 & 15 \\ 
J1709-0828 &  12.86 &  18.25 &     0.84 &  0.11 &     2.21 &   1.18 & 13 \\
J1710-4439 & 343.10 &  -2.69 &     0.16 &  0.06 &     0.27 &  13.49 & 17 \\
J1714-3857 & 348.04 &  -0.09 &     0.15 &  0.00 &     0.38 &   6.01 & 16 \\
J1717-2737 & 357.67 &   5.95 &     0.99 &  0.60 &     1.90 &   2.20 & 18 \\
J1718-3313 & 353.20 &   2.56 &     0.93 &  0.54 &     1.85 &   2.59 & 18 \\
J1719-0430 &  17.80 &  18.17 &     0.00 &  0.00 &     0.39 &   1.63 & 10 \\
J1720-7820 & 314.56 & -22.17 &     0.65 &  0.00 & $\infty$ &   0.37 & 3 \\ 
J1726-0807 &  15.52 &  14.77 &     0.35 &  0.00 &     0.81 &   1.93 & 14 \\
J1727+0429 &  27.27 &  20.62 &     0.31 &  0.00 &     0.87 &   1.84 & 5 \\
J1733+6017 &  89.12 &  32.94 &     0.39 &  0.00 &     1.38 &   1.29 & 6 \\
J1733-1313 &  12.03 &  10.81 &     0.31 &  0.17 &     0.49 &   3.96 & 14 \\
J1734-3232 & 355.64 &   0.15 &     0.00 &  0.00 &     0.24 &   4.12 & 19 \\
J1735-1500 &  10.73 &   9.22 &     1.09 &  0.00 &    10.14 &   0.52 & 16 \\
J1736-2908 & 358.79 &   1.56 &     0.66 &  0.40 &     1.09 &   4.13 & 18 \\
J1738+5203 &  79.37 &  32.05 &     1.01 &  0.49 &     3.12 &   1.91 & 6 \\
J1741-2050 &   6.44 &   5.00 &     0.41 &  0.14 &     0.70 &   2.74 & 17 \\
J1741-2312 &   4.42 &   3.76 &     0.52 &  0.18 &     1.03 &   2.17 & 18 \\
J1744-0310 &  22.19 &  13.42 &     0.72 &  0.26 &     1.94 &   1.20 & 9 \\
J1744-3011 & 358.85 &  -0.52 &     0.38 &  0.20 &     0.62 &   7.74 & 18 \\
J1744-3934 & 350.81 &  -5.38 &     0.63 &  0.28 &     1.17 &   2.42 & 15 \\
J1746-1001 &  16.34 &   9.64 &     0.32 &  0.00 &     0.60 &   2.50 & 14 \\
J1746-2851 &   0.11 &  -0.04 &     0.50 &  0.36 &     0.69 &  12.04 & 18 \\
J1757-0711 &  20.30 &   8.47 &     0.52 &  0.00 &     0.99 &   2.37 & 16 \\
J1800-0146 &  25.49 &  10.39 &     0.00 &  0.00 &     0.48 &   1.99 & 10 \\
J1800-2338 &   6.25 &  -0.18 &     0.03 &  0.00 &     0.32 &   6.11 & 19 \\
J1800-3955 & 352.45 &  -8.43 &     0.00 &  0.00 &     0.84 &   0.77 & 16 \\
J1806-5005 & 343.29 & -13.76 &     1.30 &  0.00 &    39.40 &   0.46 & 12 \\
J1809-2328 &   7.47 &  -1.99 &     0.69 &  0.49 &     1.02 &   6.06 & 18 \\
J1810-1032 &  18.81 &   4.23 &     0.35 &  0.00 &     0.79 &   2.16 & 16 \\
J1812-1316 &  16.70 &   2.39 &     0.94 &  0.61 &     1.65 &   3.80 & 17 \\
J1813-6419 & 330.04 & -20.32 &     0.00 &  0.00 &     0.54 &   1.65 & 4 \\
J1822+1641 &  44.84 &  13.84 &     1.91 &  0.77 &    26.63 &   1.98 & 6 \\
J1823-1314 &  17.94 &   0.14 &     0.72 &  0.40 &     1.37 &   4.06 & 16 \\
J1824+3441 &  62.49 &  20.14 &    68.64 &  6.86 & $\infty$ &   0.00 & 11 \\ 
J1824-1514 &  16.37 &  -1.16 &     0.00 &  0.00 &     0.51 &   3.60 & 16 \\
J1825+2854 &  56.79 &  18.03 &    73.19 &  2.59 & $\infty$ &   0.00 & 10 \\ 
J1825-7926 & 314.56 & -25.44 &     0.75 &  0.22 &     3.06 &   2.32 & 4 \\
J1826-1302 &  18.47 &  -0.44 &     0.75 &  0.49 &     1.28 &   5.74 & 16 \\
J1828+0142 &  31.90 &   5.78 &  3982.51 &  6.92 & $\infty$ &  57.13 & 8 \\ 
J1832-2110 &  12.17 &  -5.71 &     0.52 &  0.25 &     0.92 &   2.32 & 15 \\
J1834-2803 &   5.92 &  -8.97 &     0.00 &  0.00 &     0.42 &   1.61 & 16 \\
J1835+5918 &  88.74 &  25.07 &     0.15 &  0.00 &     0.32 &   7.29 & 4 \\
J1836-4933 & 345.93 & -18.26 &     0.00 &  0.00 &     0.64 &   0.98 & 11 \\
J1837-0423 &  27.44 &   1.06 &    12.01 &  2.17 & $\infty$ &   5.88 & 10 \\ 
J1837-0606 &  25.86 &   0.40 &     0.24 &  0.00 &     0.49 &   6.24 & 10 \\
J1847-3219 &   3.21 & -13.37 &     0.80 &  0.33 &     1.92 &   0.91 & 16 \\
J1850+5903 &  88.92 &  23.18 &     0.00 &  0.00 & $\infty$ &   0.15 & 5 \\ 
J1850-2652 &   8.73 & -11.76 &     0.88 &  0.16 &     2.56 &   0.81 & 15 \\
J1856+0114 &  34.60 &  -0.54 &     0.80 &  0.50 &     1.51 &   9.94 & 8 \\
J1858-2137 &  14.21 & -11.15 &     0.00 &  0.00 &     0.56 &   1.12 & 14 \\
J1903+0550 &  39.52 &  -0.05 &     0.35 &  0.18 &     0.60 &   9.00 & 7 \\
J1904-1124 &  24.22 &  -8.12 &     0.24 &  0.00 &     0.67 &   1.60 & 12 \\
J1911-2000 &  16.87 & -13.22 &     0.27 &  0.00 &     0.60 &   1.72 & 13 \\
J1921-2015 &  17.81 & -15.60 &    52.62 &  2.67 & $\infty$ &   1.01 & 14 \\ 
J1928+1733 &  52.71 &   0.07 &     0.82 &  0.43 &     2.01 &   3.68 & 8 \\
J1935-4022 & 358.65 & -25.23 &     4.67 &  1.45 &   199.30 &   2.80 & 6 \\
J1937-1529 &  23.95 & -17.12 &    82.30 &  2.69 & $\infty$ &   2.01 & 10 \\ 
J1940-0121 &  37.41 & -11.62 &     4.58 &  1.13 & $\infty$ &   1.21 & 7 \\ 
J1949-3456 &   5.25 & -26.29 &     2.83 &  1.00 &   164.05 &   1.03 & 8 \\
J1955-1414 &  27.01 & -20.56 &     0.89 &  0.37 &     2.42 &   1.51 & 8 \\
J1958+2909 &  66.23 &  -0.16 &     0.46 &  0.15 &     0.98 &   2.73 & 10 \\
J1958-4443 & 354.85 & -30.13 &    58.02 &  5.85 & $\infty$ &   0.00 & 4 \\ 
J1959+6342 &  96.61 &  17.10 &     0.00 &  0.00 &     0.49 &   1.38 & 6 \\
J2006-2321 &  18.82 & -26.26 &    30.22 &  1.86 & $\infty$ &   1.82 & 6 \\ 
J2016+3657 &  74.76 &   0.98 &     0.37 &  0.08 &     0.75 &   3.80 & 10 \\
J2020+4017 &  78.05 &   2.08 &     0.07 &  0.00 &     0.18 &  12.04 & 10 \\
J2020-1545 &  28.09 & -26.62 &     0.00 &  0.00 &     0.80 &   0.76 & 4 \\
J2021+3716 &  75.58 &   0.33 &     0.29 &  0.11 &     0.53 &   5.66 & 10 \\
J2022+4317 &  80.63 &   3.62 &     0.13 &  0.00 &     0.50 &   2.45 & 10 \\
J2025-0744 &  36.90 & -24.38 &     1.10 &  0.57 &     3.58 &   3.22 & 4 \\
J2027+3429 &  74.08 &  -2.36 &     0.00 &  0.00 &     0.28 &   2.63 & 10 \\
J2033+4118 &  80.27 &   0.73 &     0.20 &  0.00 &     0.37 &   7.41 & 10 \\
J2034-3110 &  12.25 & -34.64 &     2.88 &  0.89 &   154.84 &   0.92 & 7 \\
J2035+4441 &  83.17 &   2.50 &     0.00 &  0.00 &     0.42 &   3.28 & 10 \\
J2036+1132 &  56.12 & -17.18 &     0.12 &  0.00 &     0.77 &   1.47 & 6 \\
J2046+0933 &  55.75 & -20.23 &     0.00 &  0.00 &     0.57 &   1.81 & 7 \\
J2055-4716 & 352.56 & -40.20 &     0.99 &  0.44 &     3.71 &   1.50 & 5 \\
J2100+6012 &  97.76 &   9.16 &     0.15 &  0.00 &     0.52 &   2.29 & 6 \\
J2158-3023 &  17.73 & -52.25 &     0.40 &  0.10 &     0.87 &   1.62 & 5 \\
J2202+4217 &  92.59 & -10.44 &     0.99 &  0.43 &     3.48 &   1.16 & 9 \\
J2206+6602 & 107.23 &   8.34 &     0.27 &  0.00 &     0.63 &   2.14 & 6 \\
J2209+2401 &  81.83 & -25.65 &     0.79 &  0.00 &     4.86 &   0.63 & 7 \\
J2219-7941 & 310.64 & -35.06 &     0.00 &  0.00 &     0.51 &   1.16 & 4 \\
J2227+6122 & 106.53 &   3.18 &     0.10 &  0.00 &     0.41 &   4.33 & 6 \\
J2232+1147 &  77.44 & -38.58 &     0.64 &  0.37 &     1.16 &   1.89 & 8 \\
J2241-6736 & 319.81 & -45.02 &     0.00 &  0.00 &     1.09 &   0.73 & 3 \\
J2243+1509 &  82.69 & -37.49 &     3.42 &  0.88 &  3097.37 &   0.63 & 8 \\
J2248+1745 &  86.00 & -36.17 &     1.07 &  0.43 &     3.98 &   1.00 & 8 \\
J2251-1341 &  52.48 & -58.91 &     9.49 &  1.58 & $\infty$ &   1.47 & 7 \\ 
J2254+1601 &  86.11 & -38.18 &     0.89 &  0.60 &     1.50 &   5.11 & 8 \\
J2255+1943 &  89.03 & -35.43 &     2.31 &  0.80 &    48.58 &   0.98 & 8 \\
J2255-5012 & 338.75 & -58.12 &     0.41 &  0.00 &     1.46 &   1.70 & 4 \\
J2314+4426 & 105.32 & -15.10 &     1.72 &  0.53 &   150.35 &   1.71 & 3 \\
J2321-0328 &  76.82 & -58.07 &    10.36 &  1.63 & $\infty$ &   1.76 & 6 \\ 
J2352+3752 & 110.26 & -23.54 &    24.92 &  1.68 & $\infty$ &   2.91 & 4 \\ 
J2358+4604 & 113.39 & -15.82 &     0.00 &  0.00 &     0.33 &   1.48 & 4 \\
J2359+2041 & 107.01 & -40.58 &     0.68 &  0.06 &     2.07 &   0.97 & 8 \\

  \end{mpsupertabular}
}

\vspace{.25 in}

\section{Variability by Source Class \label{classvar}}
As can be seen from \tbl{vartbl}, the limits on the variability of an individual
source are not very strict.  By grouping similar sources together,
one can more tightly constrain the confidence regions, but the results are
meaningful only if the sources truly have the same variability.

For this purpose, the sources were divided into four principal source classes:
Unidentified sources, Pulsars, Active Galactic Nuclei (AGN), and sources which
are spatially coincident with Supernova Remnants (SNR).  The Unidentified source
class was further divided according to the source latitude, and the
AGN class was divided according to the strength of the identification as an
AGN.  These source classes were determined from the ``source ID'' (and by the
``Other Name'' category for the SNR associations).

The added $\ln {\cal L}$ vs $\tau$ curves for the different classes are
shown in \fig{bigfour} below, together with the implied 68\% and 95\% confidence
limits on $\tau$. 

\begin{figure}[tbh]
\vspace{.35 in}
\includegraphics[width = 4in]{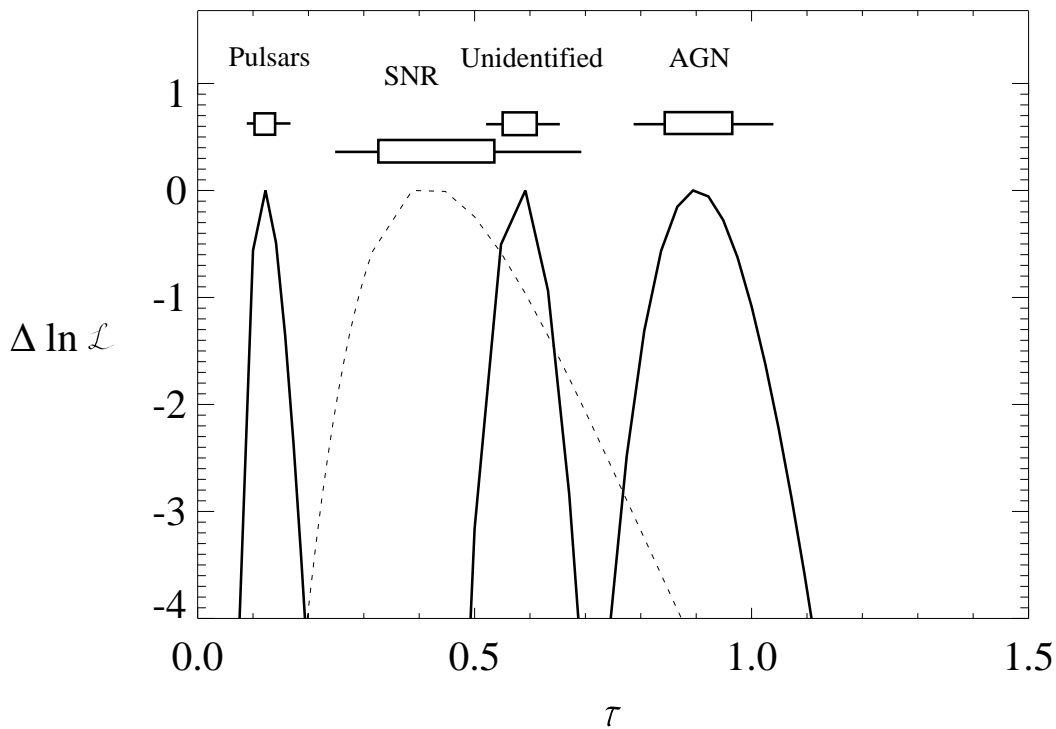}
\caption[Likelihood vs $\tau$ for four source classes]{\label{bigfour}
Log Likelihood vs Variability Index for the four principal source classes:
Pulsars, Supernova Remnant associations (dashed), Unidentified sources, and Active
Galactic Nuclei.  The 68\% and 95\% confidence regions of $\tau$ for each
class are indicated by the box and stem, respectively, under each name.
}
\end{figure}

As seen in \fig{bigfour}, the pulsars, AGN, and unidentified sources clearly
differ in their variability.  The variability of the five pulsars is about 12\%:
not significantly higher than the predicted systematic variability for a
constant sources of 10\%.

The six sources in the SNR source class are not distinguishable from the
unidentified sources, on the basis of variability, and show significantly more variation
than the pulsar class.  The generally weaker flux of these sources leads to the
large error bar for this source class.

\begin{figure}[tbhp]
\centering
\includegraphics[width = 4in]{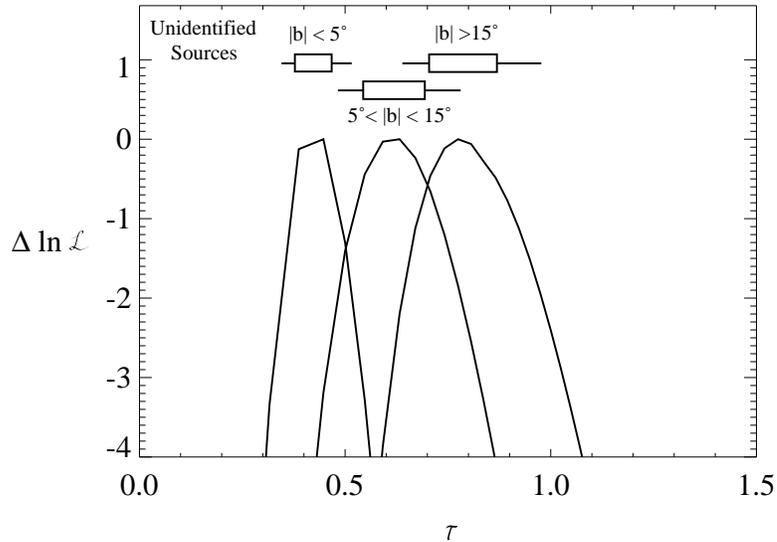}
\caption[Likelihood vs $\tau$ for the unidentified sources]{\label{unid}
Log Likelihood vs Variability Index for the unidentified sources.
The sources have been split into three classes based on source latitude, as
shown in the figure.  The 68\% and 95\% confidence regions of $\tau$ are
indicated by the box and stem as in \fig{bigfour}. 
}
\end{figure}

The change in unidentified source variability with latitude can be seen
in \fig{unid}.  It is clear from the spatial distribution of the unidentified
sources that there are multiple types of astrophysical object which are not
being identified~\cite{grenier}.  It is clear from \fig{unid} that the
different types of unidentified sources have different variabilities as well.
The high latitude sources have a variability index consistent with that of the
AGN.  The low latitude unidentified sources still exhibit variability,
at a level similar to the sources coincident with SNR.  Thus the data are not
consistent with the hypothesis that these sources are Geminga-like pulsars.
The comparatively small error regions for the unidentified source classes
stems from the number of unidentified sources.  There were 164 unidentified
sources: 49 in the low--latitude group, 39 in the mid--latitude group, and 76
in the high--latitude group.

\begin{figure}[tbhp]
\centering
\includegraphics[width = 4in]{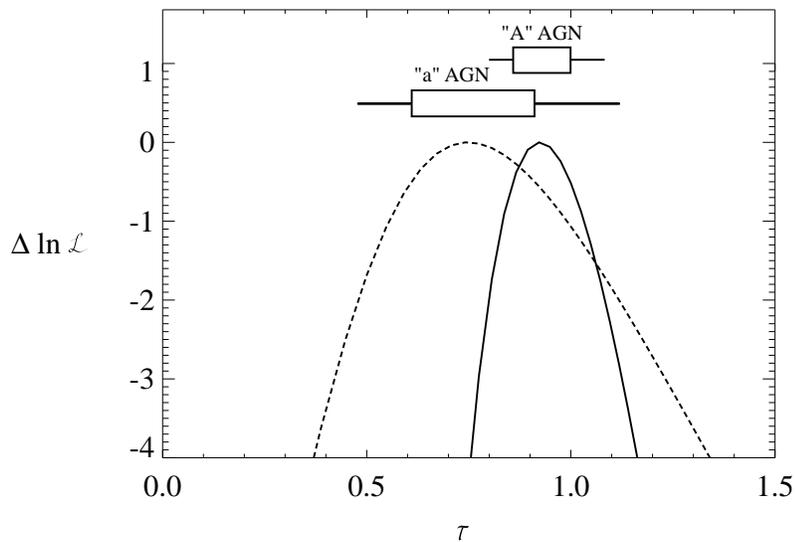}
\caption[Likelihood vs $\tau$ for the Active Galactic Nuclei]{\label{agnvar}
Log Likelihood vs Variability Index for the Active Galactic Nuclei.
The AGN have been split into two classes based on the confidence of their
identification in the Third \egret\ catalog: ``a'' for questionable identifications
(dashed), and ``A'' for confident identifications (solid).
The 68\% and 95\% confidence regions of $\tau$ are
indicated by the box and stem as in \fig{bigfour}. 
}
\end{figure}

In the Third \egret\ Catalog, 27 of the 94 sources identified as AGN are
listed as questionable identifications (and marked with an ``a'' rather
than an ``A'').  In an effort to see if many of these questionable identifications
were mis-identifications, the variability of these questionable identifications
was compared with the variability of the more confidently identified AGN.
As can be seen in \fig{agnvar}, the two classes are consistent in their
variability.  The slightly lower variability seen in the ``small a''
class may be due to a selection effect:  AGN which flare brightly are more
likely to be identified.

\section{The Low Variability Sources}

With the variability of the source classes in hand, it is possible to
see which sources can be excluded from a source class based on
the variability data.  We begin by examining those sources which
are inconsistent with being strongly variable: that is, the upper
bound of the 68\% region for $\tau$ is small.  The list of sources
where this upper bound is less than 0.3 (significantly lower than
the AGN or unidentified sources) is shown in \tbl{lowvartable}.

\begin{table}
\begin{minipage}{3 in}
\footnotesize
\begin{tabular}{|l|r@{.}l|r@{.}l|r@{.}l|r@{.}l|r@{.}l| l | l |}\hline
3EG Name & \multicolumn{2}{|c|}{$\tau$} & \multicolumn{4}{|c|}{Lower Limit} & \multicolumn{4}{|c|}{Upper Limit} & 3EG & Notes\\
 & \multicolumn{2}{|c|}{ } & \multicolumn{2}{|c}{ 95.5\% } & \multicolumn{2}{c|}{ 68\% } & \multicolumn{2}{|c}{68\%} & \multicolumn{2}{c|}{95.5\%} & Identification & \\ \hline
J0534+2200 &  0&071 &   0&018    &     0&058     &    0&104 &  0&147  &   Crab Pulsar & T\footnote[0]{T: See note in text}\\
J0633+1751 &  0&122 &   0&071    &     0&096     &    0&162 &  0&219  &   Geminga Pulsar&  \\
J2020+4017 &  0&075 &   0&000    &     0&000     &    0&181 &  0&321  &   UnID & C\footnote[0]{C: Listed as confused in 3EG}, GeV\footnote[0]{GeV: Detected above 1 GeV in (REF?)}, T \\ 
J1048-5840 &  0&000 &   0&000    &     0&000     &    0&184 &  0&403  &   UnID & C, GeV, T \\ 
J0222+4253 &  0&003 &   0&000    &     0&000     &    0&234 &  0&599  &   0219+428 & T  \\ 
J0834-4511 &  0&173 &   0&092    &     0&121     &    0&237 &  0&352  &   Vela Pulsar & T\\
J1734-3232 &  0&005 &   0&000    &     0&000     &    0&238 &  0&547  &   UnID& C, GeV \\
J0239+2815 &  0&000 &   0&000    &     0&000     &    0&242 &  0&679  &   0234+285?&   \\
J1058-5234 &  0&000 &   0&000    &     0&000     &    0&248 &  0&529  &   PSR B1055-52&  \\
J1314-3431 &  0&004 &   0&000    &     0&000     &    0&252 &  0&685  &   1313-333? & \\
J0510+5545 &  0&000 &   0&000    &     0&000     &    0&264 &  0&577  &   UnID& em\footnote[0]{em: Listed as extended morphology in 3EG} \\
J1710-4439 &  0&163 &   0&000    &     0&058     &    0&266 &  0&387  &   PSR 1706-44&  \\
J1045-7630 &  0&002 &   0&000    &     0&000     &    0&279 &  0&736  &   UnID& em, C \\
J0721+7120 &  0&083 &   0&000    &     0&000     &    0&282 &  0&536  &   0716+714& T \\
J2027+3429 &  0&005 &   0&000    &     0&000     &    0&282 &  0&753  &   UnID& em, C \\
J1621+8203 &  0&000 &   0&000    &     0&000     &    0&292 &  0&958  &   UnID& em, C \\
J0435+6137 &  0&004 &   0&000    &     0&000     &    0&299 &  0&772  &   UnID& em, C \\ \hline

\end{tabular}
\end{minipage}
\caption[Low Variability Sources]{\label{lowvartable}
Sources with low probability of being highly variable.  The sources are
sorted by the upper limit of the 68\% confidence region, and all sources where the
this limit is less than $0.3$ are shown.  The list includes all
five pulsars listed in the Third \egret\ catalog,
as well as two unidentified sources from which pulsation has been
reported (see notes in the text).  Detection above 1 GeV indicates a relatively
hard spectrum: another attribute of pulsars.  The extended morphology
and confused nature of several of the sources might indicate emission from
diffuse gas which is not accounted for in the gas map.
}
\end{table}

It is reassuring to see the bright pulsars at the top of the list;  the
possibility that others on the list might also be pulsars is intriguing.  The
sources are discussed individually below.

The position of the Crab pulsar, at the top of the list, is a bit surprising
given its listing in McLaughlin \etal\ as moderately variable~\cite{mclaughlin}.
There are two reasons for this.  The first is the difference in the methods:
McLaughlin \etal\ measure how inconsistent the source is with being
constant\footnote{More strictly, how inconsistent it is with being 6.5\% variable,
as they include a 6.5\% systematic error.}.  In this analysis, the Crab, Geminga,
and Vela pulsars are all inconsistent with being constant at the 95\% level.
The second reason is the difference in data used:  McLaughlin \etal\ used
all measurements up to to 30\deg\ from the instrument axis.  In this analysis,
we have used only data out to 25\deg, because of the larger systematic errors
at higher inclinations.

3EGJ2020+4017 is coincident with the $\gamma$-Cyg supernova remnant, and is thus
an intriguing pulsar candidate.  No pulsed signal has been detected from it,
however, and the possibility exists that it is a low variability SNR.

3EGJ1048-5840 is coincident with PSR1046-58, a high $\dot{E}/d^2$ pulsar.
Although pulsed emission was not seen in the first three years of \egret\ data~\cite{joethesis},
there has been a detection in recent data \cite{kaspi1046}.

3EGJ0222+4253 is coincident with the BL Lac 3C 66A, and is 1\deg\ from
the position of PSR0218+42, from which pulsed emission above 100 MeV has been
detected~\cite{verbunt, kuiper, hermsen}.  In the Third Catalog, the pulsar
was not detected as a separate source, and it is suggested that the
flux between 100 MeV and 1 GeV is primarily from the pulsar,
and the flux above 1 GeV is 
from 3C 66A~\cite{cat3}.  The flux measurements of this ``source'' will be
dominated by the lower energy emission, hence the low variability.

The apparent variability of the Vela pulsar is due to the presence
of several artifactual sources nearby.  In a few viewing periods, the measured
flux of Vela is low, and the flux from an artifact is quite high.  Because
nearby source fluxes were not re-fit for each point on the likelihood-vs-intensity
curve, such ``flux-leaking'' can lead to an apparently high variability.  By fixing
the fluxes of the spurious sources one could lower Vela's variability, but only if
those nearby sources were not themselves variable.  This was
not done because of the unknown biases that could result.

3EGJ0721+7120 is identified in the Third Catalog as the BL Lac 0716+714 \cite{cat3, cvm_agn, muk_agn}.
The presence of a definitely identified AGN on the list indicates a few of
the limitations of the method.  The upper end of the 95.5\% confidence region for
$\tau$ is 0.54, well below the typical AGN variability of $\sim 0.9$.
In large part, this is because this source had no dramatic flares during any
observations.  The model of the variability is that the standard deviation of
the source is constant, and this quiescent blazar had a low variability
between these observations.
In addition, the highest and lowest fluxes were each averaged
with a different observation, because of the one month averaging system
(\sect{method}, \sect{timescales}).  A
more sophisticated model might pick up longer or shorter term trends.

\section{The High Variability Sources}

From \tbl{vartbl}, 35 sources have a 68\% confidence region which has a
lower bound greater than 1.0.  These include 15 sources identified as AGN,
as well as the solar flare.  The unidentifed sources are primarily at high
latitudes, and are thus generally assumed to be AGN~\cite{cat3}.  The few
low latitude, unidentified, sources are listed in \tbl{highvartable}.

Because the $V$ statistic used by McLaughlin \etal\  indicates how
improbable it is that the observations come from a constant source, the
use of the $\tau$ statistic in this way is nearly equivalent to the
use of $V$ to find variable sources.  Indeed, when there is a close
correspondence between 2EG and 3EG sources, sources with a high $V$
also have a high lower limit on $\tau$.

\begin{table}[tbhp]
\centering
\begin{minipage}{4 in}
\footnotesize
\begin{tabular}{|l|l|l|r@{.}l|r@{.}l| r@{.}l| l |}\hline
3EG Name & $\ell$ & b & \multicolumn{2}{|c|}{$\tau$} & \multicolumn{4}{|c|}{Lower Limit} & Notes\\
 & & & \multicolumn{2}{|c|}{ } & \multicolumn{2}{|c}{ 95.5\% } & \multicolumn{2}{c|}{ 68\% } &  \\ \hline
J1828+0142 & 31.90  & 5.78   &  3982&5 & $>1$&5 & 6&92   &  em,C    \\ 
J1704-4732 & 340.10 & -3.79  &  29&9   &    1&1  & 2&67   &  C       \\ 
J1837-0423 & 27.44  & 1.06   &  12&0   &    1&1  & 2&16   &  em, C   \\ 

\hline
\end{tabular}
\end{minipage}
\caption[High Variability Sources]
{\label{highvartable}
Low latitude unidentified sources which are unlikely to have low variability.  The sources are
sorted by the lower bound of the 68\% confidence region for $\tau$.
}
\end{table}

\section{Variability Measurements with \glast}

\begin{figure}[tbh]
\centering
\includegraphics[width = 4 in]{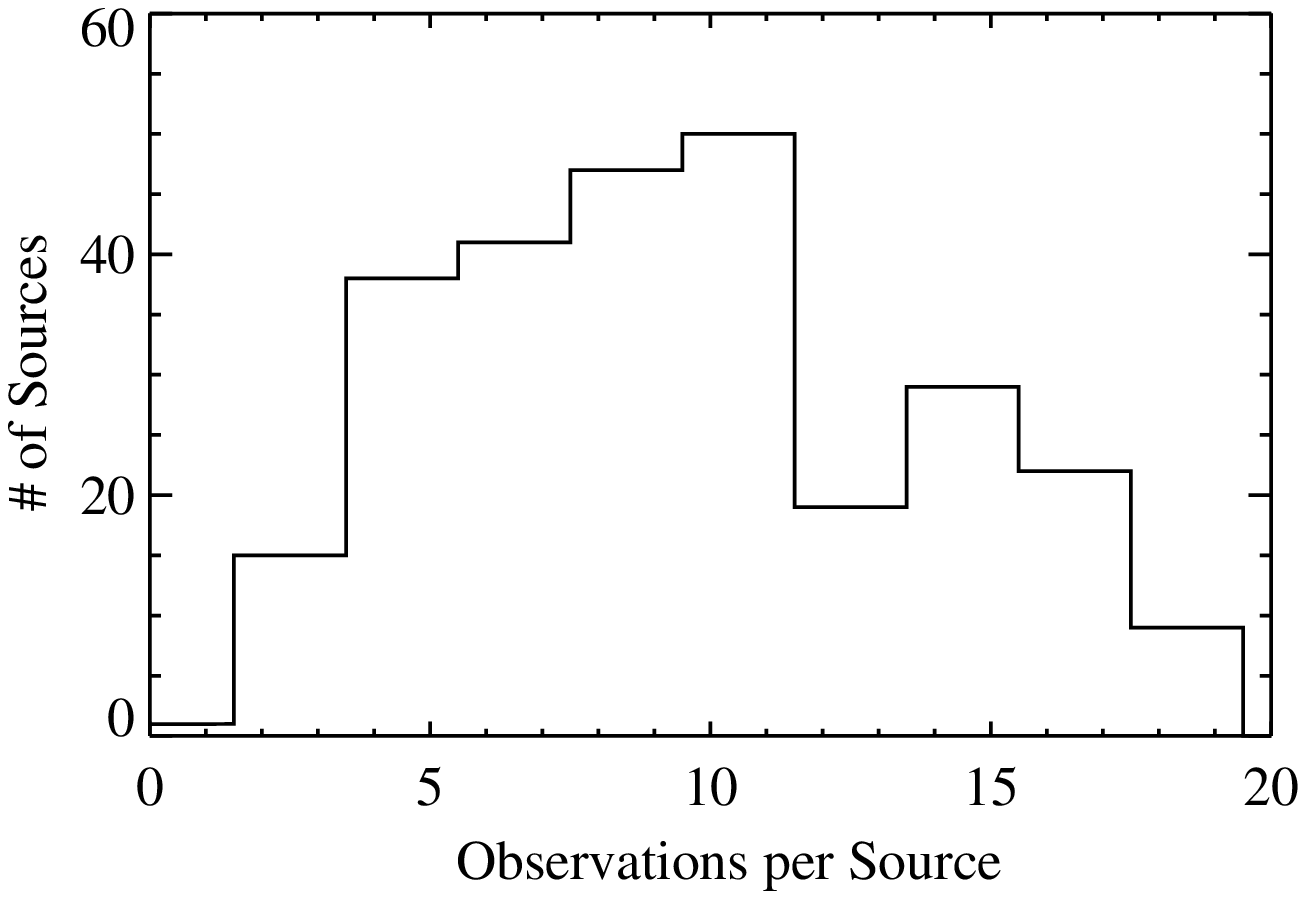}
\caption[Distribution of Number of Observations from \tbl{vartbl}]{\label{numobs}
Distribution of Number of Observations from \tbl{vartbl}.
}
\end{figure}

The two limitations on the variability measurements above are the number of
distinct observations of a source's flux, and the accuracy of those measurements.
As seen from \tbl{vartbl} and \fig{numobs}, many of the sources had few
observations, and this (together with the often large error bars in the
measurements themselves) leads to large errors in the determination of
$\tau$.
\vspace {.5 in}  

The proposed \glast\ mission will address both of these limitations, and thus
offers great promise for measurements of variability.  The large field
of view of \glast\ implies a much larger number of observations, and
the improved sensitivity
means that the error bars on $\tau$ will be much smaller.  In addition, with
the scanning mode of the \glast\ instrument, it will be possible to look at
variation on a continuous range of time intervals from several hours (for bright
sources) to months.

The improved quality and quantity of data will also permit more sophisticated
models of the variability, as discussed in \sect{variextensions}.

\section{Conclusions}

The $\tau$ statistic was calculated for the sources in the Third \egret\ Catalog,
as well as for several classes of sources.  As expected, pulsars exhibited little
variability, while the AGN showed very large variability on a month timescale.
The unidentified sources showed indications of increasing variability with
increasing latitude, indicating that they are a mixture of two or more
types of sources with different spatial distributions and variabilty.

For a few sources, the confidence regions for $\tau$ could exclude the possibility that
the source belonged to the AGN or the pulsar source class.  For most sources, however, the
confidence regions were too large to make any strong conclusions about the source.
Future measurements
should allow much better measurements of the variability of gamma ray sources.

 

\chapter{Testing the \glast\ Design}

The designs for the \glast\ instrument discussed in \sect{glastintro}
have been refined through much computer simulation.  In order to verify
the results of those simulations, as well as demonstrate the state of
the detector technology, a prototype instrument was tested in a tagged
\gammaray\ beam at the Stanford Linear Accelerator Center (SLAC) in October
1997 \cite{glastnim}.
The prototype consisted of models of the three detector systems:  the silicon
tracker, calorimeter, and the anti-coincidence detector.
The performance of each subsystem was evaluated, and the results of the
tests were compared with detailed simulations of the instrument.
The ability of the simulations to predict the performance of this prototype
validated the simulations used to to design the \glast\ instrument.

\section{The SLAC Test Beam}

The Stanford Linear Accelerator Center (SLAC) contains a two mile linac designed
to produce electrons and positrons at energies of up to 50 GeV, in bunches
containing approximately 100 billion particles.  In order to obtain the
single \gammarays\ desired for the October 1997 Beam test, the energy contained
in a few of these electrons was converted into \gammarays, then back into
electrons, then back into \gammarays.

\begin{figure}[tbhp]
\centering
\includegraphics[width = 4in]{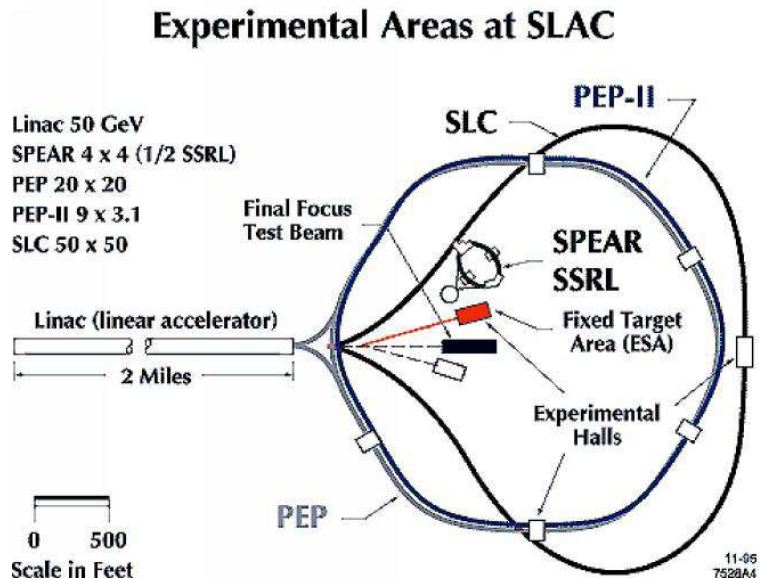}
\caption[SLAC Complex layout]{\label{slacmain}
Layout of the Stanford Linear Accelerator Center}
\end{figure}

\begin{figure}[tbhp]
\centering
\includegraphics[width = 5in]{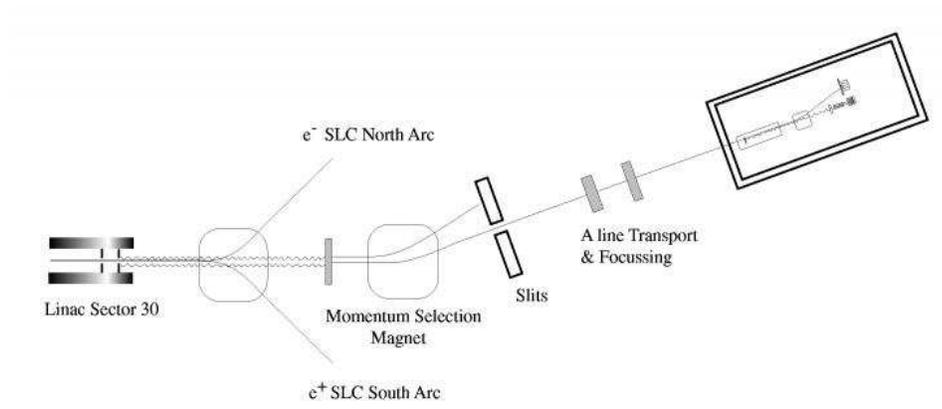}
\caption[End Station A beamline]{\label{slacbeam}
Schematic representation of the End Station A beamline}
\end{figure}

At the end of the linac (\fig{slacmain}), each bunch of electrons or
positrons is collimated with the use of a small aperture  (\fig{slacbeam}).
This collimated
beam of charged particles is then steered around the Stanford Linear Collider
(SLC) to the SLC Large Detector (SLD).  Those particles which hit the
shields around the aperture, however, produce \gammarays\ which are unaffected
by the steering magnets and continue forward.  When these \gammarays\ impact a
plate of high-Z material, they pair produce and form a beam of electrons and
positrons with a large spread of momenta.  

As illustrated in \fig{slacbeam}, electrons with a certain range of momenta are
selected from this beam.  This monoenergetic beam of electrons is then focused
and steered into End Station A (ESA).

\begin{figure}[tbhp]
\centering
\includegraphics[width = 5in]{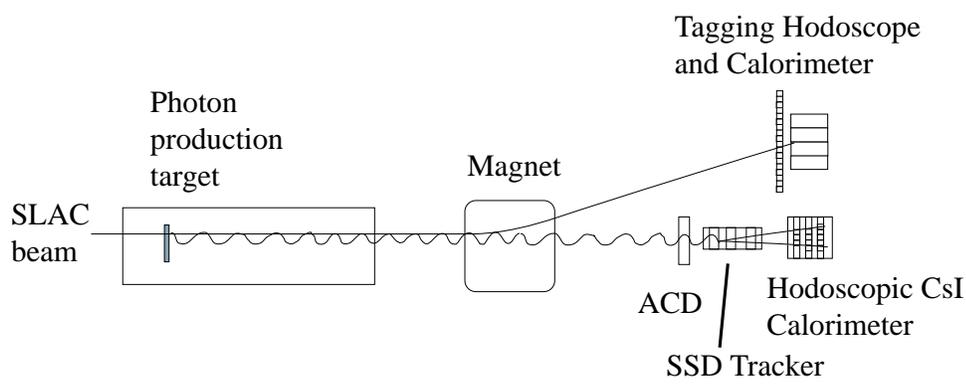}
\caption[Production of the tagged photon beam]{\label{esa}
Schematic of the tagged photon beam}
\end{figure}

Inside ESA (\fig{esa}), two possible setups existed.  This beam of electrons
could be steered into the \glast\ prototype for high energy electro-magnetic
shower studies of the calorimeter.  Or, a copper foil could be inserted into the
beam, and a sweeping magnet (called \BO) turned on.  In this configuration, the
bremsstrahlung \gammarays\ produced in the coper foil were incident on the
prototype instrument, and its capabilities as a pair conversion telescope
could be measured.

The ESA electron beam was tunable from 5 GeV to 40 GeV, and the rate of
electrons could be controlled via the collimation slits.  Most of the gamma
ray data were taken with a 25 GeV electron beam, with a rate of approximately
one electron per pulse.

The tagging hodoscope and calorimeter permitted a measurement of the energy
of the electron after the bremsstrahlung \gammaray\ was emitted, and thus
a measurement of the energy of the \gammaray, since the beam energy was known.
In addition, it permitted the rejection of events where multiple electrons
might have emitted \gammarays\ in the copper target.

\section{The \glast\ Prototype Instrument}

\subsection{The Silicon Tracker}
\begin{figure}[tbhp]
\centering
\includegraphics[width = 5in]{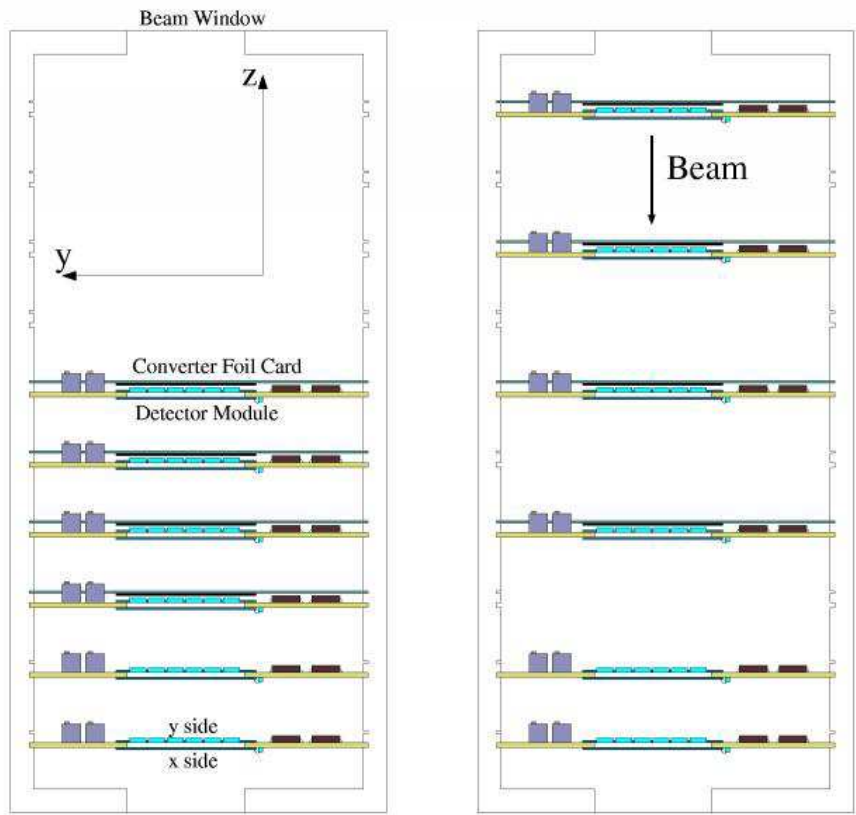}
\caption[Schematic of the prototype tracker]{\label{tracker}
Layout of the prototype tracker.  The ``pancake'' configuration
is shown on the left and the ``stretch'' configuration on the right.  Note the converter foil
cards in front of the first four planes.}
\end{figure}

The directional measurement of the $e^+ / e^-$ pair from a \gammaray\ 
conversion in the test instrument was performed in the silicon tracker.
This scaled down version of the \glast\ tracker was constructed at by
the Santa Cruz Institute for
Particle Physics at the University of California, Santa Cruz.  A schematic
layout can be seen in \fig{tracker}.  In order to measure several
different instrument configurations, the tracker was constructed to allow
the separation of planes to be changed, and for the Pb conversion sheets
to be exchanged for sheets of various thickness.

The Silicon Strip Detectors (SSD) used were $5 \times 5 \times .05$ cm high purity
Si, with a strip pitch of 236 \micron.
The number of readout chips available for these detectors was limited.
Thus each SSD had 192 instrumented strips, corresponding to an area of
$4.6 \times 5$ cm.

The two competing effects which cause error in the reconstructed direction of
a track are the multiple scattering of the track at each plane, and the
measurement error due to the finite strip pitch of the detector.
The former is due primarily to the Pb converter foils in each layer; the latter
is proportional to the strip pitch divided by the distance between planes
(as this is the angular measurement error).
By varying the separation of the tracker planes and the thickness of the
converter foils, all areas of this parameter space could be explored, and
the simulations checked.  Even though many of the configurations would never
be used for a space telescope, it is important that the Monte Carlo results
agree with experiment if the Monte Carlo is to be believed.

\subsection{The CsI(Tl) Calorimeter}

\begin{figure}[tbhp]
\centering
\includegraphics[width = 4in]{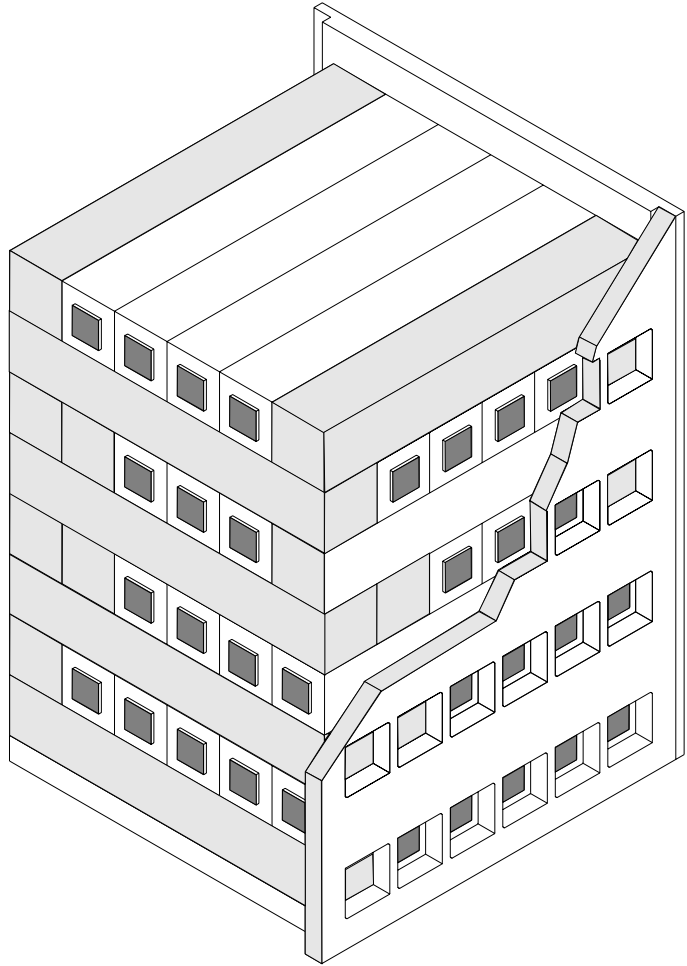}
\caption[Schematic of the prototype calorimeter]{\label{cal}
Schematic representation of the prototype Calorimeter.  The non-instrumented logs are
shown in gray.}
\end{figure}

The energy measurement by the beam test instrument was performed by a 
calorimeter prototype constructed by the
Gamma and Cosmic Ray Astrophysics Branch at the Naval Research Laboratory.
It consisted of eight layers of logs, with six of the 3 x 3 x 24 cm
logs in each layer, alternating in orientation as shown in \fig{cal}.
Only 24 of the logs were instrumented CsI(Tl), the other logs were Cu with
holes drilled through them to obtain the same number of radiation lengths per log.
The locations of the instrumented logs can also be seen in \fig{cal}.

A detailed study of the calorimeter can be found elsewhere \cite{glastnim}.
For the tracker-oriented analysis presented here, the calorimeter was used as
a monolithic block.  The energy deposited in each log was summed to find the
energy of the photon which had converted in the tracker.  This measurement
was typically more accurate than that obtained by the tagging hodoscope. 

\subsection{The Anti-Coincidence Detector}

The Anti-Coincidence Detector (ACD) was assembled by the
Laboratory for High Energy Astrophysics at Goddard Space Flight Center.
The principal role of the plastic scintillator ACD in the \glast\ mission is to serve as a means
to reject incoming charged particles: thus the utility of such a detector
in the beam test might be questioned.  The \glast\ ACD has several unique
challenges, however, and tests of this system were as important as the tests of
the other systems.

\begin{figure}[tbhp]
\centering
\includegraphics[width = 4in]{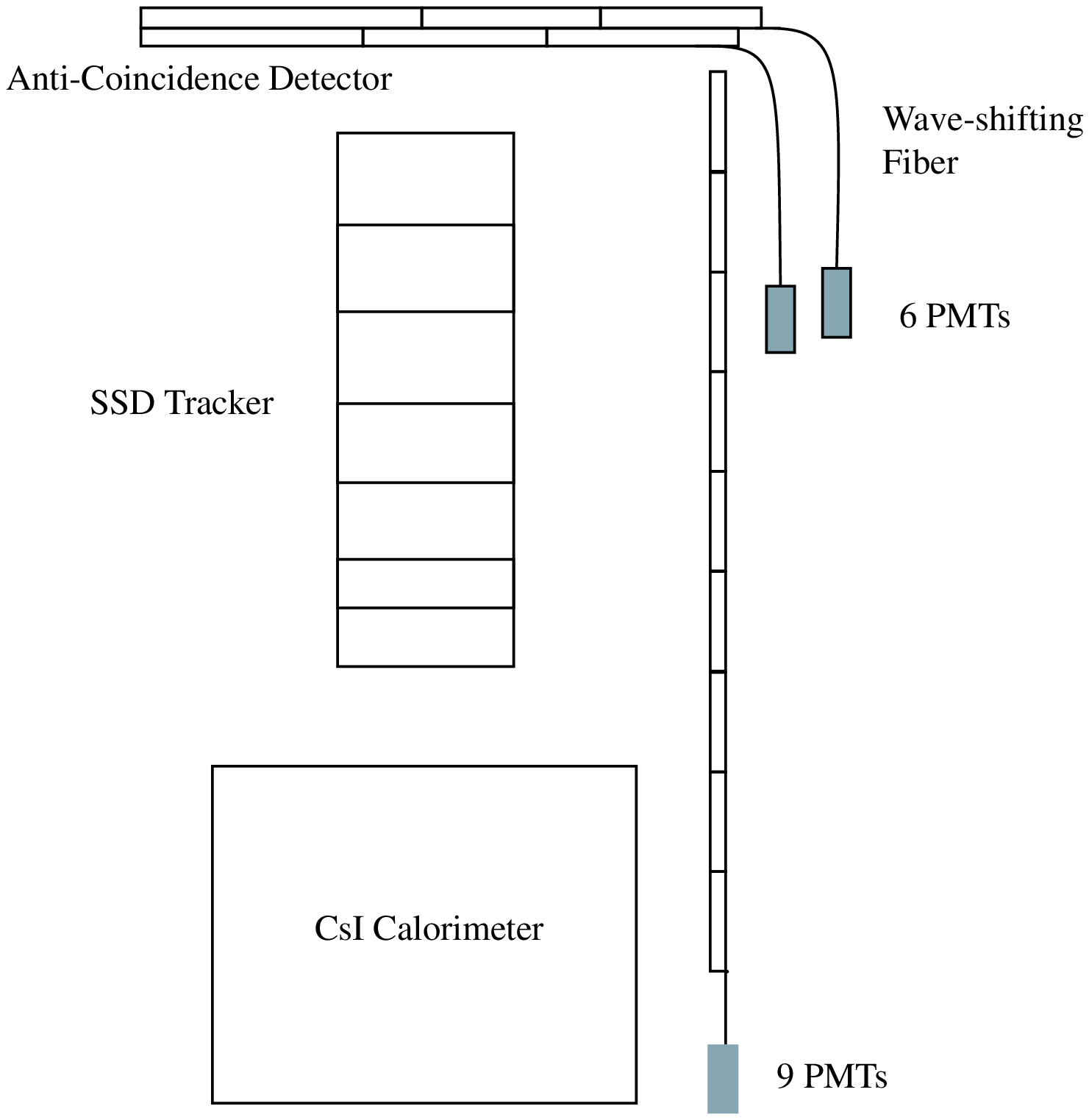}
\caption[Layout of the beam test instrument]{\label{setup}
Layout of the components of the beam test instrument.}
\end{figure}

The problem faced in the design of the ACD is that high energy gamma rays produce
a backsplash of X-rays which trigger the plastic scintilator tiles of the ACD.
Roughly half of the 10 GeV photons detectable by \egret\ are self-vetoed in this
way.
There are many possible steps to solving to this problem: the most important
is the segmentation of the ACD.  By vetoing events only when the ACD triggers
near the incoming particle trajectory, the majority of the backsplash vetos
can be eliminated.  Another possibility is the use of an ACD with two layers
of scintillator, as
was tried in the beam test (\fig{setup}).

The ACD used in the beam test was composed of 1 cm thick
paddles of Bicron BC-408 plastic scintillator, read out by waveshifting fibers
connected to photomultiplier tubes.  Nine of these paddles were placed along
the side of the instrument, and two sets of three paddles were placed in front.
This configuration permitted the measurement of backsplash over a wide area.

\section{Data taken in the October 1997 test}

\subsection{The Raw Data}

Over 400 runs collected 200 gigabytes of data during the 30 days of the
beam test.  Data were collected 24 hours a day, seven days a week during
the runs, by the 30 collaboration members who worked shifts.  The 210 million
triggers were stored on the SLAC computers \cite{perryzen}, where preliminary
filtering was done.

\subsection{Pre-Processing of the Data}

The first steps in processing the \gammaray\ runs were two cuts based on the
calorimeter and the tracker.  Either three consecutive planes in the tracker
must have fired, or more than 6 MeV must have been deposited in the calorimeter.
This eliminated 80\% of the triggers, and the reduced data were then offloaded
to other computers for more processing.

\begin{figure}[tbhp]
\centering
\includegraphics[width = 4in]{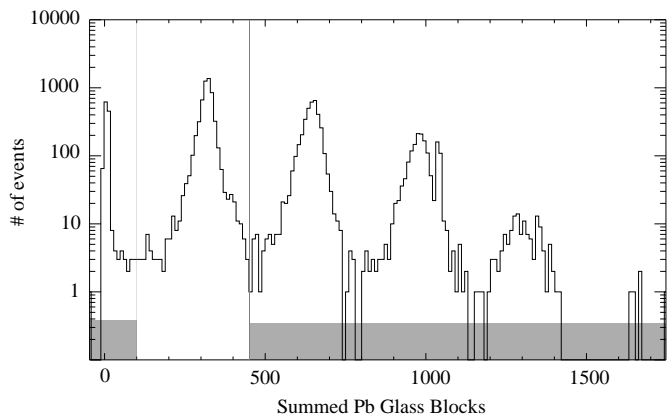}
\caption[Cut on the Pb glass hodoscope]{\label{pbcut}
Cut based on the summed signal from the Pb Glass hodoscope.
The peaks are from various numbers
of electrons in a bunch.  All events in the upper shaded region were discarded.
Events in the lower shaded region which also received less than 5 GeV in the
prototype calorimeter were also discarded.}
\end{figure}

\begin{figure}[tbhp]
\centering
\includegraphics[width = 4in]{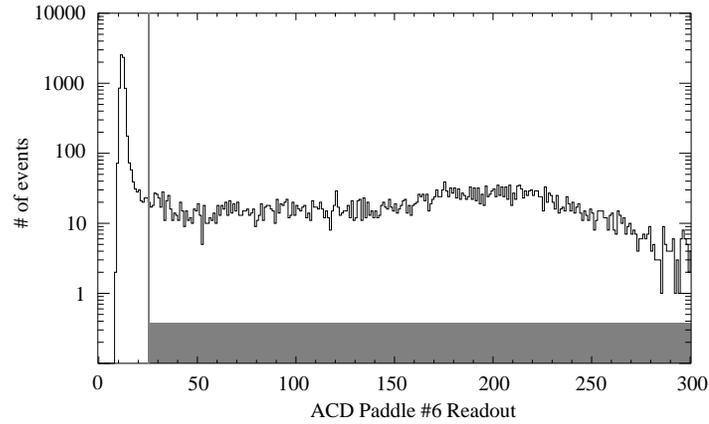}
\caption[Cut on the Anti-Coincidence Detector]{\label{acscut}
Cut based on ACD paddle \#6.  A similar cut was imposed on paddle \#2, the
other paddle directly in front of the tracker.}
\end{figure}

\begin{figure}[tbhp]
\centering
\includegraphics[width = 4in]{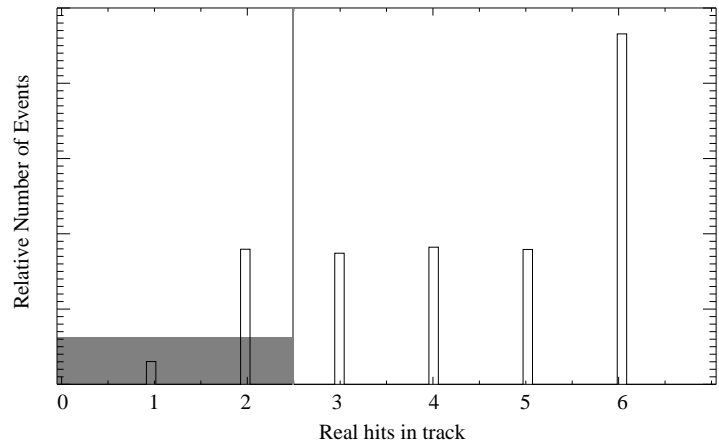}
\caption[Cut on the number of tracker hits]{\label{threehits}
Because fake hits were permitted when a track apparently went through a
dead strip, it was required that at least three of the hits in the track
be real hits.}
\end{figure}

\begin{figure}[tbhp]
\centering
\includegraphics[width = 4in]{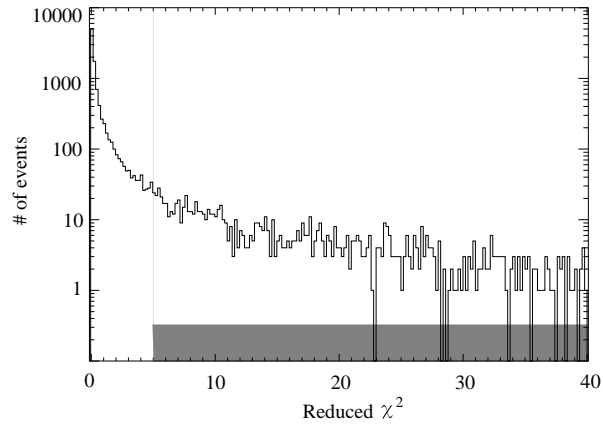}
\caption[Cut on the reduced $\chi^2$ of the track]{\label{chicut}
Both particle tracks from the pair conversion were required to pass a quality
cut: that the reduced $\chi^2$ was less than 5.}
\end{figure}

\begin{figure}[tbhp]
\centering
\includegraphics[width = 4in]{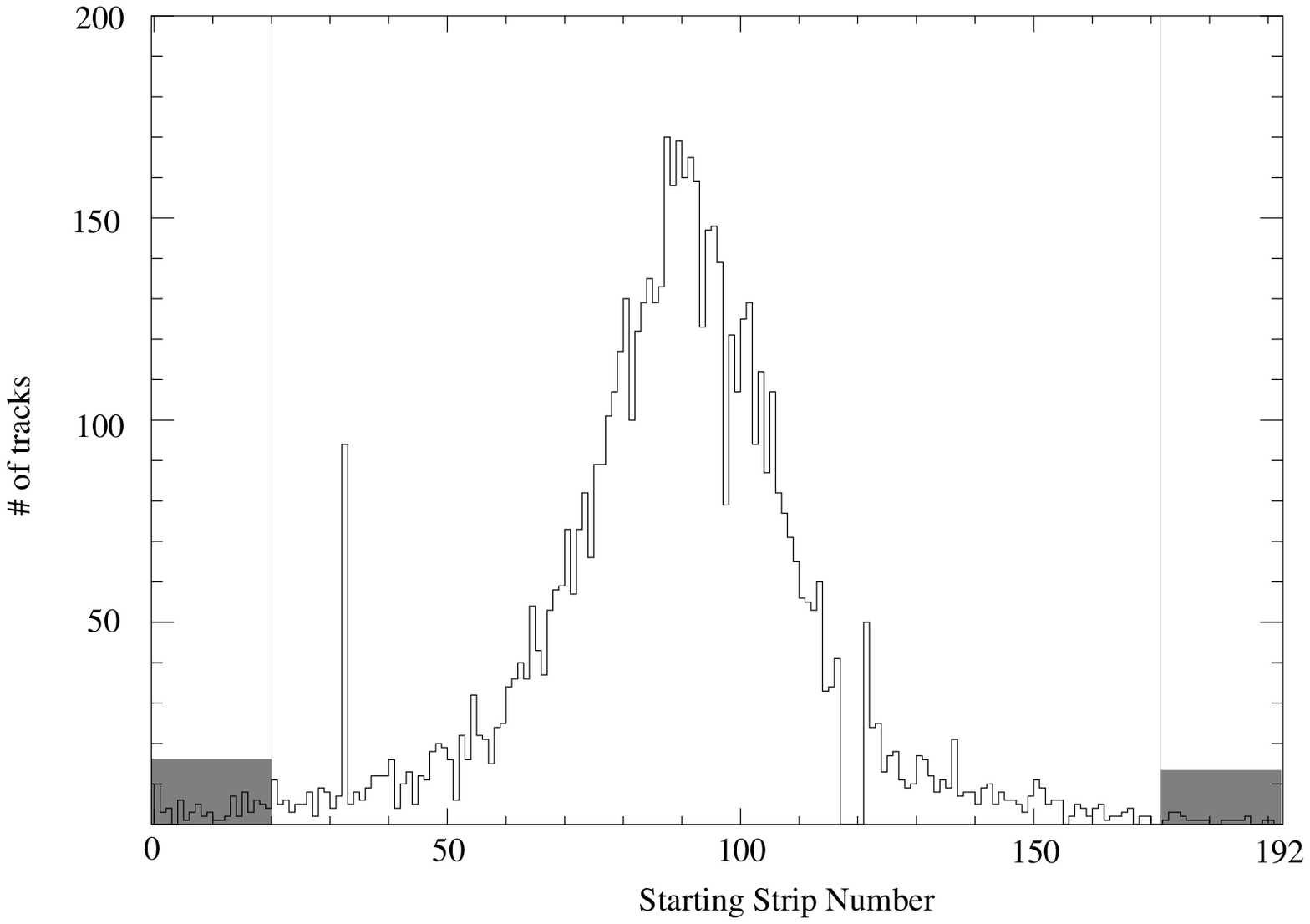}
\caption[Cut on the starting position of the track]{\label{edgecut}
To ameliorate selection effects, tracks which began near the edge of the
tracker were eliminated.  The same cut was imposed in the X and Y projections.
}
\end{figure}

Each event used in the analysis was required to pass several more stringent cuts.
First, the Pb glass blocks used in the tagging hodoscope
must have indicated that there was
only one electron in the bunch (\fig{pbcut}).
This lowered the probability of having
mutiple \gammarays\ produced at the bremstrahlung target.  Second, the
Anti-Coincidence tiles through which the \gammaray\ beam passed were
required to have less than $1/4$ of the energy deposited by a 
minimum ionizing particle (MIP) in each (\fig{acscut}).
This ensures that the gamma ray did not convert inside the scintilating tile.
Three more cuts were imposed based on the parameters of the reconstructed
tracks:  tracks must have had at least three hits regardless of the energy 
in the calorimeter (\fig{threehits}), have had a reduced
$\chi^2 < 5$ (\fig{chicut}),
and the starting position of the track must have been at least $4.7$ mm from the
edge of the tracker (\fig{edgecut}).
This last requirement lowered the probability that
a track might escape the tracker, which could bias the reconstructed track
directions.

\section{Simulations}

Simulation played two roles in the October 1997 beam test.  Simulations
of the beam setup and the instrument were crucial for determining how the beam test
should be run.  But more importantly, these simulations were compared to the beam
test results after the test, which enabled the \glast\ team to judge the quality of
the simulation package.

Monte Carlo simulations of the full \glast\ instrument have been performed with a
program called \glastsim\ developed by the \glast\ team.  This program is based on
\gismo~\cite{gismo}, an object oriented Monte Carlo package written in C++.
\gismo, in turn, incorporates several interaction packages: \egs~\cite{egs}, which
contains the electromagnetic interactions, and \gheisha~\cite{gheisha1,gheisha2}
which contains the hadronic interactions.

The simulations of the beam test instrument were performed with a modified \glastsim\ 
program.  The geometry of the instrument was incorporated, as well as the beam charactaristics.
In order to provide as realistic a simulation as possible, the starting configuration of
the simulation had a 25 GeV electron incident on a 3.5\% radiation length copper radiator.
The beam spot size and collimation were modelled as accurately as possible.  Bremsstrahlung
\gammarays\  from the radiator then illuminated the instrument, while the incident electrons
were swept away by a magnetic field, just as \BO\ did in the real experiment.

The modelled calorimeter was not always incorporated into the simulations: because
simulating the high-energy shower development could be quite compute intensive, it was
faster to gather simulated data by using an input \gammaray\ beam, with
no calorimeter.  This was done when only the tracker response for a given
\gammaray\ energy was needed.

The final modification of the \glastsim\ code was the addition to \egs\ of a realistic,
polarized cross-section for \gammaray\ pair conversion, due to Ion Yadigaroglu.
  In general, the effect of using this cross-section rather than the standard \egs\ 
pair conversion cross-section was small, but for some simulations this more sophisticated
treatment was desired.

The output of the simulation was treated identically to the beam test data.  Cuts performed
on one were also performed on the other, and strips which were found to be noisy or dead
in the beam test instrument were masked out of both datasets.  Thus, a comparison of the
results was straightforward.

\chapter{Beam Test Setup}

\section{Alignment of \glast\ Prototype \label{align}}

In order to measure the precise direction of the tracks in the detector,
it is necessary to know the position and orientation of the detector elements.
One option would be to precisely align each of the detector planes and the
calorimeter, and to maintain that alignment while the detector is handled (and
in the case of \glast\,, launched).  This would, however, be time consuming and
would impose additional constraints on the design of the structure holding the
instrument together.  A better option is to determine the actual alignment while
the detector is in use, by using straight tracks to determine the relative
positions of the detector elements.

\subsection{Finding Straight Tracks}

The first step, then, is to find the straight tracks which will be used to
align the detector.  In the beam test instrument, this was fairly straightforward:
there was a high frequency of high energy \gammaray\ events which could be used.
If, for some reason, there were not enough of these, there was also data taken
with 25 GeV electrons.  In the orbital \glast\ instrument, it will be necessary
to use high energy cosmic rays as well, as the flux of high energy \gammarays\ 
will not give sufficient data for a quick alignment.

It is necessary then, to select data which have
``clean'', straight tracks.  This can be accomplished somewhat by looking for
events which deposit significant energy in the calorimeter.  In addition, it
is necessary to do a reconstruction of the tracks, and look for outlying hits.
Tracks with many outliers can be rejected, and not used for the alignment.
There is a bit of a ``chicken-and-egg'' problem here, however: in order to
reconstruct the tracks, some alignment must be assumed.  By assuming that the
instrument is aligned at the beginning, but by not penalizing the tracks
significantly for larger scattering values, it is possible to find the tracks
for the alignment.

In addition, there is the problem of multiple hits.  Because adjacent strips
can fire even when there is a single track, there is the issue of which hit to
pick (or whether to generate a virtual hit in between the two).  In the beam test,
the strip which minimized scatter was used, which tended to straighten out tracks
to conform to the assumed alignment of the instrument.

Because of these effects, it is necessary to iterate the process of finding
tracks, and then aligning the planes based on those tracks.
At each iteration, the track selection procedure will select better tracks, and
the estimate of the position of each plane will get better.

\subsection{The Alignment Parameters}

In order to describe the position and orientation of one detector plane
relative to a fixed reference (such as the bottom plane), six parameters
are needed: three positional and three rotational.
Because a given plane of the test instrument measures only one of the X or Y
position of a track which is nearly in the Z direction, only one of the three
positional parameters is important.  For a detector plane which measures the
X position of a track, this would be the X position of the plane.  Similarly,
only one of the three rotations is important: the one about the Z axis.
In addition, the importance of the rotation depends on the size of the detector:
for the 6 cm beam test detector with pitch 236\micron, rotations
of $\sim 0.45 \deg$ are important.  The rotation can be measured by examining
the dependence of the X offset (\sect{alignalgo}) on the Y position of a track. 
Because rotations about the X and Y axes are not important, it is necessary to
define the Z axis by a means other than the orientation of the bottom plane.
If it is defined as the line from the center of the bottom tracker plane to the
center of the top tracker plane, then both the top and bottom planes are defined
to have no offset in the X or Y direction.

This situation changes somewhat for the \glast\ instrument.  Because tracks
might have a large incident angle, it becomes important to measure the Z offset
as well as the X offset for an X detector plane.  Similarly, all three rotations
become important, as they each affect the X and/or Z position of the detected hit.
Also, rotations become more important: for a \glast\ detector plane of 30 cm,
with strip pitch of 200\micron, rotations of only $\sim 0.076\deg$ will affect
the instrument performance.

\subsection{The Alignment Algorithm \label{alignalgo}}

By fitting a straight line to all of the track hits but one, a projected position
for that one hit can be found.  
If the track were perfectly straight and there were no measurement errors,
this residual would indicate the alignment
error of the plane relative to the average alignment of the tracker.
Because of the measurement error, it is necesary to use a
large number of these residuals to estimate the alignment error.

\begin{figure}[htbp]
\centering
\includegraphics[width = 6in]{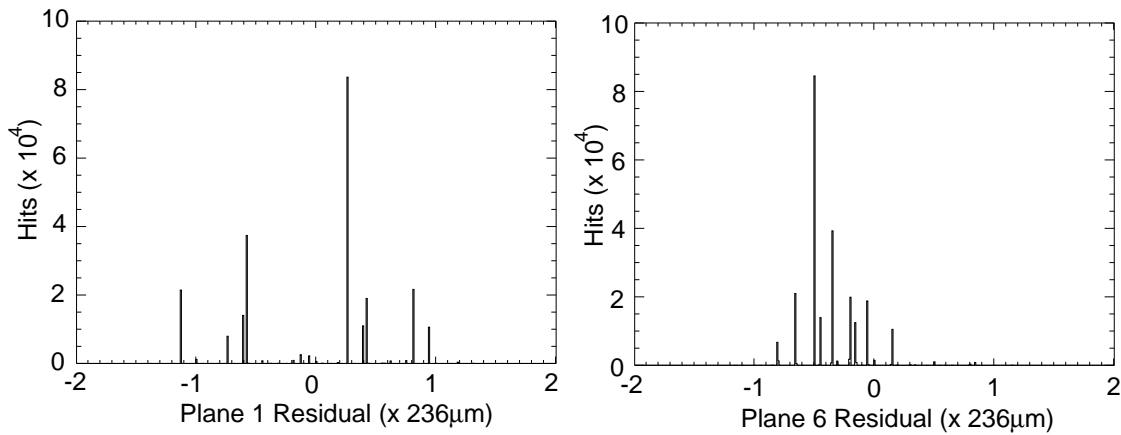}
\caption[Distribution of residuals]{\label{residspre}
The distribution of residuals for two planes in the prototype instrument.  Note
the general offset
in the plane 6 data,  and the peaks separated by one strip (236 \micron) in plane 1.
Most planes exhibit such double peaks;  they are easily understood if one visualizes the
residual distribution for on-axis tracks when only a single plane is offset. }
\end{figure}

There are several methods one might use to find this estimate from the distribution
of residuals.  The easiest might be to use the mean, median, or mode of the
distribution itself.  These all have problems however.  The mean can be unduly
influenced by outliers caused when the track finding algorithm has failed.  The
median and the mode are both affected by the non-Gaussian nature of the
measurement error:  because the measurements are discretized, the distribution
is not continuous, but instead is spiky from the strip aliasing (\fig{residspre}).

\begin{figure}[htbp]
\centering
\includegraphics[width = 6in]{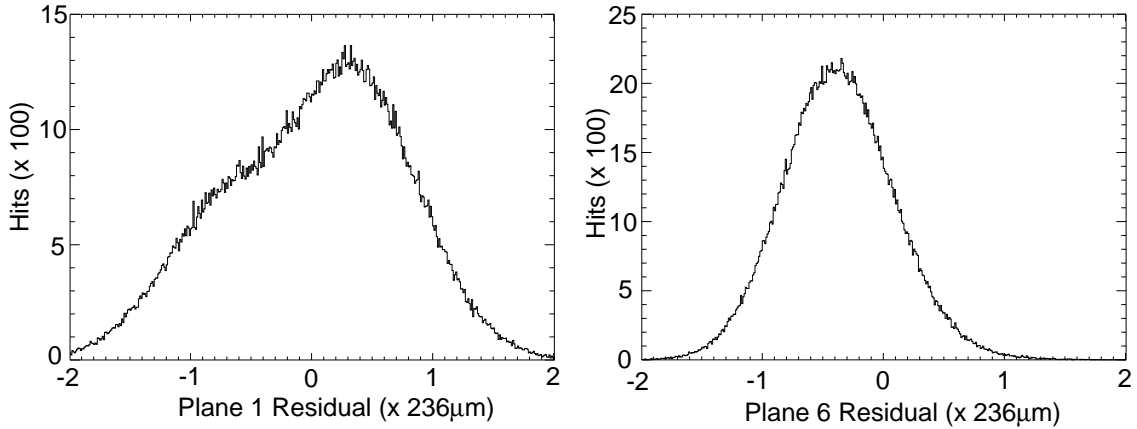}
\caption[Smoothed distribution of residuals]{\label{residspresmooth}
The distributions shown in \fig{residspre}, after the addition of Gaussian noise.
The median can now be used to measure the offset error.}
\end{figure}

By adding some Gaussian noise (with standard deviation equal to the strip pitch),
this aliasing may be dispensed with, and the median or mode may be used freely (\fig{residspresmooth}).  In cases with fewer data, the histogram could be convolved
with a Gaussian, resulting in the same curves, but with less noise in each bin.
As seen from plane 1 in \fig{residspresmooth}, there will be a difference between
using the median and the mode, due to the asymmetry in the distribution.  For the
beam test data, the median was used to determine the alignment parameters.

The assumed position of each plane can then be adjusted by the median residual
as discussed above.  As the position of the first and last plane might now
be non-zero, the positions can be translated and rotated so that the
coordinate system based on the first and last planes is maintained.

By iterating the process of finding residuals and adjusting the plane positions
until the positions converge, a consistent set of plane positions may be found
for a given set of tracks.
At this point, a new set of tracks can be found,
followed by another round of plane adjustments, until the same tracks are
found each time.

\subsection{Alignment Results \label{alignresults}}

The principal determination from the alignment calculations is that the
test beam instrument was very close to being aligned.  No significant rotation
of the planes was detected, and the planes were generally aligned to within
one strip pitch (236\micron).

\begin{figure}[htbp]
\centering
\includegraphics[width = 6in]{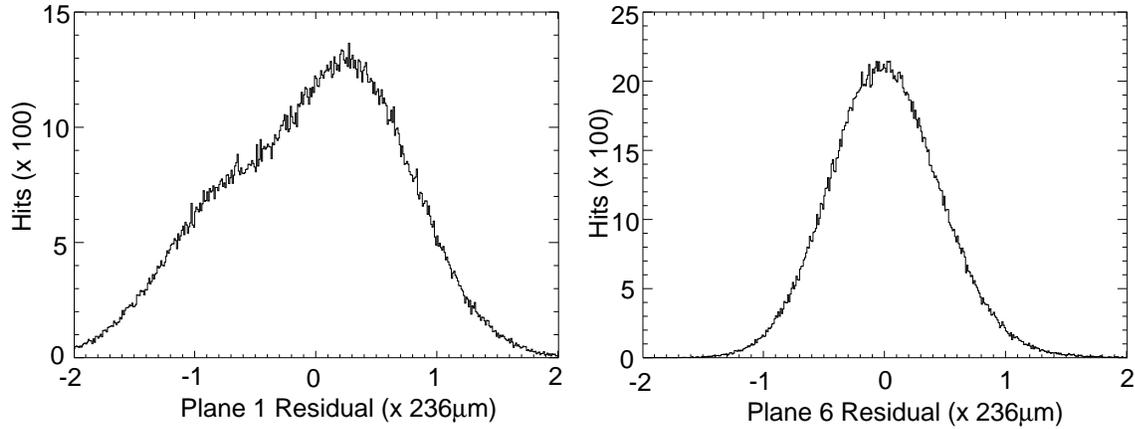}
\caption[Smoothed distribution of residuals after alignment]
{\label{postalign}
The distributions shown in \fig{residspresmooth}, after the iterated alignment
procedure discussed in \sect{alignalgo}.  The second peak (offset by one strip)
is again visible in the Plane 1 data.}
\end{figure}

The determined alignments are given in \tbl{aligntbl}, and some 
sample residual distributions after alignment are shown in \fig{postalign}.

\begin{table}
\begin{minipage}{4 in}
\footnotesize
\begin{tabular}{|l|c|r@{.}l|r@{.}l|r@{.}l|r@{.}l|r@{.}l|r@{.}l|}\hline
Configuration & Direction & \multicolumn{2}{|c|}{Plane 1}
& \multicolumn{2}{|c|}{Plane 2}
& \multicolumn{2}{|c|}{Plane 3}& \multicolumn{2}{|c|}{Plane 4}
& \multicolumn{2}{|c|}{Plane 5}& \multicolumn{2}{|c|}{Plane 6} \\ \hline
Stretch & X &  0&000 &  -0&179  &   0&176 &    0&365 &    0&237 &     0&000 \\
Stretch & Y &  0&000 &    -0&509 &   -0&264 &   0&138 &  -0&218 & 0&000 \\
Pancake & X &  0&000 & -0&92 &   0&184  &  0&103 &  0&083 &    0&000 \\
Pancake & Y &  0&000 & -0&058 &  -0&161 &  -0&134 & -0&060 &   0&000 \\ \hline
\end{tabular}
\end{minipage}
\caption[Alignments from \glast\ component beam test]
{\label{aligntbl}
Alignments for the four tested configurations in the Fall 1997 Beam Test.
Alignments are in units of strip pitch ($=236\mu m$).  Planes 1 and 6 define
the coordinate system, and thus are identically zero.
}
\end{table}

\section{The Reconstruction Algorithm \label{reconstruction}}

The reconstruction of particle tracks was done with the Kalman Filter method~\cite{kalman},
which has been widely used for particle tracking in high energy
physics~\cite{fruhwirth_phd,fruhwirth,billoir,bock90}.  The application
of these techniques to \glast\ 
has been discussed elsewhere (\cite{physicistsguide,josekalman}), as has the use for
this beam test (\cite{brianthesis}).  Thus only a brief overview of
this technique will be given here.

\subsection{The Kalman Filter}

The Kalman filter is a means to estimate some quantities (termed the ``state vector'' $= {\bf x}$)
as a function of time, using some measurements (${\bf m}$), with some errors (${\bf V}$).  The
$k$'th measurement is defined through
\[ {\bf m}_k = {\bf H}_k {\bf x}_k + {\bf \epsilon}_k, \]
where ${\bf H}_k$ is a matrix relating the state vector to the measured quantities,
and ${\bf \epsilon}_k$ represents the error in the measurement.  This error is characterized by
the covariance matrix ${\bf V}_k \equiv {\rm cov} [{\bf \epsilon}_k]$.

The remaining complication to the problem which remains to be specified is that of the evolution
of the state vector.  If the state vector were the same for all of the measurements,
it would be simple enough
to perform a least squares fit to the data.  This change of the state vector through
time is given by the evolution equation
\[  {\bf x}_{k+1} =  {\bf F}_k {\bf x}_k + {\bf w}_k. \]
Where  ${\bf F}_k$ contains the deterministic component of the evolution, and  ${\bf w}_k$ the
stochastic.  Again,  ${\bf w}_k$ is characterized by its covariance matrix: 
$ {\bf Q}_k \equiv {\rm cov} [  {\bf w}_k ]$.

It is not difficult to see how to find an estimate of the state vector at one time from a
measurement at another.  We have
\[ \left\langle {\bf x}_{k} \right\rangle = {\bf H}^{-1}_k {\bf m}_k, \]
and can use the evolution equation to find $\left\langle {\bf x}_{j} \right\rangle$ given
$\left\langle {\bf x}_{k} \right\rangle$.  Similarly, the covariance of the estimate of ${\bf x}_{k}$
given the one measurement may be found by 
\[ {\bf C}_{k} \equiv {\rm cov} [ {\bf \bar{x}}_{k} ] =
 {\bf H}^{-1}_k {\bf V}_k \left({\bf H}^{-1}_k\right)^T, \]
and by evolving ${\bf C}$ appropriately:
\[  {\bf C}_{k+1} =  {\bf F}_{k}  {\bf C}_{k}  {\bf F^T}_{k} +  {\bf Q}_{k}. \]

Note that all these estimates (so far) rely only on one measurement.  The complications arise when
we wish to use multiple measurements for an estimate.  We could simply find all of the estimates
for the different measurements independently, and combine them: this would be equivalent to a least
squares fit with no correlations between the measurements.  The correct approach, however, begins
with the realization that the state vector has no ``non-local'' evolution.  Thus, the estimate at
time $k$ should be influenced by the best estimate at time $k-1$, but should not explicitly depend
on the estimate at time $k-2$.  This dependence enters only because the best estimate at time $k-1$
does depend on the estimate at time $k-2$.

This leads us to the filtering equations.  Given that we've found the best estimate of the state
vector at time k, using the measurements at $t <= k$, how do we find the best estimate at 
$k+1$ using the measurement at $k+1$?  First, we find the best estimate by evolving the old estimate,
and then include the new measurement as we would for standard least squares.

The notation here becomes a bit awkward, as we need to denote various estimates and covariances at
different times, and also made with different data.  Let us write the estimate at time $k$, using
the data through time $k$ as ${\bf x}_{k}$.  The estimate, using the same data, at time $k+1$
we write as ${\bf x}_{k+1, \rm proj}$.  The covariance matrices are denoted similarly.

We have
\begin{equation} {\bf C}_k = \left[ {\bf C}_{k, \rm proj}^{-1} +
                                    {\bf H}_k^T {\bf V}^{-1}_k {\bf H}_k \right] ^{-1},
  \end{equation}
\begin{equation} {\bf x}_k = {\bf C}_k \left[ {\bf C}_{k, \rm proj}^{-1} {\bf x}_{k, \rm proj} +
                             {\bf H}_k^T {\bf V}^{-1}_k {\bf m}_k \right],
  \end{equation}
\begin{equation} {\bf x}_{k+1, \rm proj} = {\bf F}_{k} {\bf x}_{k},
  \end{equation} and
\begin{equation} {\bf C}_{k+1, \rm proj} = {\bf F}_{k} {\bf C}_{k} {\bf F}_{k}^T + {\bf Q}_{k}.
  \end{equation}

With the above equations, the best estimate of the state vector at some time $n$ can be found
by iteration, beginning with k=1 and progressing to $k=n$.  The only difficulty is the specification
of ${\bf x}_{1, \rm proj}$ and ${\bf C}_{1, \rm proj}$.  This is essentially the prior estimate of the
state vector.  The safest option is to pick a sensible value for ${\bf x}_{1, \rm proj}$, with a very
large error matrix ${\bf C}_{1, \rm proj}$.  As the elements of the error matrix approach infinity,
the initial value for ${\bf x}_{1, \rm proj}$ will not matter.  Equivalently, one may take
${\bf C}^{-1}_{1, \rm proj} = 0$.

\subsection{The Kalman Smoother}

What, though, if the best estimate of the state vector using later measurements as well
as earlier ones is desired?  One could simply run the Kalman filter backwards, and combine
the two (independent) estimates at time $k$.  Alternatively, one can define a sort of
``back-correction-propagator'',
\[ {\bf A}_{k} \equiv {\bf C}_{k} {\bf F}^T_{k} {\bf C}^{-1}_{k+1, \rm proj}. \]

Defining our desired estimate (using all the data) as ${\bf x}_{k, \rm smooth}$, we have
\begin{equation}
{\bf x}_{k, \rm smooth} = {\bf x}_k + {\bf A}_k ({\bf x}_{k+1, \rm smooth} - {\bf x}_{k+1, \rm proj}),
\end{equation} and
\begin{equation}
{\bf C}_{k, \rm smooth} = {\bf C}_k + {\bf A}_k ({\bf C}_{k+1, \rm smooth} -
{\bf C}_{k+1, \rm proj}) {\bf A}_k^T.
\end{equation}

These smoothing equations can be iterated from the final measurement back to the first, yielding
a set of estimates using all of the measurements.

\subsection{Limitations of the Kalman Equations}

As noted above, the Kalman formalism is equivalent to least squares fitting, and thus
is equivalent to the maximum likelihood analysis when the two stochastic terms (${\bf w}_k$ and
${\bf \epsilon}_k)$ are Normally distributed.  The benefit of the Kalman filter approach is the
speed with which estimates can be found: in ${\cal O}(N)$ time, where $N$ is the number of
measurements, with simple matrix operations (on small matrices) at each step.  The alternative, for
a straightforward least squares fit, would require inverting a correlation matrix of size $N$, which
is ${\cal O}(N^3)$.  The straightforward maximum likelihood computation would involve maximizing a
function of $ND$ variables, where D is the number of components in the state vector ${\bf x}$.

What though, of the case when the formalism breaks down?  This might happen in many different
ways:  the evolution operator ${\bf F}$ might not be linear, or the scattering term ${\bf w}$,
or the error term ${\bf \epsilon}$ might not have a Gaussian distribution.  Alternatively, these terms
might depend on the state vector.  These cases may be approached
with changes of variables, although with differing degrees of
difficulty~\cite{julier1,julier2,stampfer,regler88}.

Linearizing ${\bf F}$ is common in high energy applications, because of the curved particle tracks
in a magnetic field.  When measurement points are sufficiently close together, this linearization
is not a problem.  Note that measurements can be made arbitrarily close by inventing new measurements,
where the error on the measurement is infinite.

For the case of the beam test, no extensions to the Kalman formalism were used.
 Because tracks were necessarily
almost perpendicular to the planes (or else the track escaped the tracker, and thus was not seen),
the scattering and error matrices had little dependence on the state vector.  Similarly, although
multiple scattering at large angles is not Gaussian, this effect was minimized in the tracker, because
large angle scatters deflected out of the instrument.
\chapter{Results of the Beam Test}

\section{Characterization of the Point Spread Function}

\begin{figure}[tbhp]
\centering
\includegraphics[width = 4in]{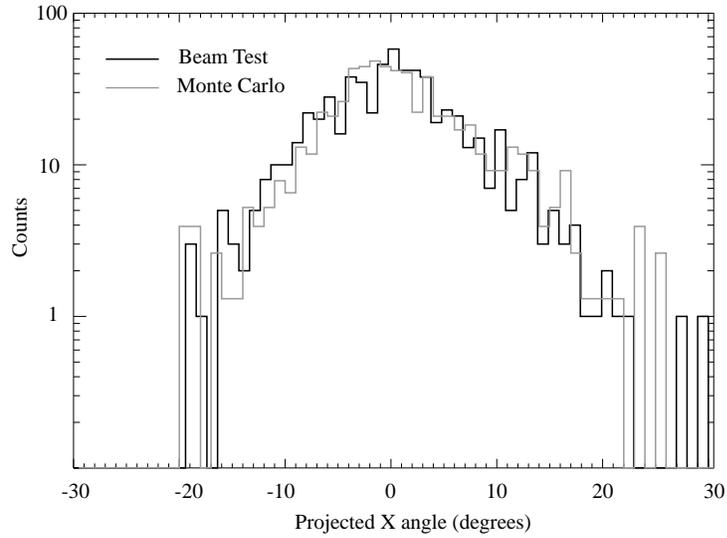}
\caption[PSF at 10--20 MeV, 2\% RL, Pancake configuration]{\label{firstpsf}
PSF for 10--20 MeV \gammarays, with 2\% R.L. Pb radiators,
in the Pancake configuration.
}
\end{figure}

\begin{figure}[tbhp]
\centering
\includegraphics[width = 4in]{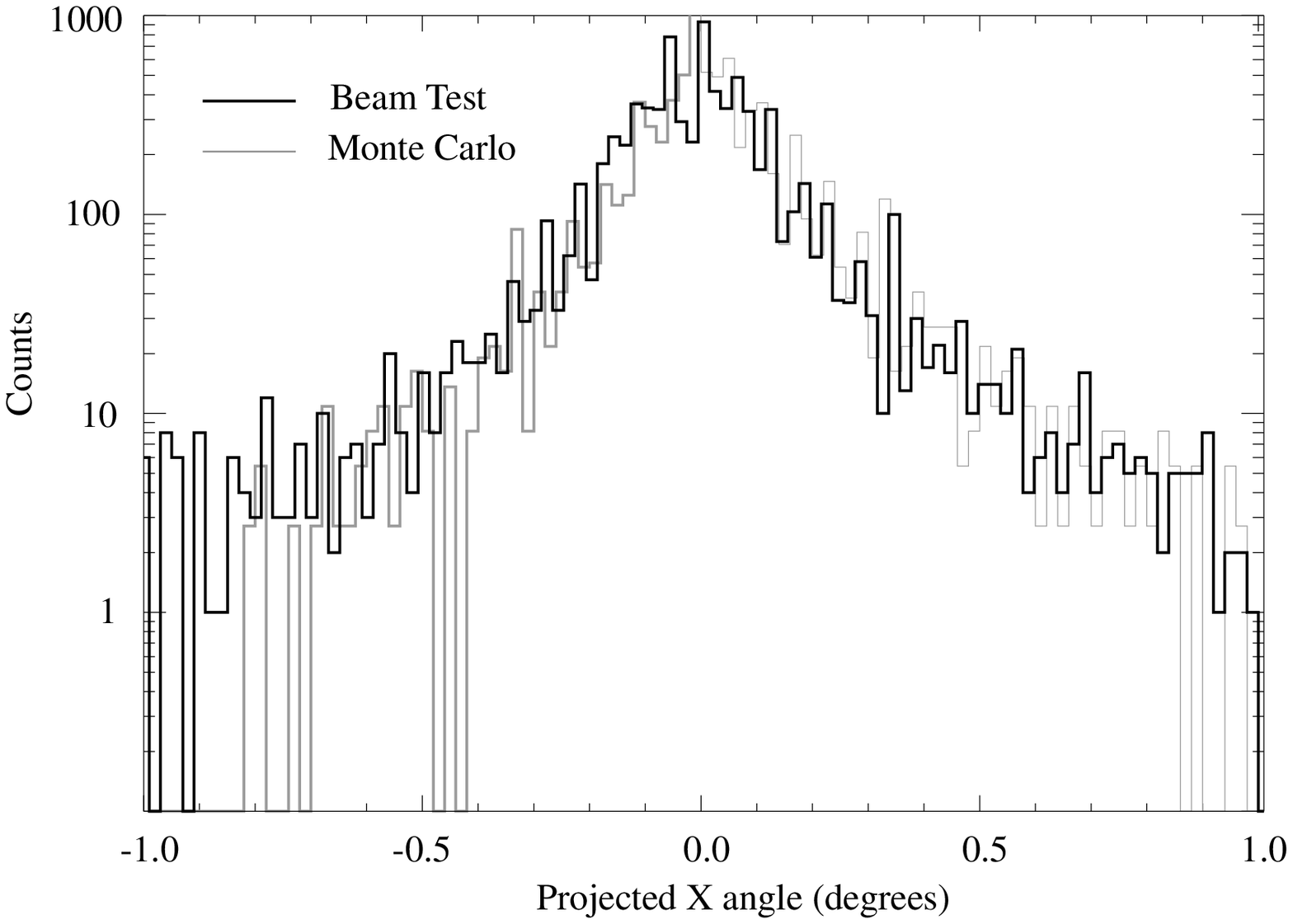}
\caption[PSF at 10--20 GeV, 4\% R.L., Pancake configuration]{\label{secondpsf}
PSF for 10--20 GeV \gammarays, with 4\% R.L. Pb radiators,
in the Pancake configuration.
}
\end{figure}

\begin{figure}[tbhp]
\centering
\includegraphics[width = 4in]{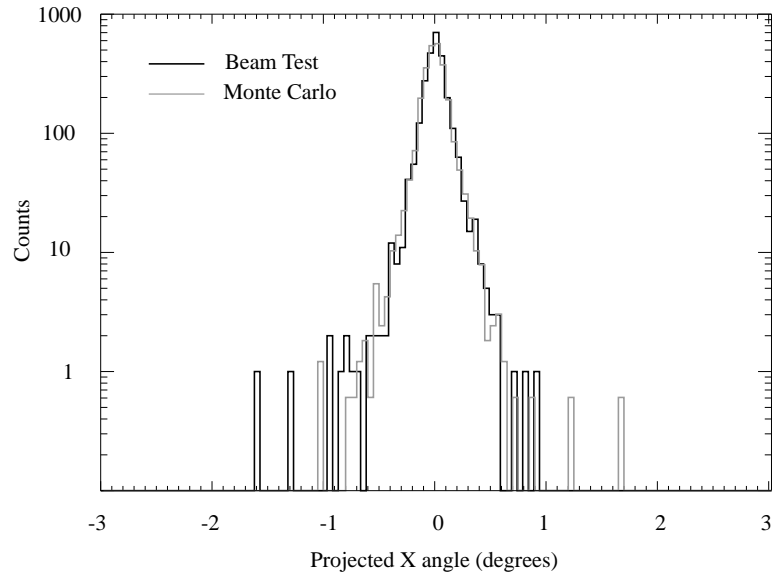}
\caption[PSF at 2--5 GeV, 0 RL, Stretch configuration]{\label{thirdpsf}
PSF for 2--5 GeV \gammarays, with no Pb radiators,
in the Stretch configuration.
}
\end{figure}

\begin{figure}[tbhp]
\centering
\includegraphics[width = 4in]{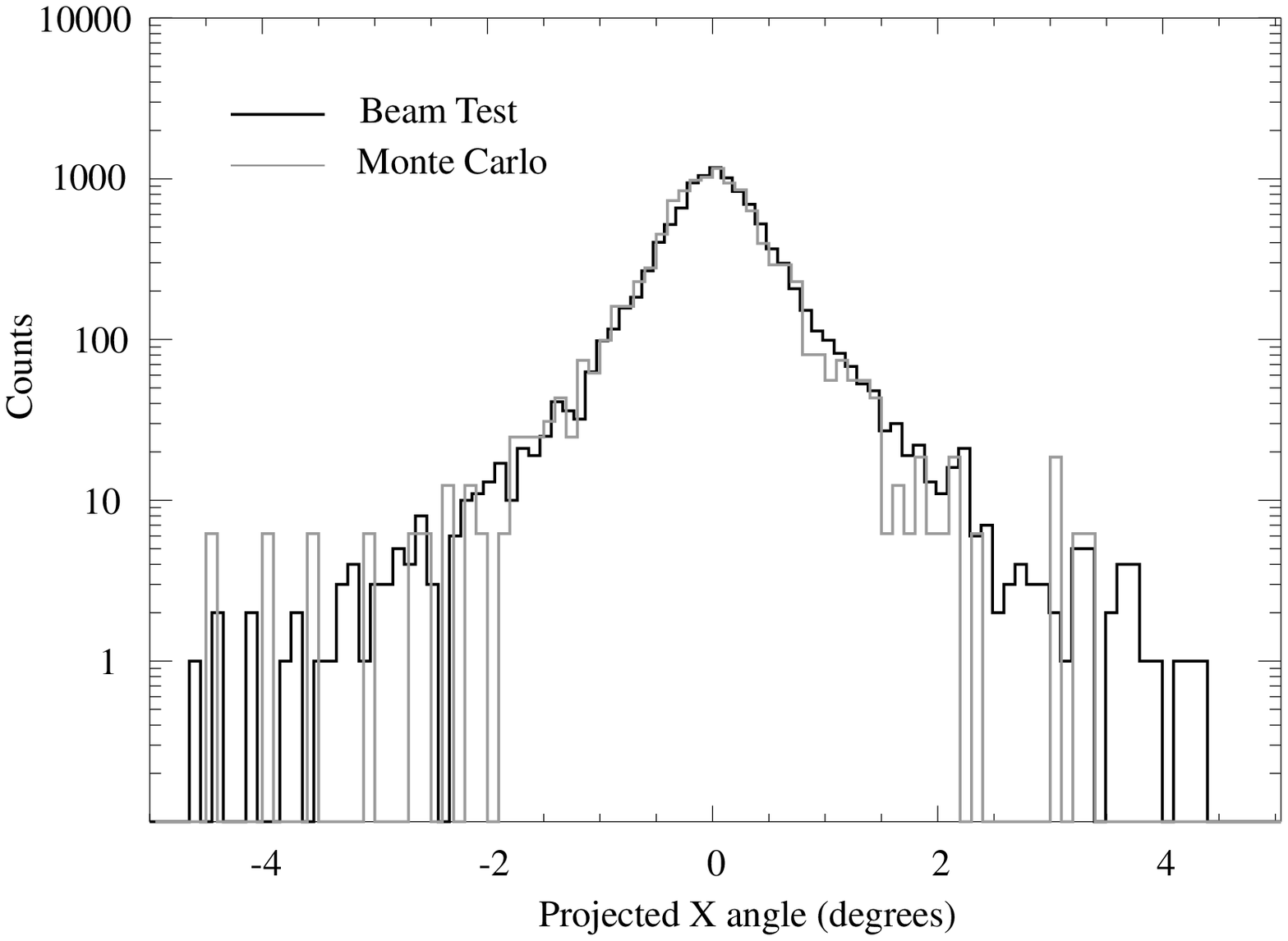}
\caption[PSF at 500--1000 MeV, 6\% RL, Stretch configuration]{\label{fourthpsf}
PSF for 500--1000 MeV \gammarays, with 6\% Pb radiators,
in the Stretch configuration.
}
\end{figure}

The principal characterization of the SSD tracker in the beam test is the Point Spread Function
(PSF), or the distribution of reconstructed directions given a collimated beam of \gammarays\ 
at a specific energy.  This distribution for beam test data is compared with that from
simulated data at several energy ranges, and for several configurations in \fig{firstpsf} --
\fig{fourthpsf}.  As can be seen, the real and simulated PSF are very
similar, indicating that the simulations do an admirable job of predicting the instrument
performance.

\section{PSF as a function of energy}

\begin{figure}[tbhp]
\centering
\includegraphics[width = 5.0in]{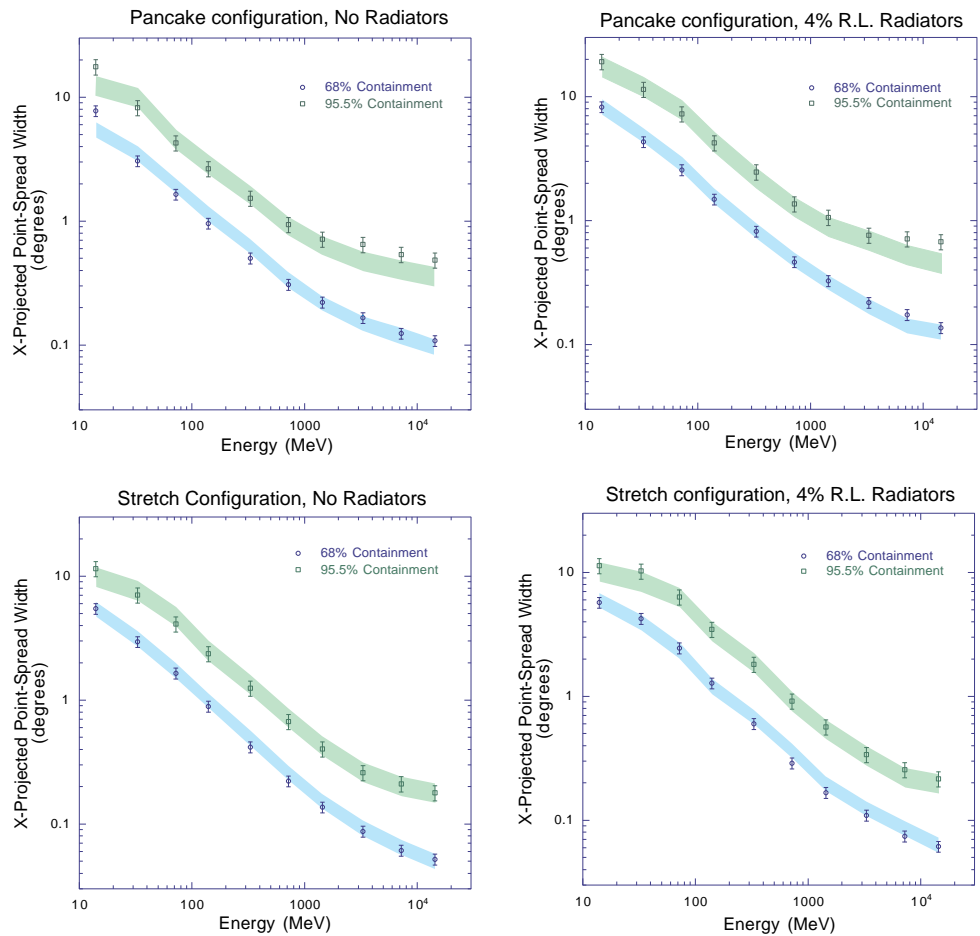}
\caption[PSF vs Energy in prototype instrument]{\label{fullpsfs}
PSF vs Energy for four different instrument configurations.
The Monte Carlo data are
represented by the shaded band.  Error bars are $2 \sigma$.
}
\end{figure}

In order to characterize the PSF as a function of energy, one might typically
fit a Gaussian to the distributions described above.  Because of the
non-Gaussian tails of the
distributions, however, it was decided to use the 68\% and 95\% containment
radii instead.
Using these two numbers to summarize the data from distributions like those
in \fig{firstpsf}, the plots in 
\fig{fullpsfs} were constructed.  Again, good agreement
is seen between the beam test data and the simulations.

\section{Conclusion}

The Fall 1997 beam test set out to demonstrate the validity of many components of the
\glast\ project: from the detector and readout technology of the three separate detectors,
to the generation of a \gammaray\ beam which can be used to calibrate the full
\glast\ instrument, to the simulations around which the detector design is being based.

As seen from the data shown in the previous sections, in all of these respects the beam test
was quite a success.  The \gammaray\ beam was well collimated, all three detectors
performed as expected, and the simulations of the \glast\ prototype instrument
proved to be quite accurate.

This verification of the \glast\ design and detector technology paves the way for the next
stage of the \glast\ project: a prototype tower to be constructed for
beam (and possibly flight) tests in the fall of this year.



\begin{singlespace}
\bibliography{Bibliography,glast,pulsars,grb,kalman}
\bibliographystyle{bbjthesis}
\end{singlespace}

\end{document}